\documentclass[final]{IEEEtran}
\usepackage[utf8]{inputenc}
\usepackage{pgfplots}
\pgfplotsset{compat=newest}
\usepackage{subfigure}
\usepackage{amsmath}
\usepackage{amsfonts}
\usepackage{graphicx}
\usepackage{cite}

\usepackage{bm}
\newcommand{\ma}[1]{\mathbf{ #1 }}         % matrix/vector

\newcommand{\compl}{\mathbb{C}}        % complex-valued numbers
\newcommand{\real}{\mathbb{R}}         % real-valued numbers

\usepackage{color}

\newcommand{\ignore}[1]{}

\ifCLASSINFOpdf

\else

\fi

\usepackage{algorithm}
\usepackage{algpseudocode}

\usepackage{url}
\newtheorem{lemma}{Lemma}
\newcommand{\MainFigureSizes}{0.7\columnwidth}
\newcommand{\revOmid}[1]{{\color[rgb]{0,0,0}#1}}
\newcommand{\response}[1]{{\color[rgb]{0,0,0}#1}}
\makeatletter

\begin{document}

\title{Hardware Impairments Aware Transceiver Design for Full-Duplex Amplify-and-Forward MIMO Relaying}
		\author{Omid Taghizadeh, \IEEEmembership{Student Member,~IEEE},~Ali Cagatay Cirik, \IEEEmembership{Member,~IEEE},~Rudolf Mathar, \IEEEmembership{Senior Member,~IEEE}
			\IEEEcompsocitemizethanks{
							\IEEEcompsocthanksitem O.~Taghizadeh and R.~Mather are with the Institute for Theoretical Information Technology, RWTH Aachen University, Aachen, 52074, Germany (email: \{taghizadeh,~mathar\}@ti.rwth-aachen.de).
				%\IEEEcompsocthanksitem A.~C.~Cirik is with the Institute for Digital Communications, School of Engineering, University of Edinburgh, Edinburgh EH9 3JL, U.K. (email:
				%a.cirik@ed.ac.uk).
						  %\IEEEcompsocthanksitem A.~C.~Cirik and L. Lampe are with the Department of Electrical and Computer Engineering, University of British Columbia, Vancouver, BC V6T 1Z4,
		%Canada  (email: \{cirik,~lampe\}@ece.ubc.ca).		
				\IEEEcompsocthanksitem A.~C.~Cirik is with the Department of Electrical and Computer Engineering, University of British Columbia, Vancouver, BC V6T 1Z4,
		Canada (email:~cirik@ece.ubc.ca).				
		\IEEEcompsocthanksitem Part of this work has been presented in ISWCS'16, the 2016 International Symposium on Wireless Communication Systems~\cite{TaYaCiMa16}.
			}} 

\maketitle

\begin{abstract}
In this work we consider a full-duplex (FD) and amplify-and-forward (AF) relay with multiple antennas, where hardware impairments of the FD relay are taken into account. Due to the inter-dependency of the transmit relay power and the residual self-interference in an AF-FD relay, we observe a \emph{distortion loop} that degrades the system performance when relay dynamic range is not high. In this regard, we analyze the relay function, and an optimization problem is formulated to maximize the signal to distortion-plus-noise ratio (SDNR) under relay and source transmit power constraints. Due to the problem complexity, we propose a gradient-projection-based (GP) algorithm to obtain an optimal solution. Moreover, a non-alternating sub-optimal solution is proposed by assuming a rank-1 relay amplification matrix, and separating the design of the relay process into multiple stages (MuStR1). The proposed MuStR1 method is then enhanced by introducing an alternating update over the optimization variables, denoted as AltMuStR1 algorithm. Numerical simulations show that compared to GP, the proposed (Alt)MuStR1 algorithms significantly reduce the required computational complexity at the expense of a slight performance degradation. Moreover, as the hardware impairments increase, or for a system with a high transmit power, the impact of applying a distortion-aware design is significant.
\end{abstract}
\IEEEpeerreviewmaketitle

\section{Introduction} \label{Intro}
\IEEEPARstart{F}{ULL-Duplex} (FD) operation, as a transceiver's capability to transmit and receive at the same time and frequency, is known with the potential to approach various requirements of future communication systems (5G), e.g., improved spectral efficiency and end-to-end latency \cite{HBCJMKL:14}. Nevertheless, such systems have been long considered to be practically infeasible due to the inherent self-interference. In theory, since each node is aware of its own transmitted signal, the interference from the loopback path can be estimated and suppressed. However, in practice this procedure is challenging due to the high strength of the self-interference channel compared to the desired communication path, up to 100 dB \cite{DMBS:12}. Recently, specialized self-interference cancellation (SIC) techniques \cite{Bharadia:14,YLMAC:11,BMK:13,khandani2013two} have provided an adequate level of isolation between transmit (Tx) and receive (Rx) directions to facilitate an FD communication and motivated a wide range of related applications, see, e.g., \cite{HBCJMKL:14,SSDBRW:14}. A common idea of the such SIC techniques is to attenuate the main interference paths in RF domain, i.e., prior to down-conversion, so that the remaining self-interference can be processed in the effective dynamic range of the analog-to-digital convertor (ADC) and further attenuated in the baseband, i.e., digital domain. While the aforementioned SIC techniques have provided successful demonstrations for specific scenarios, e.g., \cite{BMK:13}, it is easy to observe that the obtained cancellation level may vary for different realistic conditions. This mainly includes \emph{i)} aging and inaccuracy of the hardware components, e.g., ADC and digital-to-analog-convertor (DAC) noise, power amplifier and oscillator phase noise in analog domain, as well as \emph{ii)} inaccurate estimation of the remaining interference paths due to the limited channel coherence time. As a result, it is essential to take into account the aforementioned inaccuracies to obtain a design which remains efficient under realistic situations.  

In this work we are focusing on the application of FD capability on a classic relaying system, where the relay node has multiple antennas and suffers from the effects of hardware inaccuracy. An FD relay is capable of receiving the signal from the source, while simultaneously communicating to the destination. This capability, not only reduces the required time slots in order to accomplish an end to end communication, but also reduces the end to end latency compared to the known Time Division Duplex (TDD)-based half-duplex (HD) relays. In the early work by Riihonen. et. al. \cite{RWW:11} the relay operation with a generic processing protocol is modeled, and many insights have been provided regarding the multiple-antenna strategies for reducing the self-interference power. The design methodologies and performance evaluation for FD relays with decode-and-forward (DF) operation have been then studied, see e.g., \cite{DMBSR:12, XaZXMaXu:15 , ALRWW:14, ZZZPKV:14, RVRWW:15}, taking into account the effects of the hardware impairments, as well as channel estimation errors in digital domain. For the FD relaying systems with amplify-and-forward (AF) operation, single antenna relaying scenarios are studied in \cite{TaMa:AF:FD:14,TRCM:15,RWW:09_2,RTL:14,DRTSS:15,XBXL:13}. In the aforementioned works the effect of the linear inaccuracies in digital domain have been incorporated in \cite{RWW:09_2}, where the hardware imperfections from analog domain components have been addressed in \cite{TaMa:AF:FD:14,TRCM:15}, following the model in \cite{DMBS:12,DMBSR:12}, and in \cite{RTL:14} following the proposed model in \cite{FD_ExperimentDrivenCharact}. The work in \cite{RTL:14} is then extended by \cite{DRTSS:15} to enhance the physical layer security in the presence of an Eavesdropper.

While the aforementioned literature introduces the importance of an accurate transceiver modeling with respect to the effects of hardware impairments for an FD-AF relay, such works are not yet extended for the relays with multiple antennas. This stems from the fact that in an FD-AF relay, the inter-dependent behavior of the transmit power from the relay and the residual self-interference intensity results in a distortion loop effect, see Subsection II-C. The aforementioned effect results in a rather complicated mathematical description when relay is equipped with multiple antennas. As a result, related studies resort to simplified models to reduce the consequent design complexity. \revOmid{
%In \cite{KKMHPL:12, 4557197, 7378840, SKZYS:14, CP:12, ChunPark:12, SSWS:14, URW:15, 7558213, Taghizadeh2016, KKC:14, 7769216, 7515155} 
In \cite{KKMHPL:12, 4557197, 7378840, SKZYS:14, CP:12, ChunPark:12, SSWS:14, URW:15, 7558213} a multiple-antenna FD-AF relay system is studied where a perfect SIC is assumed; via estimating and subtracting the interference in the receiver \cite{KKMHPL:12, 4557197, 7378840}, or via spatial zero-forcing of the self-interference signal assuming that the number of transmit antennas exceeds the number of receive antennas at the relay \cite{SKZYS:14, CP:12, ChunPark:12, SSWS:14, URW:15, 7558213}. For the scenarios where the number of transmit antennas is not higher than the receive antennas, a general framework is proposed in \cite{7817896, 7936630}, assuming a fixed and known residual self-interference statistics, and in \cite{Taghizadeh2016, KKC:14}, where a perfect SIC\footnote{Residual self-interference is assumed to be buried in the thermal noise, following a known statistics.} is assumed via a combined analog/digital SIC scheme, on the condition that the self-interference power does not exceed a certain threshold. In \cite{7769216, 7515155} the residual self-interference signal is related to the transmit signal via a known and linear function, assuming a distortion-free hardware. A power adjustment method is proposed in \cite{7932472} for an FD-AF relay equipped with a massive antenna array, by considering the impact of limited resolution ADCs in the end-to-end performance. However, the impact of the relay's transmit/receive covariance on the residual self-interference is not considered.
%Nevertheless, it is assumed that the known self-interference signal is not canceled at the receiver, restricting the baseband SIC to the transmit/receive beamforming. 
To the best of the authors knowledge, the impact of the hardware distortions, as such extensively studied for FD-DF relaying systems \cite{DMBSR:12, XaZXMaXu:15 , ALRWW:14, ZZZPKV:14, RVRWW:15}, is not yet addressed for multiple antenna FD-AF relays.} 
\revOmid{\subsection{Contribution}%In contrast to the prior works \cite{KKMHPL:12, 4557197, 7378840, SKZYS:14, CP:12, ChunPark:12, SSWS:14, URW:15, 7558213, Taghizadeh2016, KKC:14, 7769216, 7497009,  7932472,7515155,7817896, 7936630} 
In this work, we study a multiple-input-multiple-output (MIMO) FD-AF relay, where the explicit impact of hardware distortions in the receiver and transmit chains are taken into account in the SIC process. Our goal is to enhance the instantaneous end-to-end performance via optimized linear transmit/receive strategies. The main contributions are as follows:
%In this work we study a multiple-input-multiple-output (MIMO) AF-FD relay, where the distortions resulting from hardware inaccuracies in the receiver and transmit chains are jointly taken into account. Our goal is to improve the end-to-end performance, in terms of signal-to-distortion-plus-noise ratio (SDNR), via optimized linear transmit/receive strategies. The main contributions of this paper are as follows:
\begin{itemize}
\item  Due to the joint consideration of hardware distortions in the receiver and transmit chains, we observe an inter-dependent behavior of the relay transmit covariance and the residual self-interference covariance in an FD-AF relay, i.e., the distortion loop effect. Note that this behavior may not be captured from the prior works based on simplified residual interference models, e.g., assuming a perfect SIC via estimation at the receiver [22]-[24], via transmit beamforming [25]-[30], assuming a known self-interference signal with perfect hardware [35], [36], or assuming a known (fixed) residual self-interference covariance [31]-[34]. In Section~III, the relay operation is analyzed under the effect of distortion loop, and the instantaneous end-to-end signal-to-distortion-plus-noise ratio (SDNR) is formulated in relation to the statistics of the noise and hardware impairments. 

%%%%based on simplified residual interference models due to e.g., assuming a known (fixed) residual interference statistics~\cite{7817896, 7936630,Taghizadeh2016, KKC:14}, neglecting the dependency of residual interference and the transmit/receive covariance \cite{7932472} or assuming a distortion-less hardware operation~\cite{KKMHPL:12, 4557197, 7378840, SKZYS:14, CP:12, ChunPark:12, SSWS:14, URW:15, 7558213, 7769216, 7515155}. In Section~III, the relay operation is analyzed under the effect of distortion loop, and the instantaneous end-to-end signal-to-distortion-plus-noise ratio (SDNR) is formulated in relation to the statistics of the noise and hardware impairments. 
%is then formulated under thid effect, to theIn Section~II and Section~III the relay operation is analyzed by taking into account the effects of transmit and receiver chain distortions. In particular, the SDNR perfoemance of the relaying system is obtained,  the inter-dependednt behaviour of the residual   and the resulting distortion loop effect is elaborated. Furthermore, an SDNR maximization problem is formulated.
           
\item Building on the obtained analysis, we propose linear transmit/receive strategies with the intention of maximizing the SDNR. The instantaneous CSI is utilized to control the impact of distortion, and to enhance the quality of the desired signal. This is in contrast to \cite{KKMHPL:12, 4557197, 7378840, SKZYS:14, CP:12, ChunPark:12, SSWS:14, URW:15, 7558213, Taghizadeh2016, KKC:14, 7817896, 7936630} where the dependency of the distortion statistics to the intended transmit/receive signal is ignored, to \cite{7932472} where a fixed transmit/receive strategy is assumed based on maximum ratio combining/transmission, or to \cite{7769216, 7515155} where the \emph{residual self-interference channel} is assumed to be a known design parameter. Note that the latter assumption is not practical, since the residual self-interference, by definition, is unknown to the receiver\footnote{In the other words, if the \emph{residual self-interference channel} were known, the residual self-interference signal could be subtracted in the receiver, hence, resulting in a perfect SIC.}. In this regard, an SDNR maximization problem is formulated which shows an intractable mathematical structure due to impact of the distortion loop. A gradient-projection (GP) based solution is then proposed in Section~IV to act as a benchmark for the achievable performance, however, imposing a high computational complexity.  

\item In order to reduce the design computational complexity, a sub-optimal Multi-Stage Rank-1 (MuStR1) solution is introduced in Section~V, by assuming a rank-1 relay amplification matrix and separating the design of the relay process into multiple stages. In this regard, a non-alternating algorithm is proposed by locally maximizing the resulting SDNR for each stage. Moreover, the performance of MuStR1 is improved by introducing an alternating update (AltMuStR1) at the cost of a slightly higher computational complexity compared to MuStR1. Similar to the previous parts, this approach differs from the rank-1 FD-AF relaying schemes proposed in \cite[Subsection~3.2]{Taghizadeh2016} and \cite[Section~III]{7558213}, where the impact of distortions are not considered in the design of transmit/receive strategies.   
%\item In Section~VI a special case of the defined system is studied, where the FD relay is implemented via a distributed antenna system in which the inter-relay interference can be canceled. A good example of such system is a setup where relay nodes are equipped with directive antennas, e.g., a satellite relay network, or a distributed relaying setup in mm-Wave domain where inter-relay interference is respectively canceled due to high directivity and path loss. An alternating optimization is then proposed, with a guaranteed convergence, where in each step the corresponding optimization problem is solved to optimality.  
%
%\itam In Section~VI, a sub-optimal design.. 
%
%optimization framework is proposed via quadratic approximation of the relay transmit covariance in each step. A fast convergence is observed via numerical simulations. 
%
%\item An intuitive design is then proposed in Section~VI which provides a non-alternating solution with considerable lower computation complexity. This method is then extended to provide a better performance at the expense of higher optimization iterations.  
%
%\item In Section~VII, an optimal operation of an equivalent DF relaying system is studied. This is valuable since the effect of the distortion loop can be observed by comparing the AF and DF relaying performance. 
%
%\item Finally, the behavior of the aforementioned strategies are evaluated for various system parameters via numerical simulations and the main conclusions are summarized, respectively in Section~VIII and Section~IX.
%
\end{itemize} 
Numerical simulations show that for a system with a small thermal noise variance, or a high power or transceiver inaccuracy, the application of a distortion-aware design is essential.}

\subsection{Mathematical Notation:}
Throughout this paper, column vectors and matrices are denoted as lower-case and {upper-case} bold letters, respectively. The rank of a matrix, {expectation, trace}, transpose, conjugate, Hermitian transpose, determinant and Euclidean norm are denoted by ${\text{rank}}(\cdot),\; \mathbb{E}(\cdot), \; {\text{ tr}}(\cdot), \;   (\cdot)^{ T}$, $(\cdot)^{*}$, $(\cdot)^{H}, \; |\cdot|, \; ||\cdot||_{2}$, respectively. The Kronecker product is denoted by $\otimes$. The identity matrix with dimension $K$ is denoted as ${\ma I}_K$ and ${\text{vec} }(\cdot)$ operator stacks the elements of a matrix into a vector, and $(\cdot)^{-1}$ represents the inverse of a matrix. The sets of real, real and positive, complex, natural, and the set $\{1 \ldots K\}$ are respectively denoted by $\real$, $\real^+$, $\compl$, $\mathbb{N}$ and $\mathbb{F}_K$. $\left\lfloor \mathbf{A}_i \right \rfloor_{i\in\mathbb{F}_K}$ denotes a tall matrix, obtained by stacking the matrices $\mathbf{A}_i,~i\in\mathbb{F}_K$. {The set of all positive semi-definite matrices is denoted by $\mathcal{H}$.} $\bot$ represents statistical independence. $\ma{\lambda}_{\text{max}} (\ma{A})$ calculates the dominant eigenvector of $\mathbf{A}$. $x^\star$ is the value of the variable $x$ at optimality.
%sets: N, F_K
%\lambda_max
%Add assumptions part in the relay analysis
\section{System Model} \label{sec_systemmodel}

\begin{figure*}[!t]
% ensure that we have normalsize text
\normalsize
        \includegraphics[angle=0,width=2.05\columnwidth]{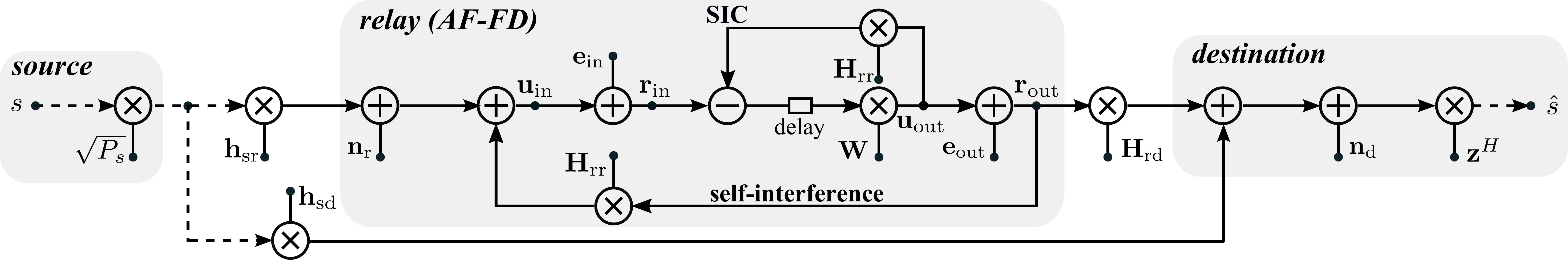}
% Restore the current equation number.
%\setcounter{equation}{19}
% IEEE uses as a separator
\hrulefill
% The spacer can be tweaked to stop underfull vboxes.
    \caption{{{The signal model in an amplify-and-forward FD MIMO relay. The impact of hardware inaccuracies form the transmitter ($\ma{e}_{\text{out}}$) and receiver ($\ma{e}_{\text{in}}$) chains is observable on the relay process. The bold arrows represent the vector signals while the dashed arrows represent the scalars. See Section~\ref{sec_systemmodel} for a detailed description.}} }
\vspace{-0pt}
\end{figure*}

We investigate a system where a single-antenna HD source communicates with an HD destination node equipped with $M_{\text{d}}$ antennas, with a help of an FD relay. The relay is assumed to have $M_{\text{t}}$ ($M_{\text{r}}$) transmit (receive) antennas, and operates in AF mode. The channels between the source and the relay, between the relay and the destination, and between the source and the destination are denoted as $\ma{h}_{\text{sr}} \in \compl^{M_{\text{r}}}$, $\ma{H}_{\text{rd}} \in \compl^{M_{\text{d}} \times M_{\text{t}}}$, and $\ma{h}_{\text{sd}} \in \compl^{M_{\text{d}}}$, respectively. The self-interference channel, which is the channel between the relay's transmit and receive ends is denoted as $\ma{H}_{\text{rr}} \in \compl^{M_{\text{r}} \times M_{\text{t}}}$. All channels are following the flat-fading model.
 
\subsection {Source-to-Relay Communication}

The relay continuously receives and amplifies the received signal from the source, while estimating and subtracting the loopback self-interference signal from its own transmitter, see Fig.~1. The received signal at the relay is expressed as 
\begin{align} \label{eq_model_r_in}
\ma{r}_{{\text{in} }}  =   \underbrace{\ma{h}_{{\text{sr}}}  \sqrt{P_{\text{s}}} s  + \ma{H}_{{\text{rr}}} \ma{r}_{{\text{out}}} + \ma{n}_{\text{r} }  }_{=: \ma{u}_{\text{in}}} +  \ma{e}_{{\text{in}}} , 
%& \underbrace{ { h}_{{\rm sr},k} \cdot s + {\tilde{h}}_{{\rm rr},k} \cdot {r}_{{\rm out},k} + {\delta}_{k} \cdot {r}_{{\rm out},k} + {m}_{k} }_{ =: { u}_{{\rm in},k} } +  {e}_{{\rm in},k}, 
%&{\ma r}_{\rm y}  = \underbrace{ {\ma h}_{\rm sr} \cdot s + {\ma H}_{\rm rr} \cdot {\ma r}_{\rm x}  + \ma{n}_{\rm r} }_{{\ma u}_1 }  +  \ma{e}_{1} ,
\end{align}
where $\ma{r}_{{\text{in}}}\in \compl^{M_{\text{r}}}$ and $\ma{r}_{{\text{out}}} \in \compl^{M_{\text{t}}}$ respectively represent the received and transmitted signal from the relay and $\ma{n}_{\text{r}} \sim \mathcal{CN} \left( \ma{0}, \sigma_{\text{nr}}^2 \ma{I}_{{M_{\text{r}}}} \right)$ represents the zero-mean additive white complex Gaussian (ZMAWCG) noise at the relay. The transmitted data symbol from the source is denoted as $s \in \compl$, $\mathbb{E}\{|s|^{2}\}=1$. $P_{\text{s}} \in \real^+$ is the source transmit power and ${\ma{u}_{{\text{in}}}} \in \compl^{M_{\text{r}}}$ represents the {\emph{undistorted}} received signal at the relay. 

The receiver distortion, denoted as \mbox{$\ma{e}_{{\text{in}}} \in \compl^{M_{\text{r}}}$}, represents the combined effects of receiver chain impairments, e.g., limited ADC accuracy, oscillator phase noise, low-noise-amplifier (LNA) distortion \cite{DMBSR:12}. 
%For a detailed elaboration on the used distortion model please refer to \cite{DMBS:12, DMBSR:12}, and system measurements in \cite{MITTX:08,MITRX:05}. 
Please note that while the aforementioned impairments are usually assumed to be ignorable for an HD transceiver, they play an important role in our system due to high strength of the self-interference path. The known, i.e., distortion-free, part of the self-interference signal is then suppressed in the receiver by utilizing the recently developed SIC techniques in analog and digital domains, e.g., \cite{Bharadia:14,BMK:13}. The remaining signal is then amplified to constitute the relay's output: 
\begin{align}
 \ma{r}_{{\text{out}}} &=  \ma{u}_{\text{out}}   +  \ma{e}_{{\text{out}}}, \;\; \ma{u}_{{\text{out}}} (t) = \ma{W} \ma{r}_{{\text{supp}}} (t - \tau),  \label{eq_model_r_out}\\
 \ma{r}_{{\text{supp}}} &= \ma{r}_{{\text{in}}} - {\ma{H}}_{{\text{rr}}} \ma{u}_{{\text{out}}}, \label{eq_model_r_supp}
\end{align}
where $\ma{r}_{{\text{supp}}} \in \compl^{M_{\text{r}}}$ and $\ma{W} \in \compl^{{M_{\text{t}}} \times {M_{\text{r}}} }$ respectively represent the interference-suppressed version of the received signal and the relay amplification matrix, $t \in \real^+$ represents the time instance\footnote{The argument indicating time instance, i.e., $t$, is dropped for simplicity for signals with a same time reference.}, and $\tau \in \real^+$ is the relay processing delay, see Subsection~\ref{model_remarks_delay}. The \emph{intended} transmit signal is denoted as $\ma{u}_{{\text{out}}} \in \compl^{M_{\text{t}}}$. Similar to the defined additive distortion in the receiver chains, the combined effects of the transmit chain impairments, e.g., limited DAC accuracy, oscillator phase noise, power amplifier noise, is denoted by \mbox{$\ma{e}_{{\text{out}}} \in \compl^{M_{\text{r}}}$}. Furthermore, in order to take into account the transmit power limitations we impose 
\begin{align}
&  \mathbb{E} \{ \|\ma{r}_{{\text{out}}}\|_2^2  \}  \leq P_{\text{r,max}}, \; P_{\text{s}} \leq P_{\text{s,max}},  \label{eq_model_P_max}
\end{align}
where $P_{\text{r,max}}$ and $P_{\text{s,max}}$ respectively represents the maximum transmit power from the relay and from the source.

\subsection{Distortion signal statistics}
\revOmid{The impact of hardware elements inaccuracy in each chain is modeled as additive distortion terms, following the FD transceiver model proposed in \cite{DMBS:12} and widely used in the context of FD system design and performance analysis, e.g., \cite{DMBSR:12, XaZXMaXu:15 , ALRWW:14, ZZZPKV:14, RVRWW:15}. The proposed model in \cite{DMBS:12} is based on the following three observations. Firstly, the collective distortion signal in each transmit/receive chain can be approximated as an additive zero-mean Gaussian term \cite{MITTX:98, MITRX:05, MITTX:08}. Secondly, the variance of the distortion signal is proportional to the power of the intended transmit/received signal. And third, the distortion signal is statistically independent to the intended transmit/receive signal at each chain, and for different chains. Please note that the the statistical independence of distortion elements holds also for an advanced implementation of an FD transceiver, assuming a high signal processing capability. This is since, any known correlation can be constructively used, and hence eliminated, to reduce the residual self-interference. However, the linear dependence of the remaining distortion signal to the signal strength varies for different SIC techniques and is considered as an approximation, see \cite[Subsections~C]{DMBS:12} for further elaboration on the used model. In the defined relaying system it is expressed as}
 %This is in particular ellaborated in \cite[]{}, considering the hardware inaccuracy in transmit chains, and in \cite[]{}   for a detailed model ellaboration and , and  .    see  the impact of  see \cite[]{DMBS:12}  
%following a zero-mean complex Gaussian process, with a variance proportional to the transmit (receive) power in the corresponding chain. 
%Following the recent studies regarding the characteristics and applications of FD transceivers (see, e.g., \cite{DMBS:12, DMBSR:12} regarding the FD transceiver modeling and applications, and \cite{MITTX:98, MITTX:08, MITRX:05, FD_ExperimentDrivenCharact} regarding the experiments on transceiver behavior) elements of the distortion signals can be closely approximated as spatially uncorrelated, zero-mean complex random process with a Gaussian distribution. This is expressed as 
\begin{align} 
\ma{e}_{{\text{in}}}  \sim \mathcal{CN} & \Big( \ma{0},  \beta \text{diag} \Big( \mathbb{E} \left\{\ma{u}_{\text{in}}\ma{u}_{\text{in}}^{{H}}\right\} \Big) \Big), \nonumber \\
&\ma{e}_{{\text{in}}} ({t}) \bot \ma{e}_{{\text{in}}}({{t}^{'}}), \;\; \ma{e}_{{\text{in}}}({t}) \bot \ma{u}_{{\text{in}}}({t}), \label{eq_model_e_in} 
\end{align}
and
\begin{align}
\ma{e}_{{\text{out}}}  \sim \mathcal{CN} & \Big( \ma{0},  {\revOmid{\kappa}} \text{diag}\Big(\mathbb{E}\left\{\ma{u}_{\text{out}}\ma{u}_{\text{out}}^{{H}}\right\} \Big) \Big), \nonumber \\
&\ma{e}_{{\text{out}}} ({t}) \bot \ma{e}_{{\text{out}}}({{t}^{'}}), \;\; \ma{e}_{{\text{out}}}({t}) \bot \ma{u}_{{\text{out}}}({t}), \label{eq_model_e_out}
\end{align}
where ${t} \neq {{t}^{'}}$ and $\beta ,\kappa \in \real^{+}$ are the receive and transmit distortion coefficients which respectively relate the \emph{undistorted} receive and transmit signal covariance to the covariance of the corresponding distortion. It is worth mentioning that the values of $\kappa, \beta$ depend of the implemented SIC scheme, and reflect the quality of the cancellation. \revOmid{For instance, for an FD massive MIMO system where analog cancelers are not used due to complexity, and the resolution of the ADC/DAC are limited to reduce the cost, the values of $\kappa,\beta$ can be determined by the used quantization bits, e.g., $\kappa \approx -6\times b$ dB, for a uniform DAC quantization with $b$ bits. However, in general, the choice of $\kappa, \beta$ depend on the implemented SIC and the used analog circuitry, e.g., the number of delay taps implemented in [6, Subsection~3.1].} Note that the defined statistics in (\ref{eq_model_e_in}), (\ref{eq_model_e_out}) indicate that unlike the traditional additive white noise model, a higher transmit (receive) signal power results in a higher transmit (receive) distortion intensity in the corresponding chain. As we will further elaborate, this effect plays an important role in the performance of an FD-AF relay. For more discussions on the used distortion model please see \cite{DMBS:12, DMBSR:12, ALRWW:14, XaZXMaXu:15}, and the references therein.

%\begin{align}
%\ma{e}_{{\rm in}}  \sim \mathcal{CN \left\{ \ma{0}, \beta \text{diag}\Big(\mathbb{E}\left\{\ma{u}_{\rm{in}}\ma{u}_{\rm{in}}^{{H}}\right \} \Big) \right\}, \label{eq_model_e_in}\\ 
%\ma{e}_{{\rm out}}  \sim \mathcal{CN} \left \{ \ma{0}, \gamma \text{diag}\Big(\mathbb{E}\left\{\ma{u}_{\rm{out}}\ma{u}_{\rm{out}}^{{H}}  \right \}  \Big) \right\}, \label{eq_model_e_out}
%\end{align}
%where $\beta ,\gamma \in \real^{+}$ are the receive and transmit distortion coefficients which respectively relate the \emph{intended} receive and transmit signal power to the variance of the corresponding distortion. Please note that the defined statistics in (\ref{eq_model_e_in}),(\ref{eq_model_e_out}) convey the intuition that unlike the traditional additive white noise model, the higher transmit/receive signal power result in higher transmit/receive distortion intensity in the corresponding chain. For further elaborations on the used distortion model please see \cite{DMBS:12, DMBSR:12, ALRWW:14,XaZXMaXu:15}, and the references therein.   

\subsection {Relay-to-destination communication}
The transmitted signal from the relay node passes through the relay to destination channel and constitutes the received signal at the destination:
\begin{align}
 \ma{y} &=  \ma{H}_{{\text{rd}}} \ma{r}_{{\text{out}}} +  \ma{h}_{{\text{sd}}}  \sqrt{P_{\text{s}}} s + \ma{n}_{\text{d}}, \;\; \hat{s}(t-\tau)= \ma{z}^{{H}}\ma{y}(t), \label{eq_model_y}
\end{align}
where $\ma{y} \in \compl^{M_{\text{d}}}$ is the received signal at the destination, and $\ma{n}_{\text{d}}\sim \mathcal{CN} \left( \ma{0}, \sigma_{\text{nd}}^2{\ma I}_{M_{\text{d}}}  \right)$ is the ZMAWCG noise. The linear receiver filter and the estimated received symbol is denoted as \mbox{$\ma{z}\in\compl^{M_{\text{{d}}}}$} and $\hat{s}$, respectively. \revOmid{Please see Table~\ref{tab_notations} for the list of used signal notations and system parameters.}
\begin{table}[!t]  
	\renewcommand{\arraystretch}{0.8}
  \caption{\revOmid{Used signal notations and system parameters}}\label{tab_notations}
  \centering
  \revOmid{\begin{tabular}[t]{||c|l||}
	\hline
   Deterministic Param. & Description \\
   \hline	 
	 $\mathbb{S}_d$    &  Set of all deterministic parameters   \\
	 $\ma{H}_{\text{rr}}$    &  Instantaneous relay-relay channel  \\
	 $\ma{H}_{\text{rd}}$    &  Instantaneous relay-destination channel  \\
	 $\ma{h}_{\text{sr}} (\ma{h}_{\text{sd}})$    &  Instantaneous source-relay (destination) channel  \\
	 %$\ma{h}_{\text{sr}}$    &  Instantaneous source-destination channel  \\
	 $P_{\text{s,max}} (P_{\text{r,max}})$      &  Maximum transmit power at the source (relay)   \\
	 $\sigma_{\text{nr}}^2 (\sigma_{\text{nd}}^2)$      &  Thermal noise variance at the relay (destination)  \\
	 %$P_{\text{r,max}}$      &    \\	
	 $\kappa (\beta)$        & Distortion coefficients for the transmit (receive) chains   \\
	 \hline
   Random Param. & Description \\
 	 \hline
	 $\mathbb{S}_r$    &  Set of all random parameters   \\
	 $s$      &  Transmit data symbol from source  \\	
	 $\ma{n}_{\text{r}} (\ma{n}_{\text{d}})$      &  Thermal noise at the relay (destination)   \\	
	 $\ma{e}_{\text{in}} (\ma{e}_{\text{out}})$      &  Receive (transmit) distortion at the relay  \\	
   %Design var. & Description \\
	 %$P_{\text{s}}$      &    \\
	 %$P_{\text{s}}$      &    \\
	\hline
  \end{tabular}}
\end{table} 

%In order to take into account the transmit power limitations in the system we impose the constraints
%\begin{align}
%\mathbb{E} \{ \|\ma{r}_{{\rm out}}\|_2^2  \}  \leq P_{\rm{r,max}}, \;\; P_{\text{s}} \leq P_{\rm{s,max}} \label{eq_model_P_max}
%\end{align}
%where $P_{\rm{r,max}}$ and $P_{\rm{s,max}}$ respectively represent the maximum transmit power from the source and the relay node. 
%\mbox{Table I} provides the description of the signal notations to better communicate the defined model. 

\subsection{Distortion loop} \label{discussion:distortionLoop}
%As mentioned
%In this part we provide an intuitive description of the aforementioned distortion loop, following the defined system model, which makes the subsequent arguments in the paper more clear. 
As the transmit power from the relay increases, the power of the error components increase in all receiver chains, see (\ref{eq_model_r_in}) in connection to (\ref{eq_model_e_in})-(\ref{eq_model_e_out}). On the other hand, these errors are amplified in the relay process and further increase the relay transmit power, see (\ref{eq_model_r_in}) in connection to (\ref{eq_model_r_supp}) and (\ref{eq_model_r_out}). The aforementioned effect causes a loop which signifies the problem of residual self-interference for the relays with AF process. In the following section, this impact is analytically studied and an optimization strategy is proposed in order to alleviate this effect.

\subsection{Remarks} \label{model_remarks}
\revOmid{\subsubsection{CSI estimation} 
In this work we assume that the channel state information (CSI), representing linear signal dependencies in the system, are known by dedicating an adequately long training sequence at the relay. Therefore, the studied framework serves best for the scenarios with a stationary channel where long training sequences can be utilized, e.g., relay channel in a static backhaul link with a directive line-of-sight (LOS) connection \cite{7306534}. Note that the acquisition of an accurate CSI is challenging due to noise, interference, as well as the impact of hardware impairments. However, an effective estimation method is presented in \cite[Subsection~III.A]{DMBSR:12} for an FD relaying setup in the presence of hardware impairments\footnote{\revOmid{A two-phase estimation is suggested to avoid interference; first, source transmits the pilot where relay is silent, thereby estimating the source-relay and source-destination channels, and then relay transmits pilot and source remain silent, hence estimating the self-interference and relay-destination channels.}}. It is observed that an accurate CSI can be obtained on the condition of employing a sufficiently long training sequence, see \cite[Equation~(9)]{DMBSR:12}, and also \cite[Equation~(10)]{DMBS:12} for a similar argument. For the scenarios where the CSI can not be accurately obtained, the results of this paper can be treated as theoretical guidelines on the effects of hardware impairments, if CSI were accurately known. }
%, and an inter-satellite relay channel where the source and destination nodes are known and follow a predictable movement, see \cite[Chaper~3]{ha1990digital}, \cite{lakhzouri2003extended}. For the scenarios where the CSI can not be accurately obtained, the results of this paper can be treated as theoretical guidelines on the effects of hardware impairments, if linear signal dependencies were correctly known. 
\revOmid{\subsubsection{Hardware impairments}
Compared to many HD scenarios the impact of hardware impairments is severe in FD transceivers due to the strong self-interference, see, e.g., specifications of SIC for 802.11ac PHY \cite{BMK:13}. This is since on one hand, the distortions originating from the transmit chains pass through a strong self-interference channel and become significant. On the other hand, the receiver chains are more prone to distortion due to the high-power received signal. In this work, we focus on the impact of hardware impairments for the FD relay transceiver, and otherwise model the inaccuracies as an additive thermal noise.} 
\subsubsection{Direct link}
In this work, we assume that the direct link is weak and consider the source-destination path as a source of interference, similar to \cite{RWW:09_2, DMBSR:12}. For the scenarios where the direct link is strong, it is shown in \cite{7105858} that the receiver strategy can be gainfully updated as a RAKE receiver \cite{864017} to temporally align the desired signal in the direct and relay links. This can be considered as a future extension of the current work.       
\subsubsection{Processing delay} \label{model_remarks_delay}
The relay output signals, i.e., $\ma{u}_{{\text {out}}}$ and $\ma{r}_{{\text {out}}}$, are generated from the received signals with a relay processing delay $\tau$, see (\ref{eq_model_r_out}). This delay is assumed to be long enough, e.g., more than a symbol duration, such that the source signal is decorelated, i.e., $s(t) \bot s(t-\tau)$ \cite{7105858, 5961159}. The zero-mean and independent statistics of the samples from data signal, i.e., $s(t)$ and $s(t-\tau)$, as well as the noise and distortion signals, are basis for the analysis in the following section.  

\section{Performance Analysis for MIMO AF Relaying with Hardware Impairments}
In this part, we analyze the end-to-end performance as a function of the relay amplification matrix, i.e., $\ma{W}$, receive linear filter at the destination, i.e., $\ma{z}$, as well as the transmit power from the source, $P_{\text{s}}$. By incorporating (\ref{eq_model_r_in}) and (\ref{eq_model_e_in})-(\ref{eq_model_e_out}) into (\ref{eq_model_r_out}) and (\ref{eq_model_r_supp}) we have
\begin{align} \label{eq_loop_U_out}
\ma{Q} = \ma{W}  \Big(   P_{\text {s}} \ma{ h}_{{\text {sr}}}\ma{ h}_{{\text {sr}}}^{{H}} &+ \sigma_{\text{nr}}^2\ma{I}_{{M_{\text {r}}}} +  \mathbb{E}\left\{\ma{e}_{\text{in}}\ma{e}_{\text{in}}^{{H}}\right \}  \nonumber \\ &\hspace{-16mm}+ {\ma{H}}_{{\text {rr}}}\mathbb{E} \{ \ma{e}_{{\text {out}}} \ma{e}_{{\text {out}}}^{{H}} \} {\ma{H}}_{{\text {rr}}}^{{H}}   \Big) \ma{W}^{{H}} \nonumber \\
= \ma{W}  \Big(   P_{\text {s}} \ma{ h}_{{\text {sr}}}\ma{ h}_{{\text {sr}}}^{{H}} & + \sigma_{\text{nr}}^2\ma{I}_{{M_{\text {r}}}} + \beta \text{diag}\big(\mathbb{E}\left\{\ma{u}_{\text{in}}\ma{u}_{\text{in}}^{{H}}\right \} \big) \nonumber \\ &\hspace{-16mm}+ \kappa  {\ma{H}}_{{\text {rr}}}\text{diag}\big(\ma{Q} \big) {\ma{H}}_{{\text {rr}}}^{{H}}   \Big) \ma{W}^{{H}},
 %& = \left|a_k \right|^2  \bigg( P_s \left|  { h}_{{\text sr},k} \right|^2 + M_k +  \mathcal{E} \{ {u}_{{\text in},k} {u}_{{\text in},k}^{*} \}  \beta_k \nonumber \\  &+ \mathcal{E} \{ {u}_{{\text out},k} {u}_{{\text out},k}^{*} \} \cdot \left( \left|{ h}_{{\text rr},k} \right|^2 \kappa_k + (1 + \kappa_k) |\delta_k|^2  \right)  \bigg), 
\end{align}
%where above identities hold considering $\beta,\kappa \ll 1$ and the terms including second (and higher) order $\beta,\kappa$ can be ignored. 
where $\ma{Q} \in \mathcal{H}$ is the covariance matrix of the undistorted transmit signal from the relay, i.e., $\ma{Q}:=\mathbb{E} \{ \ma{u}_{{\text{out}}} \ma{u}_{{\text{out}}}^{{H}} \}$. Furthermore, the undistorted receive covariance matrix can be formulated from (\ref{eq_model_r_in})-(\ref{eq_model_r_supp}) as 
\begin{align} \label{eq_loop_U_in}
 \mathbb{E} \{ \ma{u}_{{\text {in}}} \ma{u}_{{\text {in}}}^{{H}} \} & =  P_{\text {s}} \ma{ h}_{{\text {sr}}}\ma{ h}_{{\text {sr}}}^{{H}} + \sigma_{\text{nr}}^2\ma{I}_{{M_{\text {r}}}} + {\ma{H}}_{{\text {rr}}}\mathbb{E} \{ \ma{r}_{{\text{out}}} \ma{r}_{{\text {out}}}^{{H}} \} {\ma{H}}_{{\text {rr}}}^{{H}}.
 %\label{eq:approximation_power}
\end{align}
It is worth mentioning that due to the proximity of the Rx and Tx antennas on the FD device, the loopback self-interference signal is much stronger than the desired signal which is coming from a distant location, and hence constitutes the principle part in (\ref{eq_loop_U_in}). By recalling (\ref{eq_model_r_out}) and (\ref{eq_model_e_out}) the relay transmit covariance matrix can be formulated as 
\begin{align} \label{eq_loop_Q}
\mathbb{E} \{ \ma{r}_{{\text{out}}} \ma{r}_{{\text{ out}}}^{{H}} \}  = \ma{Q}+ \kappa\text{diag}\left( \ma{Q} \right),  
\end{align}
and consequently from (\ref{eq_loop_U_out}) and (\ref{eq_loop_U_in}) it follows
\begin{align}\label{eq_loop_U_out_2}
\ma{Q}   = \ma{W}  \mathcal{R}\left( \ma{Q}  \right) \ma{W}^H ,
%&  \approx \ma{W}  \Big(  P_{\text s} \ma{ h}_{{\text sr}}\ma{ h}_{{\text sr}}^{{H}} + \sigma_{\text{nr}}^2\ma{I}_{{M_{\text r}}} + \beta \text{diag}\left({\ma{H}}_{{\text rr}}  \ma{Q}  {\ma{H}}_{{\text rr}}^{{H}}\right) \nonumber \\ & + \kappa {\ma{H}}_{{\text rr}}\text{diag}\left(\ma{Q} \right) {\ma{H}}_{{\text rr}}^{{H}} \Big) \ma{W}^H, \label{eq_loop_U_out_2_approx}
\end{align} 
where 
\begin{align}\label{eq_loop_U_out_R}
\mathcal{R}\left( \ma{Q}  \right)  := P_{\text{s}} \ma{ h}_{{\text {sr}}}\ma{ h}_{{\text{ sr}}}^{{H}} & + \sigma_{\text{nr}}^2\ma{I}_{{M_{\text{ r}}}} + \beta \text{diag} \left( P_{\text {s}} \ma{ h}_{{\text {sr}}}\ma{ h}_{{\text {sr}}}^{{H}} + \sigma_{\text{nr}}^2\ma{I}_{{M_{\text {r}}}} \right) \nonumber \\
& + \beta \text{diag}\Big({\ma{H}}_{{\text {rr}}} \Big( \ma{Q} + \kappa\text{diag}\left( \ma{Q} \right) \Big) {\ma{H}}_{{\text {rr}}}^{{H}}\Big)  \nonumber \\
&+  \kappa {\ma{H}}_{{\text {rr}}}\text{diag}\left(\ma{Q} \right) {\ma{H}}_{{\text{ rr}}}^{{H}}.
%&  \approx \ma{W}  \Big(  P_{\text s} \ma{ h}_{{\text sr}}\ma{ h}_{{\text sr}}^{{H}} + \sigma_{\text{nr}}^2\ma{I}_{{M_{\text r}}} + \beta \text{diag}\left({\ma{H}}_{{\text rr}}  \ma{Q}  {\ma{H}}_{{\text rr}}^{{H}}\right) \nonumber \\ & + \kappa {\ma{H}}_{{\text rr}}\text{diag}\left(\ma{Q} \right) {\ma{H}}_{{\text rr}}^{{H}} \Big) \ma{W}^H, \label{eq_loop_U_out_2_approx}
\end{align}
%where the terms including second order distortion coefficients are ignored similar to \cite[Appendix B]{DMBSR:12}. Note that the approximation in (\ref{eq_loop_U_out_2_approx}) follows as i) the self-interference channel is much stronger than the HD links, e.g., up to 100 dB according to \cite{DMBS:12}, and ii) the distortion signals are much weaker compared to the desired transmit signal, i.e., $\beta \ll 1, \; \kappa \ll 1$, see \cite[Figure~2]{BMK:13} where the transmitter noise is measured 60 dB below the desired transmit signal level. As a result, the distortion components are only significant as they pass through the strong self-interference channel and can be ignored otherwise, as it is a usual assumption for the traditional (HD) links \cite{}.\footnotemark[2]        
 %\\
%& \approx (1+\kappa) \ma{W}  \Big(   P_{\text s} \ma{ h}_{{\text sr}}\ma{ h}_{{\text sr}}^{{H}} + \sigma_{\text{nr}}^2\ma{I}_{{M_{\text r}}} + \beta \text{diag}\big({\ma{H}}_{{\text rr}} \ma{Q} {\ma{H}}_{{\text rr}}^{{H}}\big) \nonumber \\ &\hspace{10mm}+ \frac{\kappa}{1+\kappa}  {\ma{H}}_{{\text rr}}\text{diag}\big(\ma{Q} \big) {\ma{H}}_{{\text rr}}^{{H}}   \Big) \ma{W}^{{H}}. 
 %& = \left|a_k \right|^2  \bigg( P_s \left|  { h}_{{\text sr},k} \right|^2 + M_k +  \mathcal{E} \{ {u}_{{\text in},k} {u}_{{\text in},k}^{*} \}  \beta_k \nonumber \\  &+ \mathcal{E} \{ {u}_{{\text out},k} {u}_{{\text out},k}^{*} \} \cdot \left( \left|{ h}_{{\text rr},k} \right|^2 \kappa_k + (1 + \kappa_k) |\delta_k|^2  \right)  \bigg), 
 %\label{eq:u_out_power}
%\end{align}
Note that the above derivations (\ref{eq_loop_U_out})-(\ref{eq_loop_U_out_2}) hold as the noise, the desired signal at subsequent symbol durations, and the distortion components are zero-mean and mutually independent. Unfortunately, a direct expression of $\ma{Q}$ in terms of $\ma{W}$ can not be achieved from (\ref{eq_loop_U_out_2}), (\ref{eq_loop_U_out_R}) in the current form. In order to facilitate further analysis we resort to the vectorized presentation of $\ma{Q}$. By applying the famous matrix equality \mbox{$\text{vec}(\ma{A}_1\ma{A}_2\ma{A}_3) = (\ma{A}_3^{\text{T}} \otimes \ma{A}_1 ) \text{vec}(\ma{A}_2)$,} we can write (\ref{eq_loop_Q}) as 
\begin{align} \label{eq_loop_vec_Q}
\text{vec}\left(\mathbb{E} \{ \ma{r}_{{\text {out}}} \ma{r}_{{\text {out}}}^{{H}} \}\right) =  \left( \ma{I}_{M_{\text{t}}^2} + \kappa\ma{S}_{\text{D}}^{M_{\text{t}}} \right)   \text{vec}\left( \ma{Q} \right), 
\end{align}
where $\ma{S}_{\text{D}}^M \in \{0,1\}^{ {M^2} \times {M^2} }$ is a selection matrix with one or zero elements such that $\ma{S}_{\text{D}}^{M_{\text{t}}} \text{vec}\left(\ma{Q}\right) =  \text{vec}\left(\text{diag}(\ma{Q})\right)$. Similarly from (\ref{eq_loop_U_out_2}) we obtain 
\begin{align} \label{eq_loop_vec_U_out}
\text{vec}\left(\ma{Q} \right) &= \Big( \ma{I}_{M_{\text{t}}^2} - \left( \ma{W}^{*} \otimes \ma{W} \right) \ma{C} \Big)^{-1}  \left( \ma{W}^{*} \otimes \ma{W} \right) \ma{c}, 
%\nonumber \\
%& =  \Big( \left( \ma{W}^{*} \otimes \ma{W} \right)^{-1} -  \ma{C} \Big)^{-1}   \ma{c},
\end{align}
where 
\begin{align} \label{eq_loop_A}
\ma{C}:&=  \beta\ma{S}_{\text{D}}^{M_{\text{r}}} \hspace{-0.4mm} \left( {\ma{H}}_{{\text {rr}}}^{*} \otimes {\ma{H}}_{{\text{ rr}}} \right) \hspace{-0.4mm} \left( \ma{I}_{M_{\text{t}}^2} +\hspace{-0.4mm} \kappa \ma{S}_{\text{D}}^{M_{\text{t}}} \hspace{-0.4mm} \right) \hspace{-0.4mm} + \hspace{-0.4mm} \kappa \left( {\ma{H}}_{{\text {rr}}}^{*} \hspace{-0.4mm} \otimes  \hspace{-0.4mm} {\ma{H}}_{{\text {rr}}} \right) \hspace{-0.4mm} \ma{S}_{\text{D}}^{M_{\text{t}}},\\
\ma{c}:&= \left(\ma{I}_{M_{\text{r}}^2} + \beta \ma{S}_{\text{D}}^{M_{\text{r}}}   \right)\text{vec}\left( P_{\text {s}} \ma{ h}_{{\text {sr}}}\ma{ h}_{{\text {sr}}}^{{H}} + \sigma_{\text{nr}}^2\ma{I}_{{M_{\text {r}}}} \right). \label{eq_loop_a}
\end{align}
\revOmid{The direct dependence of the relay transmit covariance matrix and $\ma{W}$ can be consequently obtained from (\ref{eq_loop_vec_U_out}) and (\ref{eq_loop_vec_Q}) as
{{\begin{align} \label{eq_loop_vec_Q_2}
 &\hspace{-3mm}\text{vec}\left(\mathbb{E} \{ \ma{r}_{{\text {out}}} \ma{r}_{{\text {out}}}^{{H}} \}\right) = \ma{\Theta} \big(  \ma{W}, \ma{H}_{\text{rr}} , \kappa, \beta \big)\text{vec}\hspace{-0.5mm}\left( \hspace{-0.5mm} P_{\text {s}} \ma{ h}_{{\text {sr}}}\ma{ h}_{{\text {sr}}}^{{H}} \hspace{-0.5mm}+  \sigma_{\text{nr}}^2\ma{I}_{{M_{\text {r}}}} \hspace{-0.5mm} \right) 
%\nonumber \\
 %& \hspace{-3mm}  \left( \ma{I}_{M_{\text{t}}^2}\hspace{-0.5mm} + \hspace{-0.5mm}\kappa\ma{S}_{\text{D}}^{M_{\text{t}}} \right) \hspace{-1mm} \Big( \ma{I}_{M_{\text{t}}^2}\hspace{-1mm} - \hspace{-1mm} \left( \ma{W}^{*} \hspace{-0.5mm} \otimes \hspace{-0.5mm} \ma{W} \right) \ma{C} \Big)^{-1}  \nonumber \\ 
%& \hspace{-1mm}\underbrace{ \;\;\;\;\;\;\;\;\; \times \left( \ma{W}^{*} \otimes \ma{W} \right)   \left(\ma{I}_{M_{\text{r}}^2} + \beta \ma{S}_{\text{D}}^{M_{\text{r}}}   \right) }_{=: \ma{\Theta} \big(  \ma{W}, \ma{H}_{\text{rr}} , \kappa, \beta \big)}\text{vec}\hspace{-0.5mm}\left( \hspace{-0.5mm} P_{\text {s}} \ma{ h}_{{\text {sr}}}\ma{ h}_{{\text {sr}}}^{{H}} \hspace{-0.5mm}+  \sigma_{\text{nr}}^2\ma{I}_{{M_{\text {r}}}} \hspace{-0.5mm} \right)\hspace{-0.5mm}, 
\end{align}}}
such that 
{{ \begin{align}
& \ma{\Theta} \big(  \ma{W}, \ma{H}_{\text{rr}} , \kappa, \beta \big) :=  \left( \ma{I}_{M_{\text{t}}^2}\hspace{-0.5mm} + \hspace{-0.5mm}\kappa\ma{S}_{\text{D}}^{M_{\text{t}}} \right) \hspace{-1mm} \Big( \ma{I}_{M_{\text{t}}^2}\hspace{-1mm} - \hspace{-1mm} \left( \ma{W}^{*} \hspace{-0.5mm} \otimes \hspace{-0.5mm} \ma{W} \right) \ma{C} \Big)^{-1}  \nonumber \\ 
& \;\;\;\;\;\;\;\;\; \;\;\;\;\;\;\;\;\;\;\;\;  \times \left( \ma{W}^{*} \otimes \ma{W} \right)   \left(\ma{I}_{M_{\text{r}}^2} + \beta \ma{S}_{\text{D}}^{M_{\text{r}}}   \right) 
\end{align}}}represents the transfer function of the relay; relating the distortion-less input, i.e., $ P_{\text {s}} \ma{ h}_{{\text {sr}}}\ma{ h}_{{\text {sr}}}^{{H}} + \sigma_{\text{nr}}^2\ma{I}_{{M_{\text {r}}}}$, to the distorted transmit covariance. It can be observed that $\ma{\Theta} \left( \ma{W}, \ma{H}_{\text{rr}} ,0,0  \right) = \ma{W}^{*} \otimes \ma{W}$, which is similar to the known FD-AF relay operation with a perfect hardware, i.e., $\kappa,\beta=0$. }   
\subsection{Optimization problem}       
In order to formulate the end-to-end link quality, we recall that the noise, the desired signal and the residual interference signals are zero-mean and mutually independent. Hence, the received signal power at the destination, after application of $\ma{z}$, can be separated as
\begin{align} 
P_{\text{des}} & = \revOmid{\mathbb{E}_{\mathbb{S}_r} } \left\{\left| \ma{z}^{{H}} \ma{H}_{{\text {rd}}} \ma{W} \ma{h}_{{\text {sr}}}  \sqrt{P_s} s\right|^2\right\} \nonumber \\ &= P_s \ma{ z}^{{H}} \ma{ H}_{{\text{rd}}} \ma{W} \ma{h}_{{\text{sr}}} \ma{ h}_{{\text {sr}}}^{{H}} \ma{W}^{{H}} \ma{H}_{{\text {rd}}}^{{H}} \ma{ z}, \label{eq_loop_P_desired}\\
P_{\text{tot}} &= \revOmid{\mathbb{E}_{\mathbb{S}_r} }\left\{\left| \ma{z}^{{H}} \left( \ma{ H}_{{\text {rd}}} \ma{r}_{\text{out}} + \ma{h}_{{\text {sd}}}  \sqrt{P_s} s+ \ma{n}_{\text{d}}\right) \right|^2\right\} =   \nonumber \\
& \ma{ z}^{{H}} \left(  \ma{ H}_{{\text {rd}}} \Big(\ma{Q} \hspace{-0.5mm} + \hspace{-0.5mm}  \kappa \text{diag}\left(\ma{Q}\right) \hspace{-0.5mm} \Big)   \ma{ H}_{{\text{ rd}}}^{{H}} \hspace{-0.5mm} + \hspace{-0.5mm} \sigma_{\text{nd}}^2 \ma{I}_{M_{\text{d}}} \hspace{-0.5mm} +\hspace{-0.5mm} P_{\text{s}}\ma{h}_{{\text{sd}}}\ma{h}_{{\text{sd}}}^H  \right) \ma{ z} ,  \label{eq_loop_P_tot} \\
%&\approx  \ma{ z}^{{H}} \ma{ H}_{{\text rd}} \ma{Q} \ma{ H}_{{\text rd}}^{{H}} \ma{ z}   + \sigma_{\text{nd}}^2\ma{ z}^{{H}}\ma{ z} \label{eq_loop_P_tot_approx} \\
P_{\text{err}} &= P_{\text{tot}} - P_{\text{des}}, \label{eq_loop_P_error}
\end{align}
\revOmid{where ${\mathbb{E}_{\mathbb{S}_r}}$ is the statistical expectation over the set of random variables $\mathbb{S}_r$, see Table~\ref{tab_notations}. Moreover,} $P_{\text{des}}$ and $P_{\text{err}}$ respectively represent the power of the desired, and distortion-plus-noise parts of the estimated signal $\hat{\ma{s}}$, and $P_{\text{tot}}:= \mathbb{E} \{ |\hat{{s}}|^2 \}$. \revOmid{It is worth mentioning that $P_{\text{tot}} \geq P_{\text{des}}$, and hence $P_{\text{err}} \geq 0$, due to the superposition of the statistically independednt noise, distortion, and signal terms at different nodes or time instances.} The corresponding optimization problem for maximizing the resulting SDNR is written as\footnote{\revOmid{Please note that the objective is the instantaneous SDNR, assuming that the instantaneous CSI is available. Due to the single stream communication, SDNR relates to the end-to-end mutual information via Shannon's formula, assuming that the desired signal, noise, and distortion signals follow a Gaussian distribution. The accuracy of the Gaussian distribution assumption may differ for different transmit signaling, as well as the used SIC scheme. However, in many related studies the Gaussian modeling of the residual interference from collective sources of impairments is considered as a good approximation, e.g., \cite{DMBS:12, DMBSR:12}.}} 
%
%Due to the defined single stream communication setup, the end-to-end rate maximization can be equivalently considered as a signal-to-interference-plus-noise (SINR) maximization problem
\begin{subequations} \label{model_original_problem}
\begin{align} 
\underset{P_{\text{s}}, \ma{z}, \ma{W} }{\text{max}} \;\;\; & \frac{P_{\text{des}}}{P_{\text{err}}}  \label{eq_loop_optimization_problem_a}\\
\text {s.t.} \;\;\; & (\ref{eq_loop_vec_U_out}), \;\; \ma{Q} \in \mathcal{H}, \;\; \text{tr}\left( \ma{Q} \right) \leq \tilde{P}_{\text{r,max}}, \;\;0 \leq P_{\text{s}} \leq P_{\text {s,max}} , \label{eq_loop_optimization_problem_b}  
\end{align}  
\end{subequations}
where (\ref{eq_loop_optimization_problem_b}) limits the feasible set of $\ma{W}$ to those resulting in a feasible $\ma{Q}$. Note that $\tilde{P}_{\text{r,max}} := \frac{{P}_{\text{r,max}}}{1+\kappa}$, and the power constraint in (\ref{eq_loop_optimization_problem_b}) follows as $\text{tr}\left( \ma{Q} + \kappa \text{diag}(\ma{Q})\right) =  (1+\kappa)\text{tr}\left( \ma{Q}\right)$. \par
As it can be observed, the optimization problem (\ref{eq_loop_optimization_problem_a})-(\ref{eq_loop_optimization_problem_b}) is a non-convex optimization problem and cannot be solved analytically, due to the structure imposed by (\ref{eq_loop_vec_U_out}). In order to approach the solution, we propose a GP-based optimization method in the following section.

\section{Gradient Projection (GP) for SDNR maximization}\label{Sec_GP}
In this part we propose an iterative solution to (\ref{eq_loop_optimization_problem_a})-(\ref{eq_loop_optimization_problem_b}) based on the gradient projection method \cite{bertsekas1999nonlinear, DMBS:12}. In this regard, the optimization variables are updated in the increasing direction of the objective function (\ref{eq_loop_optimization_problem_a}).  
\subsection{Iterative update for $\ma{W}$}
\revOmid{The update rule for $\ma{W}$ is defined following the GP method, where detailed instructions are inspired from \cite{DMBSR:12}. This includes the update of $\ma{W}$ in the steepest ascent direction, and occasional projection due to constraints violation. This is expressed as 
\begin{align} \label{eq_GP_update} 
\bar{\ma{W}}^{[l]}  &=  \mathcal{P} \left( \ma{W}^{[l]} + \delta^{[l]} \nabla \left(\text{SDNR}\right)  \right) \nonumber \\
\ma{W}^{[l+1]}  &=  \ma{W}^{[l]} + \gamma^{[l]} \left( \bar{\ma{W}}^{[l]} -  {\ma{W}}^{[l]}  \right),
\end{align}
where $\mathcal{P} \left( \cdot \right)$ represents the projection to the feasible solution space, $l$ is the iteration index, $\nabla \left( \cdot \right)$ represents the gradient with respect to $\ma{W}^{*}$, and $\delta, \gamma \in \real^+$ represent the step size variables. The update direction is obtained from the calculated gradients in (\ref{eq_nabla_P_error})-(\ref{eq_nabla_P_desired}), and the fact that 
\begin{align} \label{eq_GP_update_grad}
%\nabla_{\ma{W}^{*}} 
\nabla \left(\text{SDNR}\right)   =   \Big(\nabla \left({P_{\text{des}}}\right)  {P_{\text{err}}}  - \nabla\left({P_{\text{err}}}\right)  {P_{\text{des}}} \Big)/{{P_{\text{err}}}^2}.
\end{align}
The stepsize value $\gamma$ is chosen according to the Armijo's step size rule \cite{ARMIJO2:73}. This is expressed as 
\begin{align} 
%\nabla_{\ma{W}^{*}} 
& \text{SDNR} \left(\ma{W}^{[l+1]} \right) - \text{SDNR} \left(\ma{W}^{[l]} \right) \nonumber \\ & \geq \sigma \nu^{m} \text{tr} \left( \Big\{\nabla \left(\text{SDNR}\right) \Big\}^H \left( \bar{\ma{W}}^{[l]} - {\ma{W}}^{[l]} \right) \right), \nonumber
\end{align}
%which provides a trade-off between validity of the calculated slope and objective improvement in each step. In this regard,
where $\gamma^{[l]} = \nu^{m}$, such that $m$ is the smallest non-negative integer satisfying the above inequality, and $\nu = 0.5$, $\sigma=0.1$ and $\delta = 1$. } 

\begin{figure*}[!ht]
\normalsize
\begin{align}
%\label{eqn_slnr}
\text{vec}\left( \nabla {P_{\text{err}}}\right)^{{T}} & = 
 \text{vec}\left( \left( \ma{H}_{\text{rd}}^{{H}} \ma{z} \ma{z}^{{H}} \ma{H}_{\text{rd}} \right)^{{T}}  \right)^{{T}} \left( \ma{I}_{M_{\text{t}}^2} + \kappa\ma{S}_{\text{D}}^{M_{\text{t}}}  \right)
 \Bigg( \left[  \ma{c} + \ma{C} \big(  \ma{I}_{M_{\text{t}}^2} - \left( \ma{W}^{*}  \otimes \ma{W} \right) \ma{C}  \big)^{-1}      \left( \ma{W}^{*} \otimes \ma{W}\right) \ma{c} \right]^{{T}} 
\ \nonumber \\ 
&  \otimes \big( \ma{I}_{M_{\text{t}}^2} - \left( \ma{W}^{*} \otimes \ma{W} \right) \ma{C} \big)^{-1} \Bigg) {\ma{S}}_{\text{K}} \left(\ma{w}  \otimes \ma{I}_{ M_{\text{r}} M_{\text{t}} } \right) - \left( \ma{z}^{{T}}  \ma{H}_{\text{rd}}^{\text{*}}  \right) \otimes \left( P_{\text{s}} \ma{z}^{{H}} \ma{H}_{\text{rd}} \ma{W} \ma{h}_{\text{sr}} \ma{h}_{\text{sr}}^{{H}}  \right) {\ma{S}}_{\text{T}} , \label{eq_nabla_P_error} \\
\text{vec} \left( \nabla {P_{\text{des}}}\right)^{{T}} & = \left(\ma{z}^{{T}} \ma{H}_{\text{rd}}^{\text{*}}  \right) \otimes \left( P_{\text{s}} \ma{z}^{{H}} \ma{H}_{\text{rd}} \ma{W} \ma{h}_{\text{sr}} \ma{h}_{\text{sr}}^{{H}}  \right) {\ma{S}}_{\text{T}}, \;\; \hspace{-0.5mm}
\text{where} \hspace{-0.5mm} \;\; \ma{w}:= \text{vec}(\ma{W}),\; \hspace{-0.5mm} {\ma{S}}_{\text{K}} \hspace{-0.5mm} \in \hspace{-0.5mm} \{0,1\}^{M_{\text{r}}^2M_{\text{t}}^2\times M_{\text{r}}^2M_{\text{t}}^2},\; {\ma{S}}_{{\text{T}}} \hspace{-0.5mm} \in \hspace{-0.5mm} \{0,1\}^{M_{\text{r}}M_{\text{t}}\times M_{\text{r}}M_{\text{t}}}, \label{eq_nabla_P_desired} \\  \text{such that:} \;&\; \text{vec} \left(  \ma{W}^{*} \otimes \ma{W} \right)   = {\ma{S}}_{\text{K}} \text{vec}\left( \ma{w}^{*} \ma{w}^{{T}}\right), \;\; \text{and} \;\;  {\ma{S}}_{{\text{T}}} \text{vec}(\ma{W}) = \text{vec}(\ma{W}^{{T}}). \label{eq_selection_SK_SD}
 %\left \{ \begin{array}{c}
 %1 \;\;  {\ma \delta}^{\rm T} {\ma T}  {\ma \delta^{*}} \leq \xi \Rightarrow \;\;\; \;\;\; \;  }  \\
 %{0  \text{elsewhere}}, \end{array} \right \
%\tilde{\ma{S}} \left(\text{vec}\left( \ma{W} \otimes \ma{I}_{M_{\text{r}}} \right)  \right)\text{vec}\left( \ma{W}^{*}\right)  = \text{vec}\left( \ma{W}^{*} \otimes \ma{W}\right) 
\end{align}
\hrulefill
% The spacer can be tweaked to stop underfull vboxes.
\vspace*{-0mm}
\end{figure*}

\subsubsection{Projection rule} \label{GP_Proj_Rule}
%It is apparent that a variable update may result in a ${\ma{W}}$ that violates the problem constraints in (\ref{eq_loop_optimization_problem_b}). 
\revOmid{Once an updated ${\ma{W}}$, and the corresponding $\ma{Q}$ calculated from (\ref{eq_loop_vec_U_out}), violate the problem constrains, i.e., when $\ma{Q}$ contains a negative eigenvalue or exceeds the defined power constraint, see (\ref{eq_loop_optimization_problem_b}), it is projected in to the feasible variable space. Due to the convexity of the feasible variable space in $\ma{Q}$, similar to the suggested procedure in \cite{1237413}, we follow a projection rule which results in a minimum Euclidean distance to the feasible variable space of $\ma{Q}$, i.e., minimum Frobenius norm of the matrix difference. In order to obtain this, let $\ma{W}_{\text{old}}$ and $\ma{Q}_{\text{old}}$ be the updated relay amplification matrix from (\ref{eq_GP_update}) and the corresponding undistorted transmit covariance, calculated from (\ref{eq_loop_vec_U_out}). Moreover, let $\ma{U}_{\text{old}}\ma{\Lambda}_{\text{old}}\ma{U}_{\text{old}}^{{H}}$ be an eigenvalue decomposition of the matrix $\ma{Q}_{\text{old}}$, such that $\ma{U}_{\text{old}}$ is a unitary matrix, and $\ma{\Lambda}_{\text{old}}$ is a diagonal matrix containing the eigenvalues. The feasible relay undistorted transmit covariance matrix, i.e., $\ma{Q}_{\text{new}}$, with minimum Euclidean distance to $\ma{Q}_{\text{old}}$ is then obtained as
\begin{align} \label{eq_GP_proj_Q}
\ma{Q}_{\text{new}} \leftarrow  \ma{U}_{\text{old}} \underbrace{\left( \ma{\Lambda}_{\text{old}} - \zeta \ma{I}_{M_{\text{t}}} \right)^{+}}_{=:{\ma{\Lambda}}_{\text{new}}} \ma{U}_{\text{old}}^{{H}}, \;\; \nu \in \real,
\end{align}
where $(\cdot)^{+}$ substitutes the negative elements by zero, and $\zeta \in \real$ is the minimum non-negative value that satisfies $\text{tr}\left({\ma{\Lambda}}_{\text{new}} \right) \leq \tilde{P}_{\text{r,max}}$, see \cite[Equation~(25)-(27)]{1237413}.} The projected version of $\ma{W}_{\text{old}}$, i.e., $\ma{W}_{\text{new}}$ is then calculated as 
\begin{align}\label{eq_GP_proj_W}
\ma{W}_{\text{new}} \leftarrow \ma{Q}_{\text{new}}^{\frac{1}{2}} \ma{V} \ma{U}_{\text{r,new}}\left(\ma{\Sigma}_{\text{r,new}}\right)^{-\frac{1}{2}} \ma{U}_{\text{r,new}}^H, 
\end{align}
where $\ma{Q}_{\text{new}}^{\frac{1}{2}} = \ma{U}_{\text{old}} {\ma{\Lambda}}_{\text{new}}^{\frac{1}{2}} \ma{U}_{\text{old}}^H$, $\ma{V}$ is an arbitrary unitary matrix such that $\ma{V}\ma{V}^H = \ma{I}_{M_{\text{t}}}$, and $\ma{U}_{\text{r,new}} \ma{\Sigma}_{\text{r,new}} \ma{U}_{\text{r,new}}^H $ is the eigenvalue decomposition of ${\mathcal{R}} (\ma{Q}_{\text{new}})$, see (\ref{eq_loop_U_out_R}). 

Note that the resulting amplification matrix $\ma{W}_{\text{new}}$ consequently results in $\ma{Q}_{\text{new}}$ as the relay covariance matrix, see (\ref{eq_loop_U_out_2}), and hence belongs to the feasible set of (\ref{eq_loop_optimization_problem_b}). Moreover, the choice of $\ma{V}$ does not affect the corresponding $\ma{Q}_{\text{new}}$, and hence does not affect the feasibility. Hence it can be chosen similar to that of $\ma{W}_{\text{old}}$, with no need for modification in the projection process:
\begin{align}\label{eq_GP_proj_W_Unitary_V}
\ma{V} \leftarrow  \left(\ma{Q}_{\text{old}}\right)^{-\frac{1}{2}} \ma{W}_{\text{old}} \ma{U}_{\text{r,old}}\left(\ma{\Sigma}_{\text{r,old}}\right)^{\frac{1}{2}} \ma{U}_{\text{r,old}}^H, 
\end{align}
where $\ma{U}_{\text{r,old}} \ma{\Sigma}_{\text{r,old}} \ma{U}_{\text{r,old}}^H$ is the eigenvalue decomposition of ${\mathcal{R}} (\ma{Q}_{\text{old}})$. 

\subsection{Iterative update for $P_{\text{s}}$}
For fixed values of $\ma{W}$ and $\ma{z}$, an increase in $P_{\text{s}}$ results in an increase in the desired received power, see (\ref{eq_loop_P_desired}). On the other hand, it also results in an increase in $P_{\text{err}}$, due to the direct source-destination interference, as well as the increased received power at the relay which results in an amplified distortion effect. As a result, the impact of the choice of $P_{\text{s}}$ on the end-to-end SDNR is not clear. The following lemma provides an answer to this question.

\begin{lemma} \label{Lemma_PS_Opt}
For fixed values of $\ma{W}$ and $\ma{z}$ the resulting SDNR is a concave and increasing function of $P_{\text{s}}$. Hence, the optimum $P_{\text{s}}$ is given as 
\begin{align} \label{Lemma_PS_opt_ps_equation}
P_{\text{s}}^\star = \text{min}\left\{ P_{\text{s,max}}, \tilde{P}_{\text{s,max}}(\ma{W}) \right\}, 
\end{align}
where $\tilde{P}_{\text{s,max}}(\ma{W})$ represents the value of $P_{\text{s}}$ that results in the maximum relay transmit power, i.e., $P_{\text{r,max}}$, with $\ma{W}$ as the relay amplification matrix. 
\end{lemma}
\begin{proof}
See Appendix~\ref{appendix:PsLemmaProof}
\end{proof}

It is worth mentioning that for a setup with a weak direct link, i.e., $\|\ma{h}_{\text{sd}}\|_2 \approx 0$, we have $P_{\text{s}}^\star = P_{\text{s,max}}$ for a jointly optimal choice of $\ma{W}, P_{\text{s}}$. This is grounded on the fact that for any $P_{\text{s}} < P_{\text {s,max}}$, the joint variable update $P_{\text{s}} \leftarrow P_{\text {s,max}}$ and $\ma{W} \leftarrow \ma{W}\sqrt{\frac{P_{\text {s}}}{P_{\text {s,max}}}}$, result in the same $P_{\text{des}}$, see (\ref{eq_loop_P_desired}), while decreasing the $P_{\text{err}}$, see (\ref{eq_loop_P_error}) in connection to (\ref{eq_loop_vec_Q_2}).
%As it is discussed, $P_{\rm s}=P_{\rm {s,max}}$ provides the optimal transmit strategy at the source, for any choice of $\ma W$ and $\ma z$. 
\subsection{Iterative update for $\ma{z}$} 
It is apparent that the relay transmit covariance, and hence the constraints in (\ref{eq_loop_optimization_problem_b}) are invariant to the choice of receive linear filter. The optimal choice of $\ma{z}$ for a given $\ma{W}, P_{\text{s}}$ is obtained as
\begin{align} \label{eq_GP_z}
%\ma{z}^{\star}= \lambda_{\text{max}} \left\{ \left(\ma{H}_{\rm{rd}}\ma{Q}\ma{H}_{\rm{rd}}^{{H}} + \sigma_{\text{nd}}^2 \ma{I}_{M_{\rm d}}\right)^{-1} \left(\ma{ H}_{{\rm rd}} \ma{W} \ma{ h}_{\text{ sr} } \ma{ h}_{\text{ sr} }^H \ma{W}^H \ma{ H}_{{\text{ rd} }}^H \right)  \right\},
\ma{z}^{\star}=  \Big(\ma{H}_{\text{rd}} \big( \ma{Q} + \kappa \text{diag}\left(\ma{Q}\right) \big) \ma{H}_{\text{rd}}^{{H}} & + \sigma_{\text{nd}}^2 \ma{I}_{M_{\text{d}}} + P_{\text{s}} \ma{h}_{\text{sd}} \ma{h}_{\text{sd}}^H  \Big)^{-1} \nonumber \\ & \times \sqrt{P_{\text{s}}} \ma{H}_{{\text{rd}}} \ma{W} \ma{h}_{{\text{sr}}}.
\end{align}
The value of $\ma{z}$ is updated according to (\ref{eq_GP_z}) after the update for $\ma{W}, P_{\text{s}}$. The update iterations are continued until a stable point is achieved, or a certain number of iterations is expired, see Algorithm~\ref{alg_GP}. 
\revOmid{\subsection{Convergence}
The proposed GP algorithm leads to a necessary convergence, due to the monotonic improvement of the SDNR after each variable update and the fact that the objective is bounded from above. However, due to the non-convexity of the problem, the global optimality of the obtained solution is not guaranteed, and the converging point depends on the used initialization \cite{DMBS:12, DMBSR:12}. In Subsection~\ref{sim_alg_analysis} a numerical evaluation of the optimal performance is obtained by repeating the GP algorithm with several initializations. 
}
\revOmid{
\subsection{Computational complexity} \label{GP_CC}
In this part we study the computational complexity associated with the GP algorithm, both regarding the design, as well as the processing complexity. We base our analysis on the following assumptions. 
\begin{itemize}
\item On modern computers using math coprocessors, the time consumed to perform addition/subtraction and multiplication/division is about the same \cite{chapra1988numerical}. Hence, we measure the complexity in terms of the total floating point complex operations (FLOP)s. 
\item Matrix inversion $\ma{A}^{-1}$ where $\ma{A} \in \compl^{N\times N}$ for a positive semi-definite $\ma{A}$ requires $N^3+N^2+N$ FLOPs via Cholesky decomposition. For a non-singular (invertible), but not structured matrix $\ma{A}$ the inverse can be calculated via LU decomposition, incurring $\frac{4n^3}{3} - \frac{n}{3}$ FLOPs. The calculation of eigenvalue decomposition is associated with $8N^3/3$ FLOPs \cite{chapra1988numerical}. 
\item The complexity associated with the standard matrix/matrix or matrix/vector multiplications are given in \cite{hunger2005floating}.  
\end{itemize}
\subsubsection{Algorithm complexity}
The algorithm starts by the calculation of $\ma{C}, \ma{c}$, resulting respectively in $\mathcal{O} \left( M_{\text{r}}^2 + 5M_{\text{r}}\right)$ and $\mathcal{O} \left( M_{\text{t}}^2M_{\text{r}} + M_{\text{r}}^2M_{\text{t}} \right)$ FLOPs as the initial steps. However, the complexity of GP is dominated by the derivative (25), resulting in $\mathcal{O} \left(4/3M_{\text{t}}^6 + 4 M_{\text{t}}^4M_{\text{r}}^2 + M_{\text{t}}^4 + 3M_{\text{t}}^2M_{\text{r}}^2 \right)$ as well as the Armijo line search incurring $\mathcal{O} \left(4/3M_{\text{t}}^6 + 4 M_{\text{t}}^4M_{\text{r}}^2 + M_{\text{t}}^4 + 3M_{\text{t}}^2M_{\text{r}}^2 \right)$ FLOPs for each search iteration. Together with the calculation of $\ma{z}$ (32), the overall algorithm complexity can be  expressed as  
\begin{align}
\mathcal{O} \Big(  & \gamma_1 \left(4/3M_{\text{t}}^6 + 4 M_{\text{t}}^4M_{\text{r}}^2 + M_{\text{t}}^4 + M_{\text{d}}^3 \right) \nonumber \\ & + \gamma_2 \left(4/3M_{\text{t}}^6 + 4 M_{\text{t}}^4M_{\text{r}}^2 + M_{\text{t}}^4 \right)\Big),
\end{align}
where $\gamma_1,\gamma_2$ respectively represent the required line search, and update iterations for $\ma{W}$.

\subsubsection{Processing complexity}
The processing complexity is associated with the amplification (\ref{eq_model_r_out}). Since $\ma{W}\in\compl^{M_{\text{t}} \times M_{\text{r}}}$ is a general matrix (not necessarily symmetric or low rank), the complexity is $M_{\text{t}} M_{\text{r}}$ complex multiplications and $M_{\text{t}} (M_{\text{r}} - 1)$ addition, incurring in total $2 M_{\text{t}}M_{\text{r}} - M_{\text{t}}$ FLOPs. 

Please note that the above analysis intends to show how the bounds on computational complexity are related to different dimensions in the problem structure. Nevertheless, the actual computational load may vary in practice, due to the further structure simplifications, and depending on the used processor. A numerical study on the required computational complexity is given in Section~VI.
}

\begin{algorithm}[H]
 \small{	\begin{algorithmic}[1]
  %\SetAlgoLined
\State{$\text{Counter} \leftarrow 0$}
\Repeat   $\;\; \text{(running for multiple initializations)}$
\State{$\text{Counter} \leftarrow \text{Counter} + 1$}
\State{$\ell \leftarrow 0$}
\State{$P_{\text{s}}^{(0)} \leftarrow  P_{\text{s,max}} \times 10^{-4} $}
\State{$\ma{Q}^{(0)} \leftarrow  \text{random init., see (\ref{eq_loop_optimization_problem_b})} $}
\State{$\ma{W}^{(0)} \leftarrow  {\ma{Q}^{(0)}}^{\frac{1}{2}} \mathcal{R}^{\frac{-1}{2}}\left(\ma{Q}^{(0)}\right), \text{ see (\ref{eq_GP_proj_W}),~(\ref{eq_loop_U_out_R})}  $}
\Repeat 
\State{$\ell \leftarrow \ell + 1 $}
\State{$\ma{W}^{(\ell)} \leftarrow \text{update } \ma{W}^{(\ell)}, \text{see (\ref{eq_GP_update}),~(\ref{eq_GP_proj_W}), Section IV.B} $}
\State{$P_{\text{s}}^{(\ell)} \leftarrow  \text{update}, \text{see~(\ref{Lemma_PS_opt_ps_equation})}$} 
\State{$\ma{z}^{(\ell)} \leftarrow \text{update}, \text{see (\ref{eq_GP_z})}$}
%\If{ \text{stable SDNR value}}
\State{$\text{SDNR}^{(\ell)} \leftarrow  \text{(\ref{eq_loop_optimization_problem_a})}$}
%\State{break} 
%\EndIf
\Until {$\text{SDNR}^{(\ell)} - \text{SDNR}^{(\ell-1)} \geq c_1$    (until SDNR improves)}
\State{$\mathcal{A} \leftarrow  \text{save} \left(\text{SDNR}^{(\ell)}, \ma{W}^{(\ell)}, \ma{z}^{(\ell)}\right)$}
\Until {$\text{Counter} \leq C_1 $ }
\State{\Return{$\left(\ma{W},\ma{z},\text{SDNR}\right)\; \leftarrow \; {\text{max}} \;\; \text{SDNR}\in\mathcal{A} $}}%%%
  \end{algorithmic} } 
 \caption{\small{Iterative SDNR maximization algorithm based on GP. Number of algorithm iterations are determined by $c_1\in \real^+$ and $C_1\in\mathbb{N}$.} }  \label{alg_GP} 
 
\end{algorithm}   	
%\section{Iterative Covariance Shaping via Quadratic Approximation}\label{Sec_QAP}
%\input{main_Quadratic_Approximation_Covariance_Shaping}
\section{An Intuitive Approach: Distortion-Aware Multi-Stage Rank-1 Relay Amplification (MuStR1)} \label{section:channel_norm_1}
\begin{figure}[!t] 
% ensure that we have normalsize text
\normalsize
        \includegraphics[angle=0,width=0.99\columnwidth]{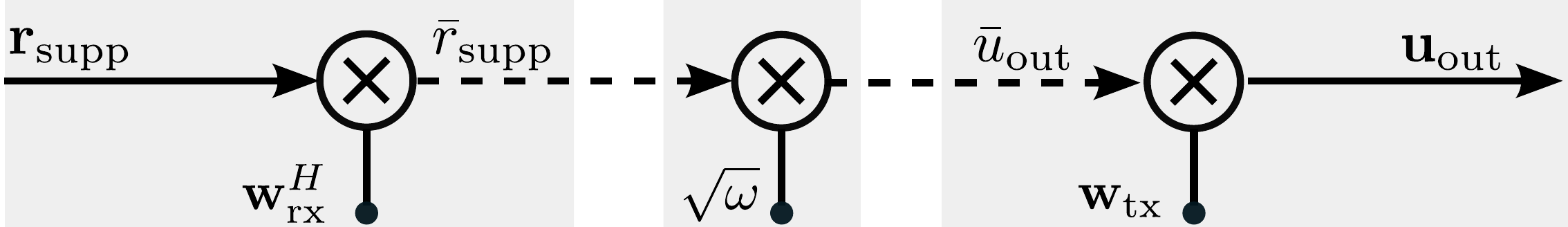}       
% Restore the current equation number.
%\setcounter{equation}{19}
% IEEE uses as a separator
\hrulefill
% The spacer can be tweaked to stop underfull vboxes.
    \caption{{\footnotesize{The relay amplification is divided into three parts: $\ma{w}_{\text{rx}}$ and $\ma{w}_{\text{tx}}$ respectively represent the reception and transmit filters, and $\omega$ represents the scaling factor. The bold arrows represent the vector signals while the dashed arrows represent the scalars. The overall process can be described as $\ma{W}=\omega\ma{w}_{\text{tx}}\ma{w}_{\text{rx}}^H$.}}}
\vspace{-0pt}
\label{fig:three_part_W}
\end{figure}

The proposed method in Section~\ref{Sec_GP} directly deals with the SDNR as optimization objective, which also leads to the maximization of end-to-end mutual information for Gaussian signal codewords. Nevertheless, the proposed procedure imposes a high computational complexity, due to the number of the required iterations. In this section we introduce a simpler design by considering a rank-one relay amplification matrix. Note that the near-optimality of rank-one relay amplification matrices for single stream communication has been established, see the arguments in \cite[Subsection~3.2]{Taghizadeh2016} and \cite[Section~III]{7558213}. Nevertheless, in the aforementioned works, the impacts of transmit/receive distortion have not been considered in the FD transceiver. A rank-one relay amplification process is expressed as
%Nevertheless, the proposed procedure imposes a high computational complexity, due to the number of the required iterations. In this section we introduce a simpler design, by focusing on the ratio between the desired and the distortion channel norms which is observed separately from the relay transmit and receiver ends. Note that similar norm-maximization approaches for other relaying scenarios, while considered as sub-optimal techniques, have proven effective in terms of performance and complexity, see \cite{RH:09}, \cite{RH:10}. In the current approach we divide the role of the relay amplification matrix into three stages
\begin{align} \label{eq_W_three_parts}
\ma{W}=\sqrt{\omega} \ma{w}_{\text{tx}}\ma{w}_{\text{rx}}^H,
\end{align}
where $\ma{w}_{\text{rx}}\in \compl^{M_{\text{r}}} , \|\ma{w}_{\text{rx}}\|_2=1$ and $\ma{w}_{\text{tx}} \in \compl^{{M_{\text{t}}}},  \|\ma{w}_{\text{tx}}\|_2=1$, respectively act as the receive and transmit linear filters at the relay, while $\omega \in \real^+$ acts as a scaling factor, see Fig.~\ref{fig:three_part_W}. The idea is to separately design the transmit (receive) filters to maximize the SDNR at each segment. Afterwards, the value of $\omega$ is optimized.    
%The intermediate signals $\tilde{\ma{r}}_{\text{supp}}\in \compl$ and $\tilde{\ma{u}}_{\text{out}}\in \compl$ represent the relay signal after the receive filter and prior to the transmit filter process, respectively. It is worth mentioning that similar ideas for partitioning the design of the relay process is proposed in \cite{RWW:11,HU2:09}, where the impact of the transmit (receive) distortion, in relation to the transmit (receive) signal at the relay is not considered. 
A detailed role and design strategy for the aforementioned parts is elaborated in the following.
%\begin{align} \label{eq_W_three_parts}
%\ma{W}=\omega\ma{w}_{\text{tx}}\ma{w}_{\text{rx}},
%\end{align}
%where $\ma{w}_{\text{rx}}\in \compl^{ d \times {M_{\text{r}}} }$ and $\ma{W}_{\text{tx}}\in \compl^{ {M_{\text{t}}} \times d }$ respectively act as the receive and transmit linear filters at the relay, while $\omega \in \real^+$ acts as a scaling factor, see Fig.~\ref{fig:three_part_W}. Moreover, $d \in \mathbb{N}$ is the rank of $\ma{W}$, and $d \leq \text{min}\left({M_{\text{r}},{M_{\text{t}} \right)$. The intermediate signals $\tilde{\ma{r}}_{\text{supp}}$ and $\tilde{\ma{u}}_{\text{out}}$ are defined as the relay signal after the receive filter and prior to the transmit filter process, respectively. It is worth mentioning that similar ideas for partitioning the relay process is proposed in \cite{RWW:11,HU2:09}, where the impact of the transmit (receive) distortion, in relation to the transmit (receive) signal at the relay is not considered. A detailed role and design strategy for the aforementioned parts is elaborated in the following.   
      
\subsection{Design of $\ma{w}_{\text{tx}}$}
The role of $\ma{w}_{\text{tx}}$ is to direct the relay transmit beam towards the destination, while imposing minimal distortion on the receiver end of the relay and destination nodes. For this purpose we define the following optimization problem
\begin{align} \label{eq_ch_norm_W2_OP} 
\underset{\ma{w}_{\text{tx}},  \| \ma{w}_{\text{tx}}\|_{2} = 1}{\text{max}} \;\;\; & \frac{\mathbb{E}\{ \| \ma{H}_{\text{rd}}  \ma{u}_{\text{out}}  \|_{2}^2 \}  } { \mathbb{E}\{ \| \ma{H}_{\text{D,tx}} \ma{u}_{\text{out}} \|_{2}^2 \} + \text{tr}\left(P_{\text{s}}\ma{h}_{\text{sd}}\ma{h}_{\text{sd}}^H + \sigma_{\text{nd}}^2 \ma{I}_{M_{\text{d}}} \right) },
\end{align}
where the nominator is the desired received power from the relay-destination path, and $\ma{H}_{\text{D,tx}}$ is the equivalent distortion channel observed from the relay transmitter, see Appendix~\ref{appendix:DistortionChannels}. 
%\begin{align} \label{eq_ch_norm_H2} 
%\ma{H}_{\text{D,tx}} := \left \[ \begin{array}{c}
 %\sqrt{\gamma}  \lfloor \ma{H}_{\text{rr}} \ma{\Gamma_i} \rfloor_{ i  \in  \mathbb{F}_{M_{\text{t}}} }\\
 %\sqrt{\beta} \ma{\Gamma}_i \ma{H}_{\text{rr}}\end{array} \right \]
%\end{align}
 %see Appendix~\ref{appendix:DistortionChannels}. 
The optimization problem (\ref{eq_ch_norm_W2_OP}) can be hence formulated as\footnote{For the calculation of the spatial filters $\ma{w}_{\text{tx}}$ and $\ma{w}_{\text{rx}}$, it is assumed that the relay operates with maximum power, to emphasize the impact of hardware distortions. The relay transmit power is afterwards adjusted by the choice of $\omega$. An alternating adjustment of the spatial filters with the optimized relay power is later discussed in Subsection~\ref{sec_AltMuStR1}.}  
\begin{align} \label{eq_ch_norm_W3_OP} 
\underset{\ma{w}_{\text{tx}},  \| \ma{w}_{\text{tx}}\|_{2} = 1}{\text{max}} \;\;\; & \frac{ \ma{w}_{\text{tx}}^H \left(\ma{H}_{\text{rd}}^H  \ma{H}_{\text{rd}}\right) \ma{w}_{\text{tx}}  }   {  \ma{w}_{\text{tx}}^H \left(\ma{H}_{\text{D,tx}}^H  \ma{H}_{\text{D,tx}} + N_{\text{tx}} \ma{I}_{M_{\text{t}}} \right) \ma{w}_{\text{tx}}  }, 
\end{align}
which holds a generalized Rayleigh quotient structure~\cite{li2014rayleigh}, and $N_{\text{tx}}:= \left(M_{\text{d}}\sigma_{\text{nd}}^2 + P_{\text{s}}\|\ma{h}_{\text{sd}}\|_2^2\right)/{P_{\text{r,max}}}$. The optimal transmit filter is hence obtained as 
\begin{align} \label{eq_ch_norm_W2_solution} 
\ma{w}_{\text{tx}}^{\star} :=  \lambda_{\text{max}} \left\{  \left(\ma{H}_{\text{D,tx}}^H  \ma{H}_{\text{D,tx}} +  N_{\text{tx}} \ma{I}_{M_{\text{t}}} \right)^{-1}  \left(\ma{H}_{\text{rd}}^H  \ma{H}_{\text{rd}}\right)\right\}.
\end{align} 

%the ratio between the transmit channel norm, i.e., channel between $\tilde{\ma{u}}_{\text{out}}$ and the destination, to the norm of the equivalent distortion channel, i.e., the channel relating the statistics of $\tilde{\ma{u}}_{\text{out}}$ to the received distortion power at ${\ma{r}}_{\text{supp}}$. The resulting optimization problem can be hence formulated as 
%\begin{align} \label{eq_ch_norm_W2_OP} 
%\underset{\ma{W}_2}{\text{max}} \;\;\; & \frac{\| \ma{H}_{\text{rd}} \ma{W}_2\|_{\text{fro}}}{\| \ma{H}_2 \ma{W}_2 \|_{\text{fro}}}, \;\; \text{s.t.} \;\;  \| \ma{W}_2\|_{\text{fro}} = 1, 
%\end{align}
%where $\ma{H}_2$ is defined as
%\begin{align} \label{eq_ch_norm_H2} 
%\ma{H}_2 := \left \lfloor \begin{array}{c}
 %\sqrt{\gamma}  \ma{H}_{\text{rr}} \ma{\Gamma_i}\\
 %\sqrt{\beta} \ma{\Gamma_i} \ma{H}_{\text{rr}}\end{array} \right \rfloor_{i \in \{1 \cdots M_{\text{r}}\}}
%\end{align}
%see Appendix~\ref{appendix:DistortionChannels}. Note that similar to (\ref{eq_ch_norm_W1_OP}), the objective remains constant with a scaling of $\ma{W}_2$ and the constraint in (\ref{eq_ch_norm_W2_OP}) is added to ensure a norm-bounded solution. The optimal $\ma{W}_2$ can be consequently achieved as  
%\begin{align} \label{eq_ch_norm_W2_solution} 
%\text{vec} \left(\ma{W}_2^{\star} \right) :=  \lambda_{\text{max}} \left\{ \ma{I}_{M_{\text{r}}} \otimes  \left[\left( \ma{H}_2^{H}\ma{H}_2 \right)^{-1} \ma{H}_{\text{rd}}^{H} \ma{H}_{\text{rd}}  \right] \right\}.
%\end{align} 
\subsection{Design of $\ma{w}_{\text{rx}}:$} The role of $\ma{w}_{\text{rx}}$, is to accept the desired received signal from the source, while rejecting the received distortion-plus-noise terms at the relay. Similar to (\ref{eq_ch_norm_W2_OP}) this is expressed as 
\begin{align} \label{eq_ch_norm_W_rx_OP} 
\underset{\ma{w}_{\text{rx}}}{\text{max}} \;\;\; & \frac{ \ma{w}_{\text{rx}}^H \left(\ma{h}_{\text{sr}}  \ma{h}_{\text{sr}}^H \right) \ma{w}_{\text{rx}}  }   {  \ma{w}_{\text{rx}}^H \left( \ma{\Phi} + \sigma_{\text{nr}}^2 \ma{I}_{M_{\text{r}}}  \right) \ma{w}_{\text{rx}}  }, \;\; \text{s.t.} \;\;  \| \ma{w}_{\text{rx}}\|_{2} = 1, 
\end{align}
where 
\begin{align}  \label{eq_ch_norm_W_rx_solution_phi}
\ma{\Phi} &= \kappa  P_{\text{r,max}} \ma{H}_{\text{rr}}\text{diag}\left( \ma{w}_{\text{tx}} \ma{w}_{\text{tx}}^H \right) \ma{H}_{\text{rr}}^H \nonumber \\ &+ \beta P_{\text{r,max}}  \text{diag}\left( \ma{H}_{\text{rr}}  \ma{w}_{\text{tx}} \ma{w}_{\text{tx}}^H  \ma{H}_{\text{rr}}^H \right)  
\end{align}
approximates the covariance of the received distortion signal at the relay. The optimal solution to $\ma{w}_{\text{rx}}$ can be hence obtained as 
 \begin{align} \label{eq_ch_norm_W_rx_solution} 
\ma{w}_{\text{rx}}^{\star} :=  \lambda_{\text{max}} \left\{  \left( \ma{\Phi} + \sigma_{\text{nr}}^2 \ma{I}_{M_{\text{r}}}  \right)^{-1}  \left(\ma{h}_{\text{sr}}  \ma{h}_{\text{sr}}^H \right) \right\} .
\end{align}  

\subsection{Design of $\omega$} \label{Chnormratio_design_omega}
\begin{figure}[t]  
\hspace{0.0cm} \subfigure[${\omega}_0 \leq {\omega}_{\text{max}} $]{\includegraphics[angle = 0, width = 0.48\columnwidth]{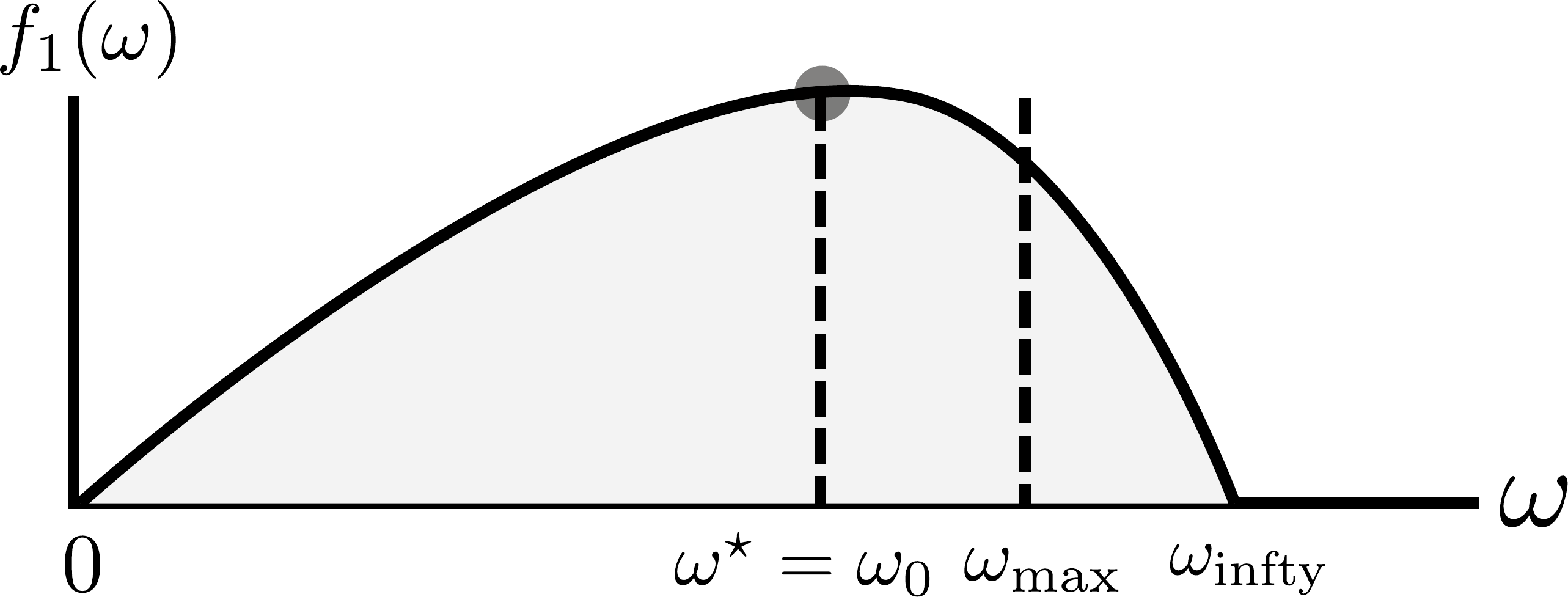}} 
\subfigure[${\omega}_0 >{\omega}_{\text{max}}$]{\includegraphics[width = 0.48\columnwidth]{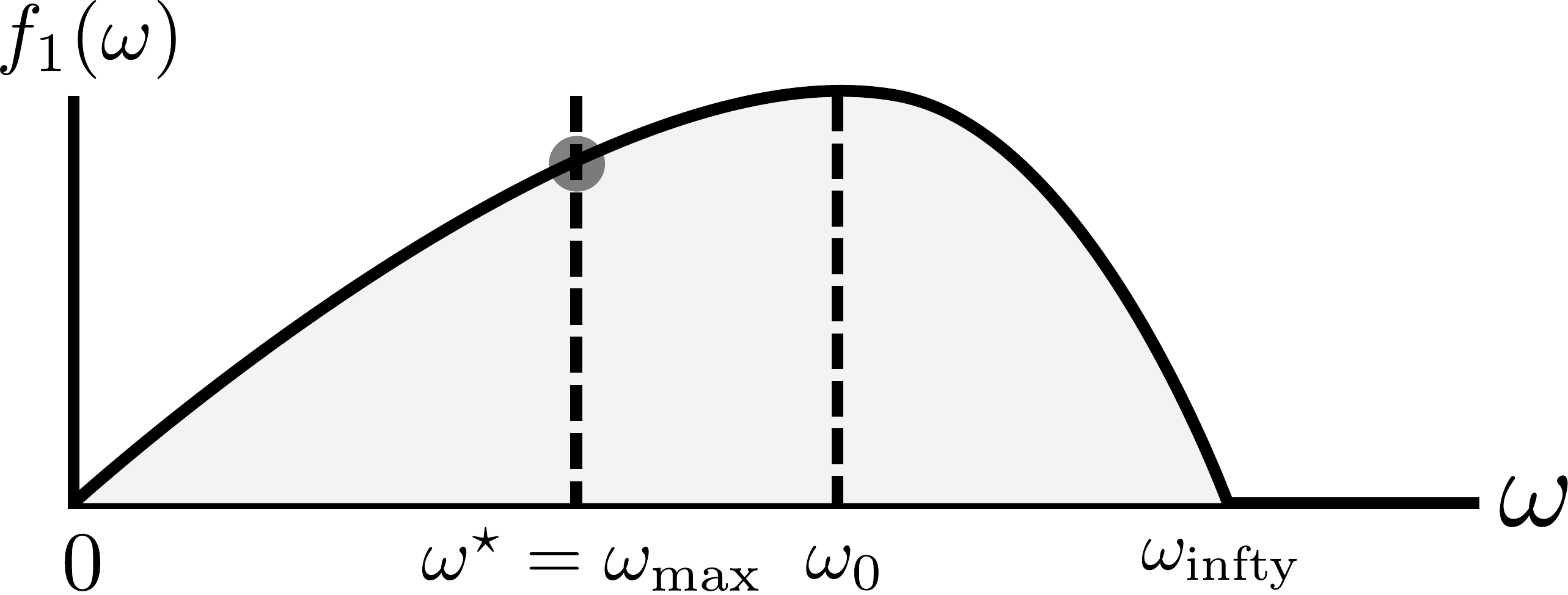}} 
%\LoadClass[onecolumn]{article}
%\addtolength{\linewidth}{3.6in}
\caption{{\footnotesize{Possible situations of ${\omega}_0$ with respect to ${\omega}_{\text{max}}$, considering the feasible region of ${\omega}$. The dark circle indicates the position of the optimum point.}}}
\label{fig:extremum}
\vspace{-7pt}
\end{figure}

The role of $\omega$ is to adjust the amplification intensity at the relay. This plays a significant role, considering the fact that even with optimally designed spatial filters, i.e., $\ma{w}_{\text{tx}}$ and $\ma{w}_{\text{rx}}$, a weak amplification at the relay reduces the desired signal strength at the destination, resulting to a low signal-to-noise ratio. On the other hand, a strong amplification may lead to instability and (theoretically) infinite distortion transmit power, i.e., low signal-to-distortion ratio. Hence, similar to (\ref{model_original_problem}), we focus on maximizing the end-to-end SDNR, assuming that $\ma{w}_{\text{tx}}$ and $\ma{w}_{\text{rx}}$ are given from the previous parts. The end-to-end SDNR and the transmit power from the relay corresponding to a value of $\omega$ are respectively approximated as $f_1$ and $f_2$ such that
\begin{align} \label{eq_ch_norm_omega_equation_sinr} 
f_1 ({\omega}) =  {{a_{\text{d}}} {\omega}} \left( a_0 + \sum_{k \in \mathbb{F}_K } a_k {{\omega}}^k  \right)^{-1},
\end{align} 
and
\begin{align} \label{eq_ch_norm_omega_equation_power} 
f_2 ({\omega}) =    \sum_{k \in \mathbb{F}_K } b_k {{\omega}}^k ,
\end{align} 
where 
\begin{align} 
{a}_{\text{d}} &= P_{\text{s}} \ma{d}_{M_{\text{d}}}^T \left( \ma{H}_{\text{rd}}^* \otimes \ma{H}_{\text{rd}}  \right) \tilde{\ma{W}} \text{vec}\left( \ma{h}_{\text{sr}}  \ma{h}_{\text{sr}}^H\right) , \label{eq_ch_norm_coeffs_1} \\
a_0 &= \ma{d}_{M_{\text{d}}}^T \text{vec} \left( \sigma_{\text{nd}}^2 \ma{I}_{M_{\text{d}}} + P_{\text{s}} \ma{h}_{\text{sd}} \ma{h}_{\text{sd}}^H \right) , \label{eq_ch_norm_coeffs_2} \\
a_1& =  -a_{\text{d}} + \ma{d}_{M_{\text{d}}}^T \left(\ma{H}_{\text{rd}}^{*} \otimes \ma{H}_{\text{rd}}  \right) \left( \ma{I}_{M_{\text{t}}^2} + \kappa\ma{S}_{\text{D}}^{M_{\text{t}}} \right) \tilde{\ma{W}} \ma{c} , \\
a_k &=  \ma{d}_{M_{\text{d}}}^T \left(\ma{H}_{\text{rd}}^{*} \otimes \ma{H}_{\text{rd}}  \right) \left( \ma{I}_{M_{\text{t}}^2} + \kappa\ma{S}_{\text{D}}^{M_{\text{t}}} \right) \nonumber \\
& \quad\quad\quad\quad\quad \times  \left(\tilde{\ma{W}}  \ma{C} \right)^{k-1}  \tilde{\ma{W}} \ma{c} , \; k \in \{2 \ldots K \}, \label{eq_ch_norm_coeffs_3} \\
b_k &= \ma{d}_{M_{\text{t}}}^T \left( \ma{I}_{{M_t}^2} + \kappa \ma{S}_{\text{D}}^{M_{\text{t}}} \right) \left( \tilde{\ma{W}}  \ma{C} \right)^{k-1}\tilde{\ma{W}}\ma{c} , \; k \in \mathbb{F}_K, \label{eq_ch_norm_coeffs_4} 
\end{align} 
see Appendix~\ref{appendix:coefficients} for more details. In the above expressions $K$ is the approximation order and $\ma{d}_M\in\{0,1\}^{M^2}$ is defined such that $\text{tr}(\ma{A}) = \ma{d}_M^T \text{vec}(\ma{A})$, $\ma{A} \in \compl^{M \times M}$. Furthermore $\ma{C},\ma{c},\tilde{\ma{W}}$ are respectively defined in (\ref{eq_loop_A}), (\ref{eq_loop_a}) and in Appendix~\ref{appendix:coefficients}.  
%Note that the above approximations become accurate as \mbox{$K \rightarrow \infty$}. Hence, higher accuracy can be achieved at the expense of higher design complexity\footnotemark[1]\footnotetext[1]{Our numerical study shows that no significant performance gain is observed after $K=3$. More elaboration is provided on this issue in Appendix~\ref{appendix:coefficients}.}. Our goal is to maximize (\ref{eq_ch_norm_omega_equation_sinr}) within the feasible range of $\bar{\omega}$. 
The corresponding optimization problem can be written as 
\begin{subequations} \label{eq_ch_norm_omega_optproblem} 
\begin{align} 
\underset{{\omega}}{\text{max}} \;\;\; & f_1 ({\omega})   \\ 
\text{s.t.} \;\;\; & 0\leq {\omega} \leq \text{min}\{ {\omega}_{\text{infty}}, {\omega}_{\text{max}}\} = {\omega}_{\text{max}},
\end{align}
\end{subequations} 
where ${\omega}_{\text{max}}$ corresponds to the relay amplification that results in a tight transmit power constraint, i.e., $f_2 ({\omega}_{\text{max}}) = {P}_{\text{r,max}}$. Moreover, ${\omega}_{\text{infty}}$ is the smallest pole of $f_2 ({\omega})$ in the real positive domain which yields to the instability of the relay distortion loop, i.e., infinite relay transmit power. We observe that while $f_1 ({\omega})$ remains positive and differentiable in the domain $[0, {\omega}_{\text{infty}} )$, we have $\underset{{\omega}\rightarrow 0}{f_1} ({\omega}) \rightarrow 0$ and $\underset{{\omega}\rightarrow {\omega}_{\text{infty}}}{f_1} ({\omega}) \rightarrow 0$. This concludes the existence of (at least) one local maximum point in this domain, see Fig.~\ref{fig:extremum} for a visual description. By setting the derivative of (\ref{eq_ch_norm_omega_equation_sinr}) to zero, we conclude that the resulting extremum points are necessarily located such that
\begin{align} \label{eq:channelnorm_der} 
a_0 = \sum_{k\in\mathbb{F}_K } (k-1) a_k {{\omega}}^k. 
\end{align}
While the left side of (\ref{eq:channelnorm_der}) is a constant, the right side of the equality is monotonically increasing with respect to ${\omega}\in \real^+$, as $a_k \geq 0, \forall k$. This readily results in the exactly one extremum point in the positive domain of ${\omega}$\footnote{Note that both $f_2 ({\omega}_{\text{max}}) = {P}_{\text{r,max}}$ or (\ref{eq:channelnorm_der}) result in exactly one solution for $\omega$ in $\real^+$, as $a_k,b_k \geq 0, \forall k$. In this regard, values of ${\omega}_{\text{max}}$ and ${\omega}_{0}$ can be obtained via a bi-section search, or can be obtained in closed-form for small values of $K$, i.e., $K\leq 3$, as a known polynomial root.}. Hence, the optimality occurs either at the obtained extremum point, i.e., a local maximum in the domain $[0, {\omega}_{\text{max}} )$, see Fig.~\ref{fig:extremum}-a, or at the point where the relay transmit power constraint is tight, see Fig.~\ref{fig:extremum}-b. The optimum ${\omega}$ can be hence formulated as  
\begin{align} \label{eq_ch_norm_omega_solution} 
{\omega}^{\star} = \text{min} \{ {\omega}_0 , {\omega}_{\text{max}}  \},
\end{align}
where ${\omega}_0$ is the only solution of (\ref{eq:channelnorm_der}) in the positive domain.

\subsection{Design of $\ma{z}$}
The optimal design of $\ma{z}$ is given in (\ref{eq_GP_z}) as a closed form expression. Note that the solution of $\ma{z}$ is dependent on the choice of other optimization variables. However, as the proposed designs for the other optimization variables do not depend on $\ma{z}$, there is no need for further alternation among the design parameters. The Algorithm~\ref{algorithm:channel_norm_max} defines the required steps.   
\begin{algorithm}[H]
 \small{	\begin{algorithmic}[1]
  %\SetAlgoLined
\State{$P_{\text{s}} \leftarrow  P_{\text{s,max}} $ }
%\State{$\ma{H}_1 \leftarrow  \text{ calculate equivalent Rx distortion channel, see (\ref{eq_ch_norm_H1})} $   }
\State{$\ma{w}_{\text{tx}} \leftarrow  \text{ calculate Tx filter, see (\ref{eq_ch_norm_W2_solution})} $   }
\State{$\ma{w}_{\text{rx}} \leftarrow  \text{ calculate Rx filter, see (\ref{eq_ch_norm_W_rx_solution})} $       }
\State{$\omega \leftarrow  \text{adjust amplification intensity, see Subsection~\ref{Chnormratio_design_omega}} $   }
%\State{$a_{\text{d}},{a}_k,{b}_k,\; k \in \{1 \cdots K \} \leftarrow \text{ see (\ref{eq_ch_norm_coeffs_1})-(\ref{eq_ch_norm_coeffs_4})} $}
%\State{$ {\omega}^{\star} \leftarrow  \text{calculate equivalent Tx filter, see (\ref{eq_ch_norm_omega_solution})} $}
%\State{$ \ma{W} \leftarrow  {\omega}^{\star}\ma{W}_1^{\star}\ma{W}_2^{\star}\text{, see (\ref{eq_W_three_parts})} $}
\State{$\ma{z} \leftarrow \text{ see (\ref{eq_GP_z})}$}
%\Repeat   $\;\; \text{(running for multiple initializations)}$
%\State{$\text{Counter} \leftarrow \text{Counter} + 1$}
%\State{$\ell \leftarrow 0$}
%\State{$\ma{Q}^{(0)} \leftarrow  \text{random init., see (\ref{eq_loop_optimization_problem_b})} $}
%\State{$\ma{W}^{(0)} \leftarrow  {\ma{Q}^{(0)}}^{\frac{1}{2}} \ma{R}^{\frac{-1}{2}}\left(\ma{Q}^{(0)}\right), \text{ see (\ref{eq_GP_proj_W})-(\ref{eq_GP_R_define})}  $}
%\Repeat 
%\State{$\ell \leftarrow \ell + 1 $}
%\State{$\ma{W}^{(\ell)} \leftarrow \text{update } \ma{W}^{(\ell)}, \text{see (\ref{eq_GP_proj_W}), Section IV.B} $}
%\State{$\ma{z}^{(\ell)} \leftarrow \text{update }, \text{see (\ref{eq_GP_z})}$}
%%\If{ \text{stable SINR value}}
%\State{$\text{SINR}^{(\ell)} \leftarrow  \text{(\ref{eq_loop_optimization_problem_a})}$}
%%\State{break} 
%%\EndIf
%\Until {$\text{SINR}^{(\ell)} - \text{SINR}^{(\ell-1)} \geq c_1$    (until SINR improves)}
%\State{$\mathcal{A} \leftarrow  \text{save} \left(\text{SINR}^{(\ell)}, \ma{W}^{(\ell)}, \ma{z}^{(\ell)}\right)$}
%\Until {$\text{Counter} \leq C_1 $ }
%\State{\Return{$\left(\ma{W},\ma{z},\text{SINR}\right)\; \leftarrow \; {\text{max}} \; \text{SINR}\in\mathcal{A} $}}%%%
\State{ \Return{$\left(\ma{z}, \; \ma{W}=\omega\ma{w}_{\text{tx}}\ma{w}_{\text{rx}}^H\right) $}} %%%
  \end{algorithmic}  
 \caption{\small{Distortion Aware Multi-Stage Rank-1 Relay amplification (MuStR1) for SDNR maximization. The solution is obtained with no alternation among the optimization variables.} }   \label{algorithm:channel_norm_max}  } 
 %\label{algorithm:channel_norm_max}
\end{algorithm}  
 
\subsection{Alternating enhancement of MuStR1 (AltMuStR1)} \label{sec_AltMuStR1}
The proposed MuStR1 design is accomplished with no alternation among the optimization variables, see Algorithm~\ref{algorithm:channel_norm_max}. In this part we provide an alternating enhancement of MuStR1 which results in an increased performance, at the expense of a higher computational complexity. In order to accomplish this purpose, similar to (\ref{eq_loop_P_desired}) and (\ref{eq_loop_P_error}) we focus on the signal and power values at the destination, after the application of $\ma{z}$. In this respect, the values of $\ma{w}_{\text{tx}}, \ma{w}_{\text{rx}}, \ma{z}$ and $\omega$ will be calculated as a joint alternating optimization. This is done by replacing $\ma{H}_{\text{rd}}$ with $\ma{z}^H\ma{H}_{\text{rd}}$ in the design of $\ma{w}_{\text{tx}}$, and $\ma{d}^T$ with $\left(\ma{z}^{T}\otimes \ma{z}^{H} \right)$ in (\ref{eq_ch_norm_coeffs_1})-(\ref{eq_ch_norm_coeffs_3}). The steps~$2$-$6$ in Algorithm~\ref{algorithm:channel_norm_max} are then repeated until a stable point is achieved, or a maximum number of iterations is expired. The performance of the proposed (Alt)MuStR1 algorithms in terms of the resulting communication rate, convergence, and computational complexity are studied via numerical simulations in Subsection~\ref{sim_alg_analysis}. In particular, it is observed that the performance of the AltMuStR1 algorithm reaches close to the performance of GP, with a significantly lower computational complexity.  

\revOmid{
\subsection{Convergence}
Due to the proposed SDNR approximation, as well as the sub-optimal solutions for $\ma{w}_{\text{tx}},\ma{w}_{\text{tx}}$ at each iteration, the convergence of the AltMuStR1 algorithm is not theoretically guaranteed. However, it is observed via numerical simulations in Subsection\ref{sim_alg_analysis}, that the algorithm shows a fast average convergence, with a close performance to the proposed GP method for a wide range of system parameters. 

\subsection{Computational complexity}
In this part we study the computational complexity associated with AltMuStR1, following simular arguments as given in Subsection~\ref{GP_CC}.
\subsubsection{Algorithm complexity}
%The complexity of the algorithm is dominated by the calculation of the transmit and receive filters, $\ma{w}_{\text{tx}}, \ma{w}_{\text{rx}}$. Note that the constant scalars (\ref{eq_ch_norm_coeffs_3}) and (\ref{eq_ch_norm_coeffs_3}) can be calculated efficiently due to the rank-1 relay amplifications. This can be observed by considering  
%\begin{align}
%\tilde{\ma{W}} = (\ma{w}_{\text{tx}} \ma{w}_{\text{rx}}^H)^* \otimes (\ma{w}_{\text{tx}} \ma{w}_{\text{rx}}^H) = ( \ma{w}_{\text{tx}}^* \otimes \ma{w}_{\text{tx}} ) ( \ma{w}_{\text{rx}}^T \otimes \ma{w}_{\text{rx}}^H ), 
%\end{align}
%%and 
%\begin{align}
%\tilde{\ma{W}} (\ma{H}_{\text{rr}}^* \otimes \ma{H}_{\text{rr}}) & = \Big((\ma{w}_{\text{tx}} \ma{w}_{\text{rx}}^H)^* \otimes (\ma{w}_{\text{tx}} \ma{w}_{\text{rx}}^H)\Big)  (\ma{H}_{\text{rr}}^* \otimes \ma{H}_{\text{rr}})  \nonumber \\ &=  \Big( (\ma{w}_{\text{tx}} \ma{w}_{\text{rx}}^H)^*  \ma{H}_{\text{rr}}^* \Big) \otimes \Big( (\ma{w}_{\text{tx}} \ma{w}_{\text{rx}}^H) \ma{H}_{\text{rr}} \Big).
%\end{align}
%which turns the sequences of matrix multiplications in (45) and (46) into a sequence of vector multiplications. 
%is useful in the calculation of the $\tilde{\ma{W}} \ma{C}$ terms and results, for instance, in the complexity of $\mathcal{O} \left(2 M_{\text{t}}^2 + 2 M_{\text{t}}M_{\text{r}} + (k-1)(3 M_{\text{t}}^2-1) \right)$ for the calculation of $b_k$ (46). 
The calculation of $\ma{w}_{\text{rx}}$ is dominated by the expressions (\ref{eq_ch_norm_W_rx_solution_phi}), (\ref{eq_ch_norm_W_rx_solution}). Exploiting the rank-1 structure of the matrix $\ma{h}_{\text{sr}}\ma{h}_{\text{sr}}^H$ incurring $\mathcal{O} \left( M_{\text{r}}^3 + 2 M_{\text{r}}^2M_{\text{t}}\right)$ FLOPs. Similarly, the calculation of $\ma{w}_{\text{tx}}$ is dominated by (\ref{eq_ch_norm_W2_solution}), resulting in $\mathcal{O} \left(11 M_{\text{t}}^3/3 + 2 M_{\text{t}}^2M_{\text{r}} + 2 M_{\text{t}}^2M_{\text{d}}\right)$ FLOPs. Together with the calculation of $\ma{z}$ this requires in total
{\small{\begin{align}
\mathcal{O} \Big( \gamma_3  \left( M_{\text{r}}^3  + 11 M_{\text{t}}^3/3  + 2 M_{\text{r}}^2M_{\text{t}} + 2 M_{\text{t}}^2M_{\text{r}} + 2 M_{\text{t}}^2M_{\text{d}} +  M_{\text{d}}^3  \right) \Big)
\end{align}}}
FLOPs, where $\gamma_3$ is the number of the AltMuStR1 algorithm iterations.

\subsubsection{Processing complexity}
Since $\ma{W}$ follows a general rank-1 structure, the amplification is simplified to only $M_{\text{t}}+M_{\text{r}}+1$ complex multiplications, and $M_{\text{r}}-1$ summations, totaling $M_{\text{t}}+2M_{\text{r}}$ FLOPs. 
}

\section{Comparison with Decode-and-Forward Relaying Under Residual Self-Interference} \label{FD_DF}
In the previous sections we studied the behavior of an AF-FD relay under residual self-interference, from the aspects of performance analysis and design methodologies. While an AF relaying is known to provide a simple strategy, it particularly suffers from the aforementioned distortion loop especially when the residual interference power is not ignorable, see~Subsection~\ref{discussion:distortionLoop}. In this part, we provide a similar study for a setup where the relay operates with DF protocol, in order to act as a comparison benchmark. In a DF relay, the discussed distortion loop is significantly alleviated as the decoding process eliminates the inter-dependency of the received residual interference intensity to the relay transmit signal power, see \cite[Section~V]{TRCM:15} for a similar discussion regarding single antenna relays. Note that optimization strategies for FD relays with DF process have been discussed in the literature, see~\cite{XaZXMaXu:15, DMBSR:12,ALRWW:14,RVRWW:15}, for different relaying setups and assumptions. In order to adopt the available literature to our setup, we follow a modified version of the approach given in \cite{ALRWW:14, RVRWW:15}, where an SDNR maximization problem is studied at a MIMO DF relay. Note that the proposed design in \cite{ALRWW:14,RVRWW:15} considers a maximization of the SDNR at the relay input/output, taking advantage of spatial and temporal filters. In our system, we focus on the resulting performance from the source to the destination, after the application of the receive filter $\ma{z}$, and hence including $P_{\text{s}}$ and $\ma{z}$ as optimization variables. Afterwards, similar to \cite{RVRWW:15}, we propose a joint optimization where in each step one of the variables is updated to the optimality. A detailed model update and design strategy is given in the following. 

\subsection{DF Relaying} We define $\ma{v}_{\text{in}} \in \compl^{M_{\text{r}}}$ as linear filter at the relay input, $\hat{s}_{\text{r}} \in \compl$, $\mathbb{E}\{|\hat{s}_{\text{r}}|^{2}\}=1$ as the decoded symbol, and $\ma{v}_{\text{out}} \in \compl^{M_{\text{t}}}$ as the beamforming vector to transmit the decoded data symbol, see Fig.~\ref{fig:DF_relay}. The end-to-end rate maximization can be equivalently formulated as 
\begin{subequations} \label{eq_DF_Op_array} 
\begin{align}  
\underset{\ma{v}_{\text{in}}, \ma{z}, \ma{Q}\in \mathcal{H}, P_{\text{s}} }{\text{max}} \;\; & \text{min} \left( \zeta_{\text{sr}}, \zeta_{\text{rd}} \right), \;\; \label{eq_DF_Op_a} \\ 
\text{s.t.} \;\;\;\;\;\; &\text{tr}\left( \ma{Q} \right) \leq \tilde{P}_{\text{r,max}}, \;P_{\rm {s}} \leq P_{\text{s,max}},\; \text{rank}(\ma{Q}) = 1,\label{eq_DF_Op_b}
\end{align}
\end{subequations}
where the rank-one constraint is imposed to ensure the structure $\ma{Q}:=\mathbb{E} \{ \ma{u}_{{\rm out}} \ma{u}_{{\rm out}}^{{H}} \} = \ma{v}_{\rm out} \ma{v}_{\rm out}^{H}$. The values $\zeta_{\text{sr}}$ and $\zeta_{\text{rd}}$, respectively represent the resulting SDNR in source-to-relay and relay-to destination links which can be calculated as 
\begin{align} 
\zeta_{\text{sr}} & \approx \frac{ P_{\text{s}}\ma{v}_{\text{in}}^H  \ma{h}_{{\rm sr}} \ma{ h}_{{\rm sr}}^H \ma{v}_{\text{in}} }{ \hspace{-0.5mm} \ma{v}_{\text{in}}^H  \Big( \sigma_{\text{nr}}^2 \ma{I}_{M_{\text{r}}} \hspace{-0.9mm} + \hspace{-0.9mm} \kappa \ma{H}_{\text{rr}} \text{diag} (\ma{Q}) \ma{H}_{\text{rr}}^H  \hspace{-0.9mm} +  \hspace{-0.9mm} \beta \text{diag} \left(\ma{H}_{\text{rr}} \ma{Q} \ma{H}_{\text{rr}}^H  \right) \hspace{-0.9mm} \Big) \ma{v}_{\text{in}}} ,\label{eq_DF_SINRsr}  \\
\zeta_{\text{rd}} & = \frac{ \ma{z}^H  \ma{H}_{\text{rd}} \ma{Q}  \ma{H}_{\text{rd}}^H \ma{z} }{\ma{z}^H  \Big(\sigma_{\text{nd}}^2 \ma{I}_{M_{\text{d}}}  + \kappa \ma{H}_{\text{rd}} \text{diag} (\ma{Q}) \ma{H}_{\text{rd}}^H + P_{\text{s}} \ma{h}_{\text{sd}}\ma{h}_{\text{sd}}^H\Big)  \ma{z}}. \label{eq_DF_SINRrd} 
\end{align}  
where the approximation in (\ref{eq_DF_SINRsr}) follows considering the fact that the self-interference signal constitutes the dominant part of the power in $\ma{u}_{{\rm in}}$, and can be regarded as the only cause of distortion signals at the relay, see also (\ref{eq_loop_U_in}). 
%It is easy to observe that at the source power constraint is necessarily tight at the optimality, as both $\zeta_{\text{sr}}$ and $\zeta_{\text{rd}}$ are monotonically increasing with respect to $P_{\text{s}}$. 
Note that (\ref{eq_DF_Op_array}) is not a convex optimization problem, and also can not be solved in a closed form. Nevertheless, recognizing the generalized Rayleigh quotient structure in (\ref{eq_DF_SINRsr}), (\ref{eq_DF_SINRrd}) we obtain 
\begin{align} 
\ma{z}^{\star} & =   \Big( \sigma_{\text{nd}}^2 \ma{I}_{M_{\text{d}}}  + \kappa \ma{H}_{\text{rd}} \text{diag} (\ma{Q}) \ma{H}_{\text{rd}}^H  +  P_{\text{s}} \ma{h}_{\text{sd}}\ma{h}_{\text{sd}}^H\Big)^{-1} \ma{H}_{\text{rd}} \ma{v}_{\text{out}} , \label{eq_DF_z_star}\\
\ma{v}_{\text{in}}^{\star} & =  \Big( \sigma_{\text{nr}}^2 \ma{I}_{M_{\text{r}}}  + \kappa \ma{H}_{\text{rr}} \text{diag} (\ma{Q}) \ma{H}_{\text{rr}}^H +  \beta \text{diag} \left(\ma{H}_{\text{rr}} \ma{Q} \ma{H}_{\text{rr}}^H  \right) \Big)^{-1} \nonumber \\ & \;\;\;\;\;\;\;\;\;\; \times \ma{h}_{\text{sr}}\sqrt{P_{\text{s}}}   , \label{eq_DF_v_in_star}
\end{align}  
where $\ma{z}^{\star}$, $\ma{v}_{\text{in}}^{\star}$ represent the optimum $\ma{z}$, $\ma{v}_{\text{in}}$ for a fixed $\ma{Q}$ and $P_{\text{s}}$. In order to obtain the optimal $\ma{Q}, P_{\text{s}}$, the problem in (\ref{eq_DF_Op_array}) can be re-formulated as 
\begin{subequations} \label{eq_DF_opt_Ps_Q}
\begin{align} 
\underset{ P_{\text{s}}, \zeta \in \real^+, \ma{Q}\in \mathcal{H} }{\text{max}} \;\; & \zeta \label{eq_DF_Op2_A}  \\
\text{s.t.} \;\;\;\;\;\;\; & \zeta \leq \zeta_{\text{sr}} , \; \zeta \leq \zeta_{\text{rd}} , \label{eq_DF_Op2_B}  \\
&\ma{Q}\in \mathcal{H}, \; \text{tr}\left( \ma{Q} \right) \leq \tilde{P}_{\rm{r,max}}, \; \text{rank}(\ma{Q}) = 1, \label{eq_DF_Op2_C} 
\end{align}
\end{subequations} 
where the values of $\ma{v}_{\text{in}}$ and $\ma{z}$ are fixed. By temporarily relaxing the rank constraint, the above problem can be written as a convex feasibility check over $\ma{Q} , P_{\text{s}}$, for each value of $\zeta$ as 
\begin{subequations} \label{eq_DF_Op3_array}
\begin{align} 
{\text{find}} \;\; & \ma{Q}, P_{\text{s}} \label{eq_DF_Op3_a}  \\
\text{s.t.} \;\; & \text{tr}\left( \ma{Q} \right) \leq \tilde{P}_{\text{r,max}},\;\;  P_{\text{s}} \geq 0,\;  \ma{Q}\in \mathcal{H}, \label{eq_DF_Op3_d} \\
%&  \label{eq_DF_Op3_c}  \\
&   \text{tr}\left( {\ma{\Psi}}_{\text{sr}} +  {\ma{\Phi}}_{\text{sr}} \ma{Q}  \right) \geq 0, \; \text{tr}\left( {\ma{\Psi}}_{\text{rd}} +  {\ma{\Phi}}_{\text{rd}} \ma{Q}  \right) \geq 0, \label{eq_DF_Op3_b}   
\end{align}
\end{subequations} 
where
\begin{align} \label{eq_DF_PsiPhi} 
{\ma{\Psi}}_{\text{rd}} &:=  - \zeta  \ma{z}\ma{z}^H \left( \sigma_{\text{nd}}^2 \ma{I}_{M_{\text{d}}} + {P}_{\text{s}} \ma{h}_{{\text{sd}}} \ma{h}_{{\text{sd}}}^H \right) \\
{\ma{\Psi}}_{\text{sr}} &:= \ma{v}_{\text{in}}\ma{v}_{\text{in}}^H \left( {P}_{\text{s}} \ma{h}_{{\text{sr}}} \ma{h}_{{\text{sr}}}^H  - \zeta \sigma_{\text{nr}}^2 \ma{I}_{M_{\text{r}}}  \right), \\
{\ma{\Phi}}_{\text{rd}} &:=  \ma{H}_{{\text{rd}}}^H \ma{z} \ma{z}^H \ma{H}_{{\text{rd}}} - \zeta  \kappa \text{diag} \left( \ma{H}_{{\text{rd}}}^H \ma{z}\ma{z}^H \ma{H}_{{\text{rd}}} \right)  , \\
{\ma{\Phi}}_{\text{sr}} &:= -\zeta  \left( \kappa \text{diag} \left( \ma{H}_{{\text{rr}}}^H \ma{v}_{\text{in}}\ma{v}_{\text{in}}^H \ma{H}_{{\text{rr}}} \right) + \beta \ma{H}_{{\text{rr}}}^H \text{diag} \left( \ma{v}_{\text{in}}\ma{v}_{\text{in}}^H \right) \ma{H}_{{\text{rr}}} \right).
\end{align}
Note that the iterations of the defined feasibility check will be continued, following a bi-section search procedure until an optimum $\zeta$ is obtained with sufficient accuracy. Fortunately, for the special structure of (\ref{eq_DF_Op3_array}) as a complex-valued semi-definite program with three linear constraints, we can always achieve an optimal rank-$1$ $\ma{Q}\in \mathcal{H}$, following \cite[Theorem~3.2]{PalomarSDPRank}. The optimal $\ma{v}_{\text{in}}$ is hence obtained as 
\begin{align} \label{eq_DF_v_out} 
\ma{v}_{\text{out}}^\star = {\ma{Q}^{\star}}^{\frac{1}{2}},   
\end{align}
where $\ma{Q}^{\star}$ represents the obtained $\ma{Q}$ for the highest feasible $\zeta$. 
\revOmid{\subsubsection{Algorithm initialization} \label{alg_DF_initi}
The initial value of $\ma{Q}$ is obtained by setting $\ma{v}_{\text{out}}$ as the dominant eigenvector of $\ma{H}_{\text{rd}}$, resulting in a maximum-ratio transmission, and assuming a maximum transmit power at the relay. The initial values of $\ma{v}_{\text{in}}, \ma{z}$ are then obtained from (\ref{eq_DF_z_star}) and (\ref{eq_DF_v_in_star}). Moreover, the feasible values of $\zeta$ is necessarily located in the region $[0, \zeta_{\text{max}}]$ where 
\begin{align} \label{eq_DF_zeta_max} 
\zeta_{\text{max}} := \text{min}  \left\{ \frac{{P}_{\text{s,max}} \|\ma{h}_{{\rm sr}}\|_2^2}{\sigma_{\text{nr}}^2} ,  \frac{ {P}_{\text{r,max}} \lambda_{\text{max}} \left\{ \ma{H}_{{\text{rd}}}\ma{H}_{{\text{rd}}}^H\right\} }{\sigma_{\text{nd}}^2} \right\}
\end{align}
corresponds to the minimum of the individual link qualities, assuming $\beta=\kappa=0$, and will be used as the initial bounds for the aforementioned bi-section search.
\subsubsection{Convergence and optimality gap}
The solutions in (\ref{eq_DF_z_star})-(\ref{eq_DF_v_out}), result in a necessary improvement of $\zeta$, since in each step the defined sub-problem is solved to optimality. Moreover, since the performance of the studied relay system is bounded from above, see~(\ref{eq_DF_zeta_max}), the defined iterative update results in a necessary convergence. However, the global optimality of the obtained solution may not be guaranteed, since the problem is not a jointly convex optimization problem. Hence, the resulting solution depends on the used initialization. On order to observe this, we have repeated Algorithm~\ref{algorithm:DF} with several random initializations in Fig.~\ref{fig_DF_init}, in comparison to the initialization defined in Subsection~\ref{alg_DF_initi}. It is observed that the proposed initialization in Algorithm~\ref{algorithm:DF} reaches close to the best achievable performance via several initializations. Hence, it is used as the performance benchmark for an achievable SDNR in an FD-DF relaying system.     
\begin{figure}[!t] 
\begin{center}
        \includegraphics[angle=0,width=\MainFigureSizes]{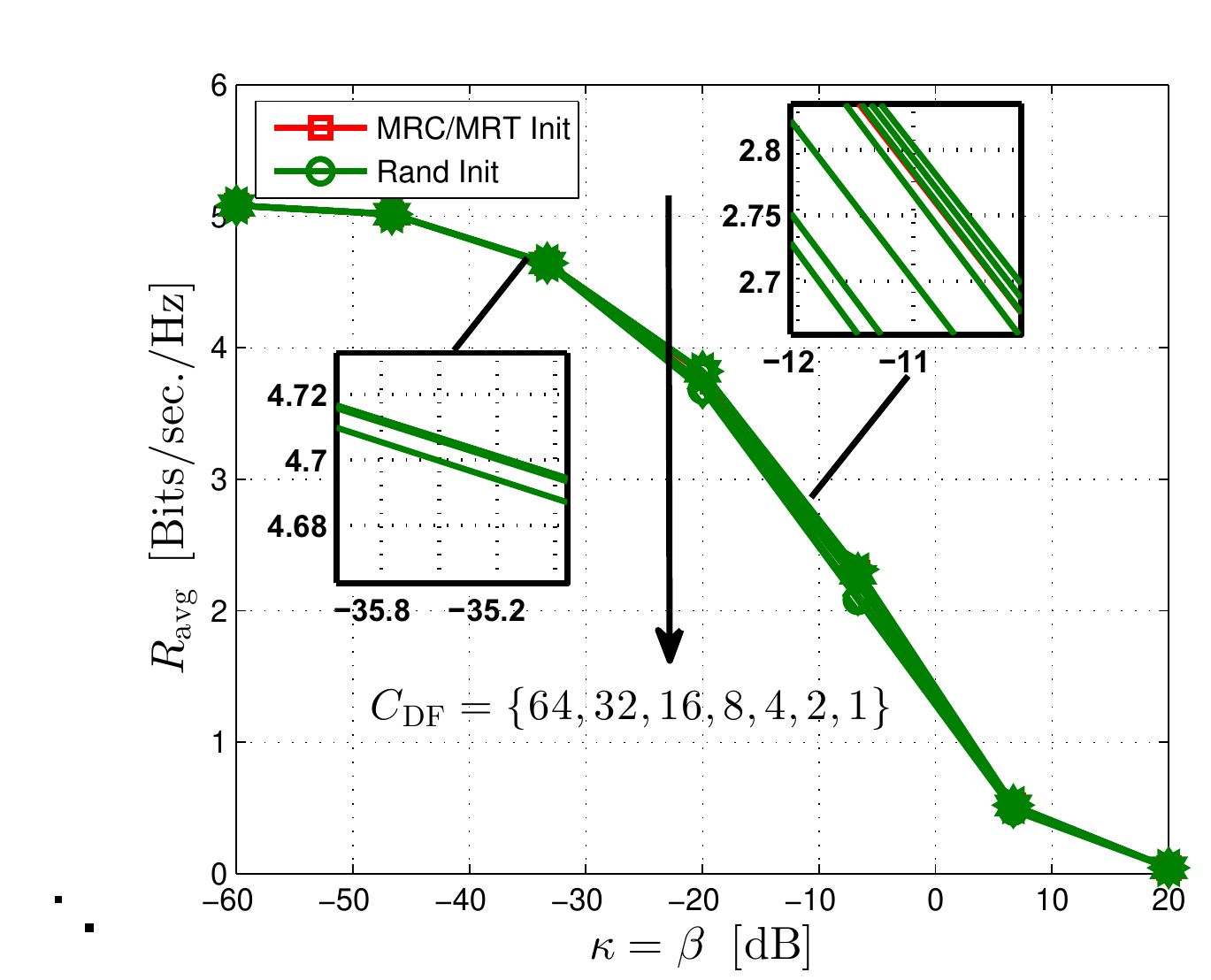}
				        %\fbox{model_rect6.pdf}
    \caption{Evaluation of the DF relaying performance with multiple random initializations, compared to the used MRC/MRT initialization. $C$} \label{fig_DF_init}
    \end{center} \vspace{-0mm} 
\end{figure}
}

\begin{algorithm}[H]
 \small{	\begin{algorithmic}[1]
  %\SetAlgoLined
\State{$P_{\rm s} \leftarrow  P_{\text{s,max}} \times 10^{-2}  $}
\State{$\text{counter}_{1} \leftarrow 0$}
\Repeat    $\;\; \text{(main optimization steps)}$
\State{$\text{counter}_{1} \leftarrow \text{counter}_{1} + 1$}
\State{$\ma{v}_{\text{in}} \leftarrow \ma{v}_{\text{in}}^{\star}, \;\; \text{see~(\ref{eq_DF_v_in_star})}$}
\State{$\ma{z} \leftarrow \ma{z}^{\star}, \;\; \text{see~(\ref{eq_DF_z_star})}$}
\State{$\text{counter}_{2} \leftarrow 0$}
\State{$\zeta_{\text{min}} \leftarrow 0$}
\State{$\zeta_{\text{max}} \leftarrow \text{see~(\ref{eq_DF_zeta_max})} $}
\Repeat   $\;\; \text{(bi-section search steps)}$
\State{$\zeta \leftarrow (\zeta_{\text{min}} + \zeta_{\text{max}})/2$}
\If{\text{(\ref{eq_DF_Op3_a})-(\ref{eq_DF_Op3_d}) is feasible}}
\State{$\zeta_{\text{min}} \leftarrow \zeta$}
\Else 
\State{$\zeta_{\text{max}} \leftarrow \zeta$}
\EndIf
\Until {$\text{counter}_2 \leq C_2 $ }
\State{$\zeta^{\text{counter}_1}  \leftarrow (\zeta_{\text{min}} + \zeta_{\text{max}})/2$}
\State{$P_{\text{s}} \leftarrow P_{\text{s}}^{\star}, \;\; \text{see~(\ref{eq_DF_opt_Ps_Q})}$}
\State{$\ma{v}_{\text{out}} \leftarrow \ma{v}_{\text{out}}^{\star}, \;\; \text{see~(\ref{eq_DF_v_out})}$}
\Until {$\text{counter}_1 \leq C_{\text{DF}} \;\; \text{or}\;\; \zeta^{\text{counter}_1} - \zeta^{\text{counter}_1-1} \leq {c}_1$ }
\State{\Return{$\left(P_{\text{s}},\ma{z},\ma{v}_{\text{in}},\ma{v}_{\text{our}},\right)$}}%%%
  \end{algorithmic}  
 \caption{\small{An iterative rate maximization algorithm for decode-and-forward relaying. $C_{\text{DF}} ,C_2 \in \mathbb{N}$ respectively represent the number of the main optimization steps and the bi-section process, and ${c}_1 \in \real^+$ determines the stability condition.} }  \label{algorithm:DF} } 
\end{algorithm}

\begin{figure}[t]  
{\includegraphics[angle = 0, width = 0.99\columnwidth]{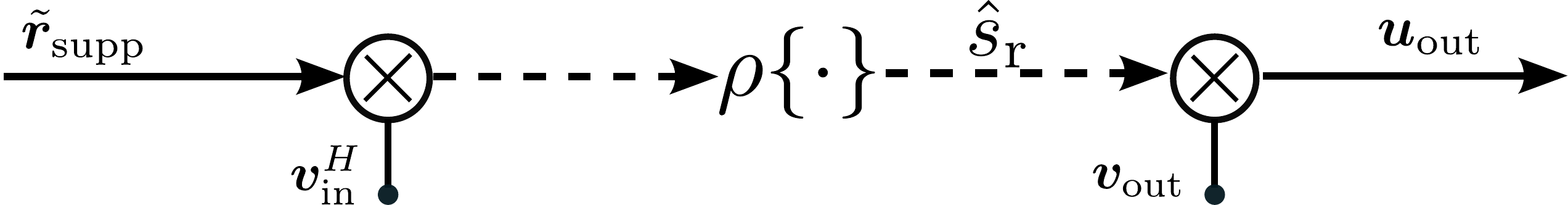}} 
%\LoadClass[onecolumn]{article}
%\addtolength{\linewidth}{3.6in}
\caption{\footnotesize{A schematic of a decode-and-forward relaying process. $\rho\{\cdot\}$ and $\hat{s}_{\text{r}}$ represent the decoding process, and the estimate of the transmitted symbol, $s$, at the relay. The receive and transmit spatial filters are represented as $\ma{v}_{\text{in}}$ and $\ma{v}_{\text{out}}$, respectively. The bold arrows represent the vector signals while the dashed arrows represent the scalars. }} 
\label{fig:DF_relay}
\end{figure}
\section{Simulation Results} 
%\subsection{Computational Complexity}
\begin{table*}[!t] \label{}
  \renewcommand{\arraystretch}{1.2}
  \caption{Default setup parameters}\label{tab:DefaultSetup}
  \centering
  \begin{tabular}[t]{|l||c|c|c|c|c|c|c|c|c|}
    %\firsthline
		    \hline
      Parameter & $P_{\text{max}} := P_{\text{s,max}} = P_{\text{r,max}}$ & $\sigma_{\text{n}}^2 := \sigma_{\text{nr}}^2 = \sigma_{\text{nd}}^2$ & $\kappa=\beta$ & $ M := M_{\text{t}} = M_{\text{r}}=M_{\text{d}}$ & $\rho_{\text{rr}}$ & $\rho_{\text{sr}} = \rho_{\text{rd}}$ & $\rho_{\text{sd}}$ & $K_R$ \\
    \hline
   Value & $1~[Watt]$ & $-40~\text{[dBW]}$ & $-40~\text{[dB]}$ & $4$ & $1$ & $-30~\text{[dB]}$ & $-60~\text{[dB]}$ & $10$ \\
    %\lasthline
		    \hline
  \end{tabular}
\end{table*}

In this section we evaluate the behavior of the studied FD AF relaying setup via numerical simulations. In particular, we evaluate the proposed GP design in Section~\ref{Sec_GP}, as well as the (Alt)MuStR1 algorithms in Section~\ref{section:channel_norm_1}, under the impact of hardware inaccuracies, in comparison with the available relevant methods in the literature. We assume that $\ma{h}_{\text{sr}}, \ma{H}_{\text{rd}}$ and $\ma{h}_{\text{sd}}$ follow an uncorrelated Rayleigh flat-fading model, where $\rho_{\text{sr}},\, \rho_{\text{rd}}$ and $\rho_{\text{sd}}\in \real^+$ represent the path loss. For the self-interference channel, we follow the characterization reported in \cite{FD_ExperimentDrivenCharact}. In this respect we have $\ma{H}_{\text{rr}} \sim \mathcal{CN}\left( \sqrt{\frac{\rho_{\text{rr}} K_R}{1+K_R}} \ma{H}_0 , \frac{1}{1+K_R} \ma{I}_{M_{\text{t}}} \otimes \ma{I}_{M_{\text{r}}} \right)$ where $\rho_{\text{rr}}$ represents the self-interference channel strength, $\ma{H}_0$ is a deterministic term\footnote{For simplicity, we choose $\ma{H}_0$ as a matrix of all-$1$ elements.} and $K_R$ is the Rician coefficient. For each channel realization the resulting performance in terms of the communication rate, i.e., $\log_2 (1+ \text{SDNR})$, is evaluated by employing different design strategies and for various system parameters. \revOmid{The overall system performance in terms of the average rate, i.e., $R_{\text{avg}}$, is then evaluated via Monte-Carlo simulations by employing $500$ channel realizations.} Unless explicitly stated, the defined parameters in Table~\ref{tab:DefaultSetup} are used as our default setup.
 %$M:= M_{\text{t}} = M_{\text{r}} = M_{\text{d}} = 4$, $\kappa=\beta = -40~\text{[dB]}$, $P_{\text{max}} := P_{\text{s,max}} = P_{\text{r,max}} = 1$, $\sigma_{\text{n}}^2:= \sigma_{\text{nr}}^2 = \sigma_{\text{nd}}^2 = -40~\text{[dB]}$, $\rho_{\text{rr}} = 1$, $\rho_{\text{sr}} = \rho_{\text{rd}}= -30~\text{[dB]}$, $\rho_{\text{sd}}= -60~\text{[dB]}$, $K_R = 10$.

\subsection{Algorithm analysis} \label{sim_alg_analysis}  
In this subsection we study the behavior of the proposed iterative algorithms, i.e., GP and (Alt)MuStR1, in terms of the convergence speed, approximation accuracy, and computational complexity. Moreover, we study the gap of the proposed GP method with the optimality, with the help of an extensive numerical simulation.  

\subsubsection{Convergence}  
Both GP and AltMuStR1 operate based on iterative update of the variables, until a stable point is obtained. A study on the convergence behavior of these algorithms are necessary, in order to verify the algorithm function, and also as a measure of the required computational effort.

In Fig.~\ref{fig:ConvergenceComplexity}~(a) the average convergence behavior of the GP algorithm is depicted. At each iteration, the obtained performance is depicted as a percentage of the final performance after convergence, where $\text{SDNR}^\star$ is the $\text{SDNR}$ at the convergence. It is observed that the convergence speed differs, for different values of transceiver inaccuracy. This is expected, as a higher distortion results in the complication of the mathematical structure by signifying non-quadratic terms, see (\ref{eq_loop_vec_Q_2}), and requires the application of the projection procedure (Subsection~\ref{GP_Proj_Rule}) more often. It is observed from our simulations that the algorithm requires $10^2$ to $10^4$ number of iterations to converge, depending on the value of the distortion coefficients $\kappa,\beta$.

In Fig.~\ref{fig:ConvergenceComplexity}~(b) the average convergence behavior of the proposed AltMuStR1 is depicted. Note that unlike GP, AltMuStR1 operates based on the local increase of SDNR in the defined relaying segments, see Section~\ref{section:channel_norm_1}. Hence, a theoretical guarantee on the monotonic increase of the overall SDNR in each iteration is not available. Nevertheless, the numerical evaluation shows that the algorithm converges within much fewer number of iterations, with an effective increase in each step. Similar to GP, a higher value of $\kappa,\beta$ results in a slower convergence.  

\subsubsection{Computational complexity}
Other than the required number of iterations, the computational demand of an algorithm is impacted by the required per-iteration complexity. In Fig.~\ref{fig:ConvergenceComplexity}~(c), the required CPU time of the proposed algorithms are depicted as the number of antennas increase. The reported CPU time is obtained using an Intel Core i$5-3320$M processor with the clock rate of $2.6$ GHz and $8$ GB of random-access memory (RAM). As our software platform we have used MATLAB $2013$a, on a $64$-bit operating system. While the GP method is considered as the performance benchmark, the proposed (Alt)MuStR1 algorithms show a significant advantage to the GP algorithm in terms of computational complexity.

\subsubsection{Approximation accuracy} 
The proposed (Alt)MuStR1 algorithms are based on the used approximation (\ref{eq_ch_norm_coeffs_3}) and (\ref{eq_ch_norm_coeffs_4}). In this regard, the choice of the approximation order $K$ leads to a trade-off between algorithm complexity and the resulting performance. In Fig.~\ref{fig:ConvergenceComplexity}-(d) the exact, and approximated SDNR values are depicted for a pessimistic case of $\kappa = 0.1$. By repeating such experiments, we have decided on $K=5$ as a good balance between approximation accuracy and the resulting complexity.        

\subsubsection{GP optimality gap} \label{sim:GP_OptimalityGap}
Via the application of the gradient ascend the proposed GP converges with a monotonically increasing objective. Nevertheless, the global optimality of the resulting stationary point may not be guaranteed, due to the possibility of a local extrema. In Fig.~\ref{fig:ConvergenceComplexity}~(e) the performance of GP algorithm is evaluated with multiple random initializations, where the best converging point is considered as the algorithm solution. It is observed that the occurrence of the non-optimum solutions is more likely for higher values of $\kappa, \beta$, as a larger number of initial points results in a better performance. Nevertheless, it is observed that no significant performance improvement is observed by employing more than $C_1 = 10$ number of random initial points. We consider the obtained performance of the GP algorithm with $C_1 = 10$ as our performance benchmark for FD AF relaying hereinafter. 
%
%In this regard, we have initialized the algorithm with multiple feasible initializations, and observed the performance gap. where the best result is considered as point to observe the impact on the resulting optimum solution.   

\subsubsection{Rank profile}
\begin{figure*}[!t]  
\hspace{0.0cm} \subfigure[Convergence-GP]{\includegraphics[angle = 0, width = 0.666\columnwidth]{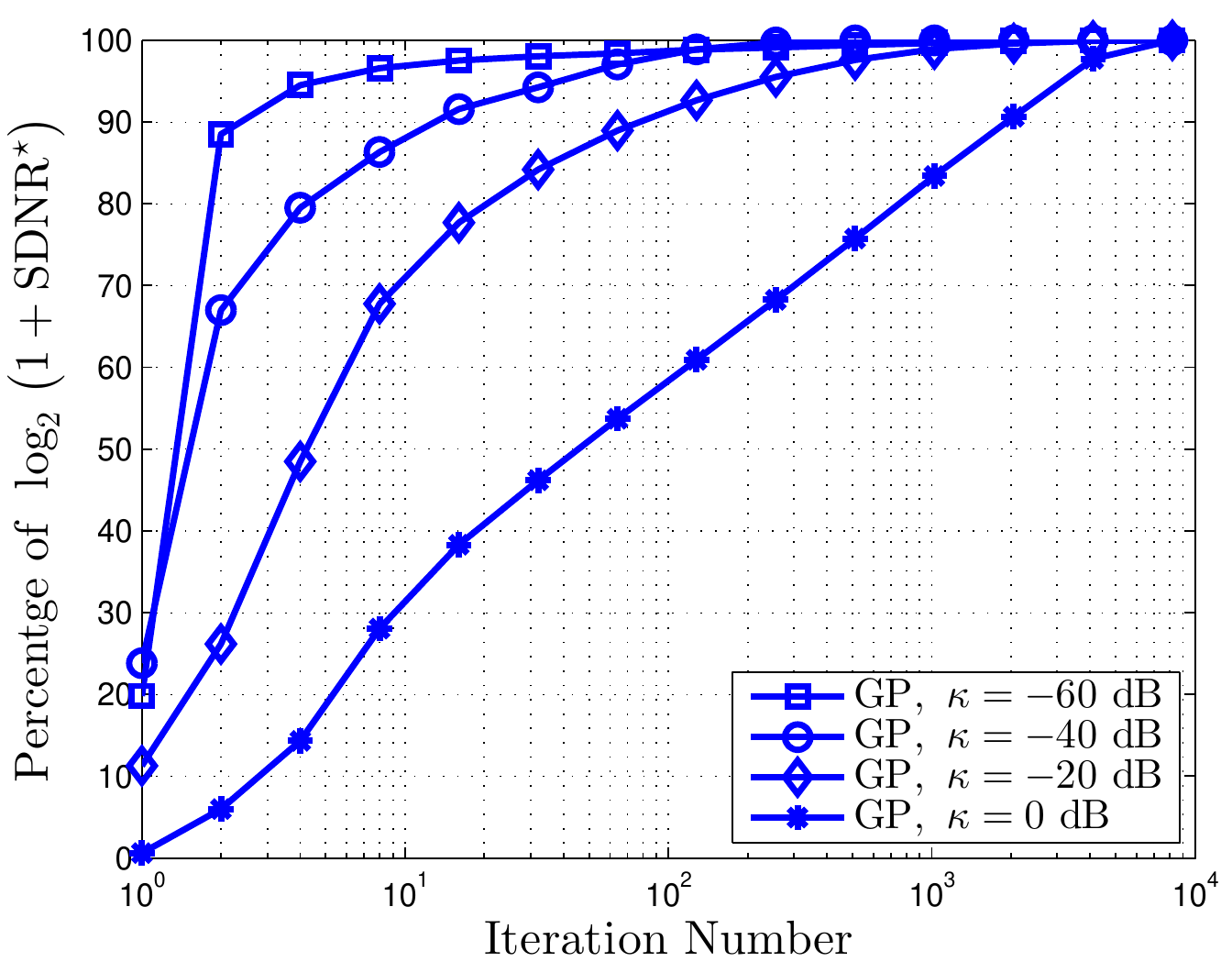}} \label{fig_conv1}
\subfigure[Convergence-AltMuStR1]{\includegraphics[width = 0.666\columnwidth]{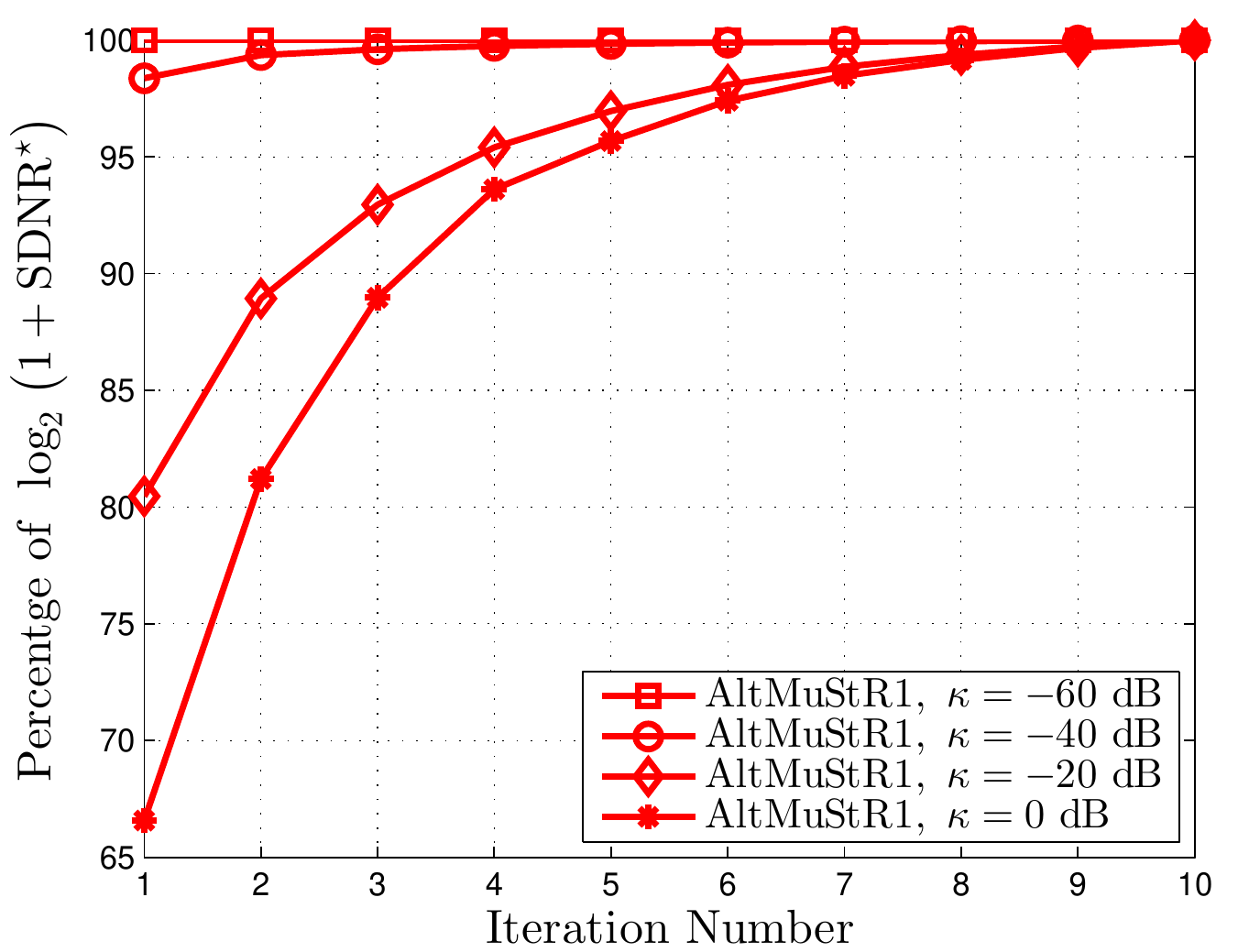}} \label{fig_conv2}
\subfigure[CPU~Time]{\includegraphics[width = 0.666\columnwidth]{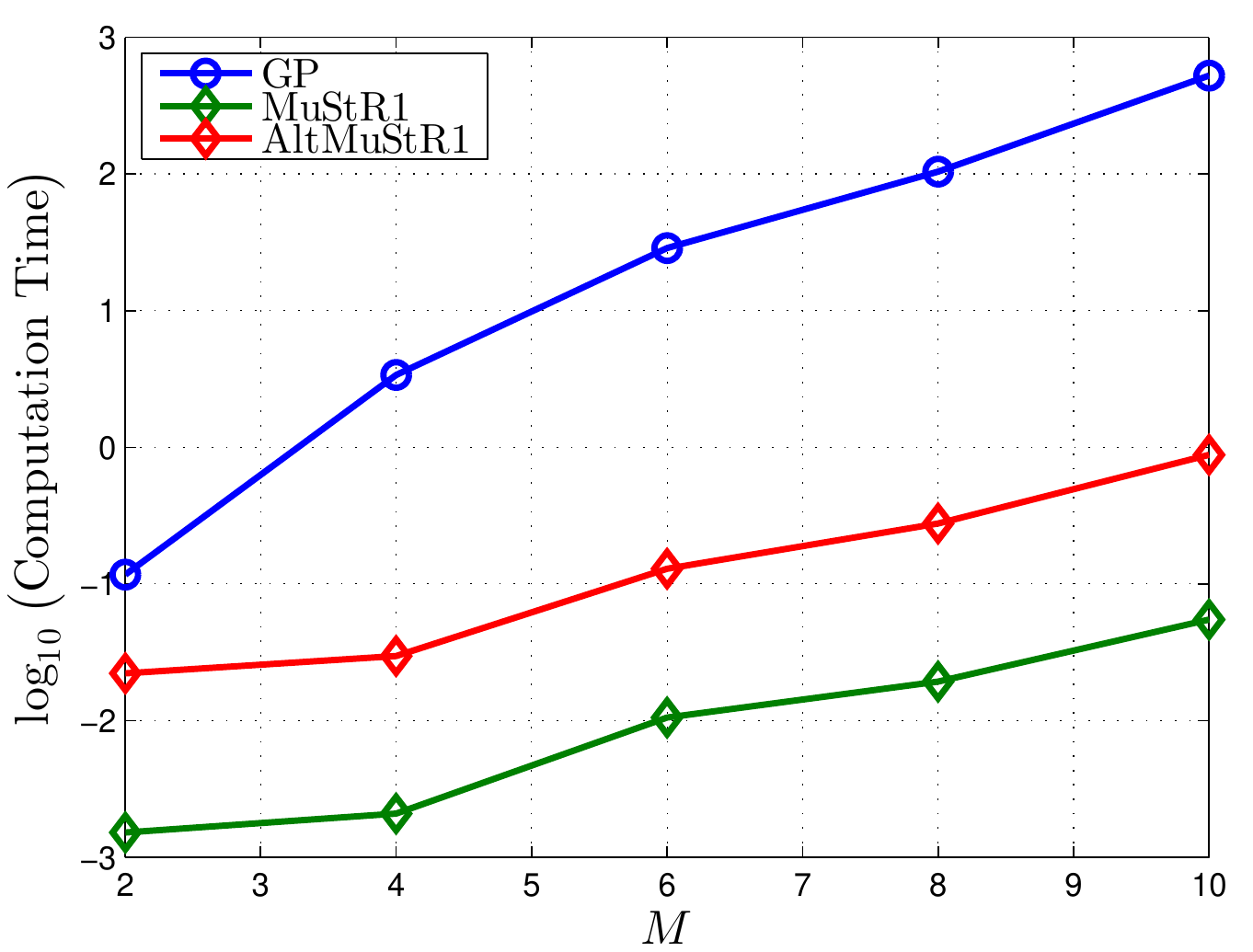}} \label{fig_conv3}
\hspace{0.0cm} \subfigure[SDNR Approximation]{\includegraphics[angle = 0, width = 0.666\columnwidth]{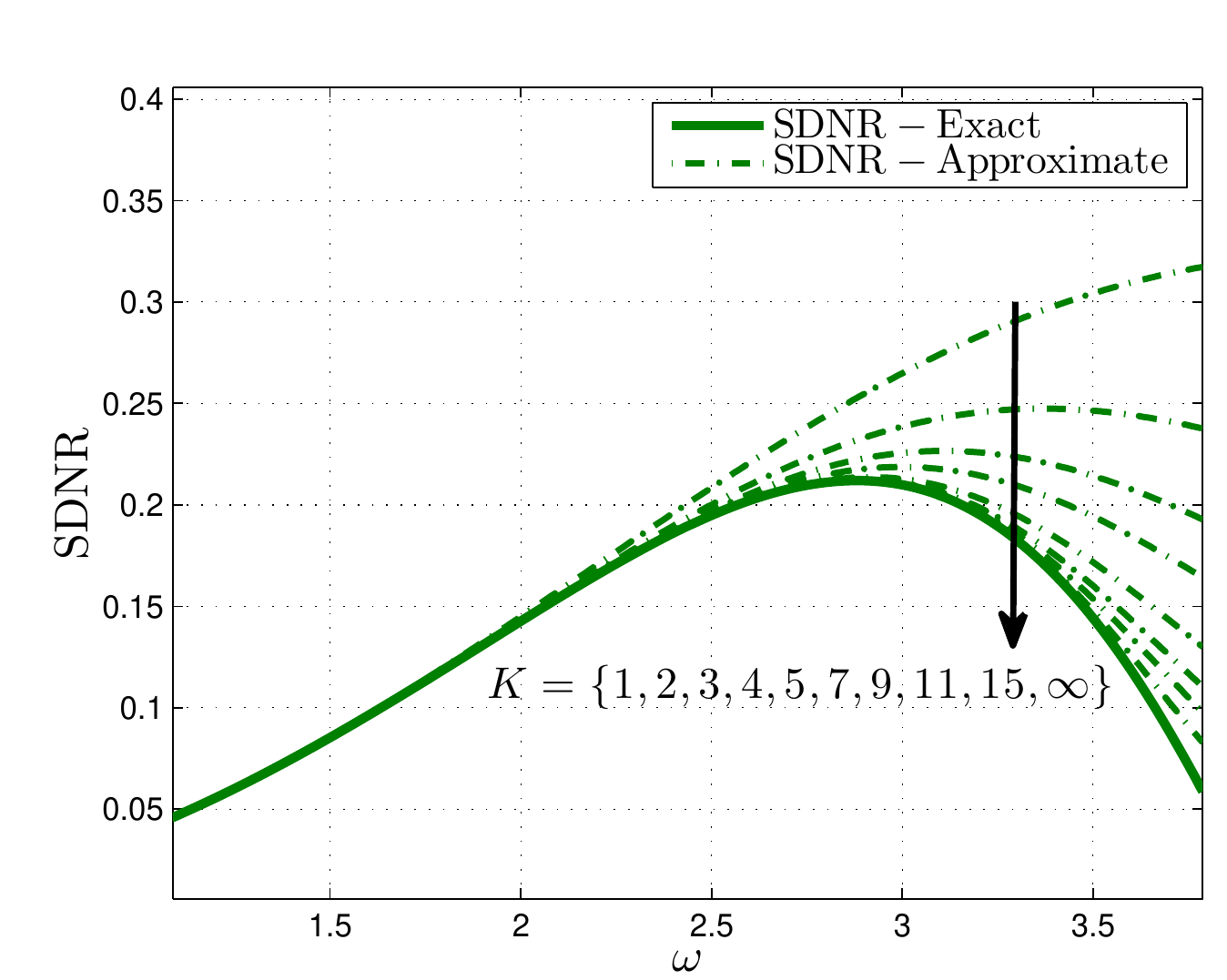}} \label{fig_conv1}
\hspace{0.0cm} \subfigure[GP-Initialization]{\includegraphics[width = 0.666\columnwidth]{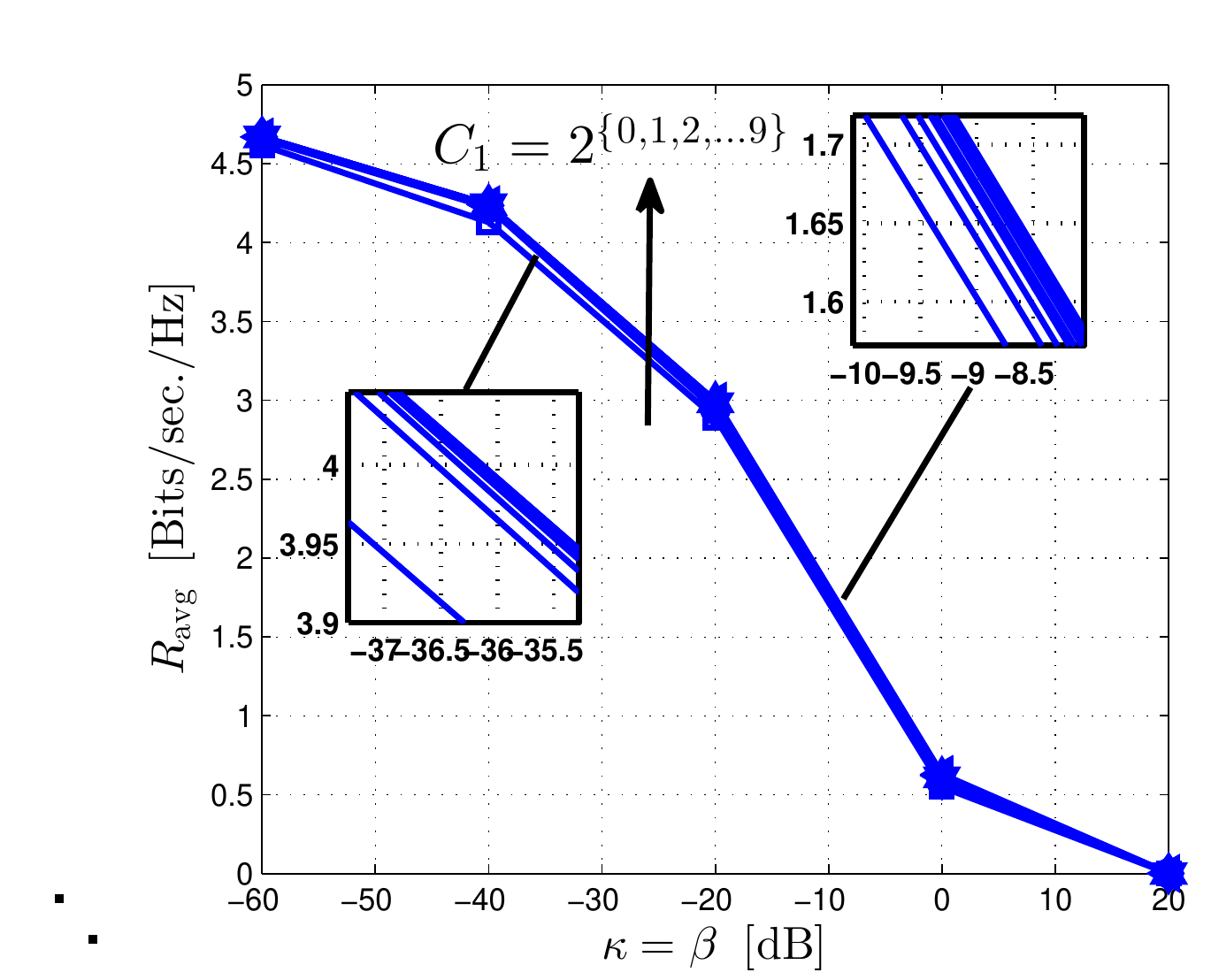}} \label{fig_conv2}
\hspace{0.2 cm}\subfigure[Singular energy distribution]{\includegraphics[width = 0.666\columnwidth]{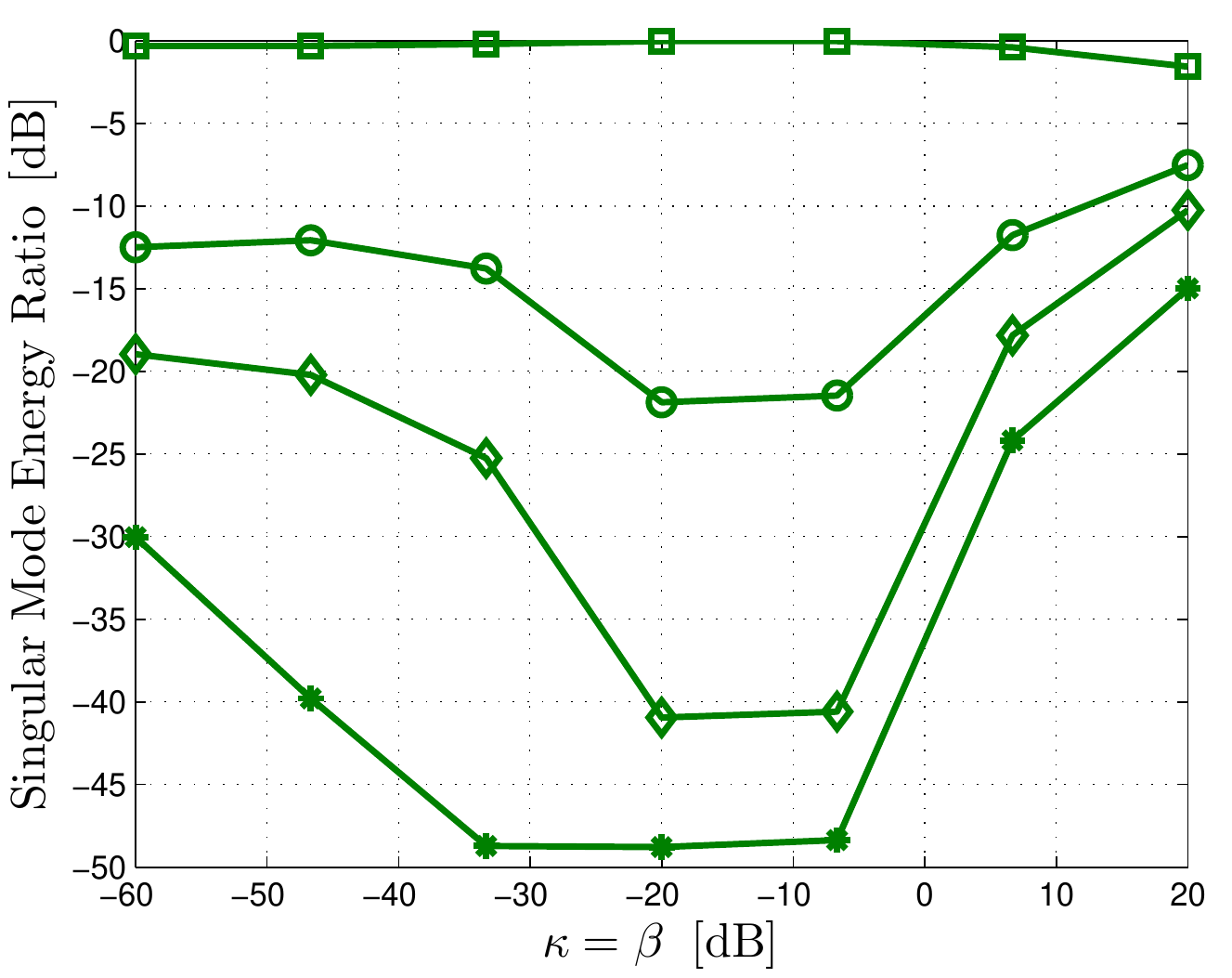}} \label{fig_conv3}
\caption{\revOmid{Numerical algorithm analysis, including the average convergence behavior of the proposed GP algorithm (a), the AltMuStR1 algorithm (b), a comparison on computational complexity~(c), accuracy of the SDNR approximation (d), impact of algorithm initialization on the GP method (e), and the singular mode energy profile for the obtained $\ma{W}$ from the GP method (f).}}
\label{fig:ConvergenceComplexity}
\vspace{-7pt}
\end{figure*}

In the proposed (Alt)MuStR1 method, a rank-1 relay amplification is assumed. Hence, it is interesting to observe how the solution to the GP method behaves in terms of the matrix rank. In Fig.~\ref{fig:ConvergenceComplexity}~(f) the energy distribution of the singular values of the obtained relay amplification from the GP method is depicted. It is observed that for most of the distortion conditions, the highest singular value holds almost all of the energy, indicating an approximately rank-1 property.

\begin{figure*}[!t]  
\hspace{-0.3cm}\subfigure[$R_{\text{avg}}$~vs.~Distortion]{\includegraphics[angle = 0, width = 0.666\columnwidth]{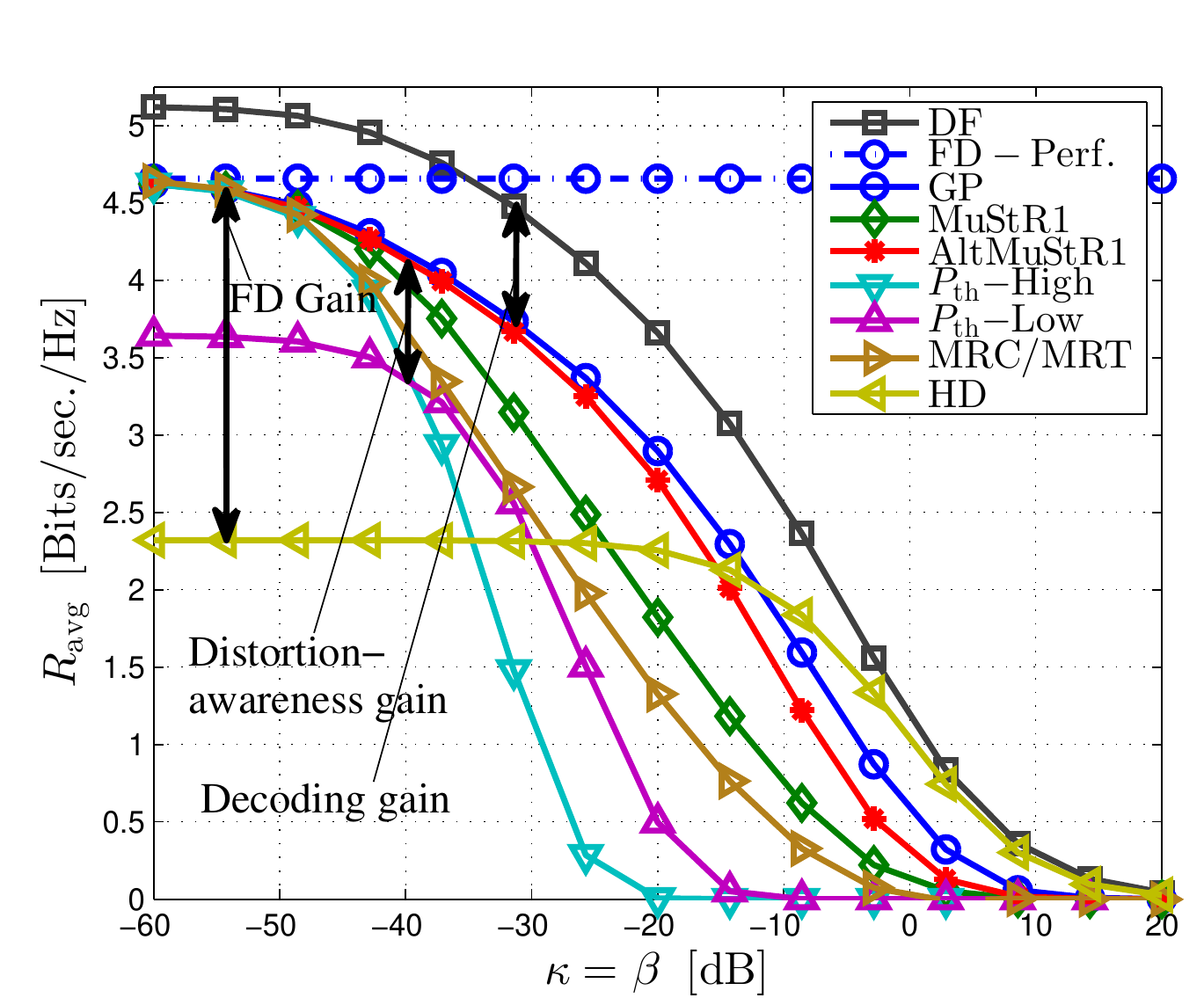}} \label{fig_conv1}
\hspace{-2.2cm}\subfigure[$R_{\text{avg}}$~vs.~Noise]{\includegraphics[width = 0.666\columnwidth]{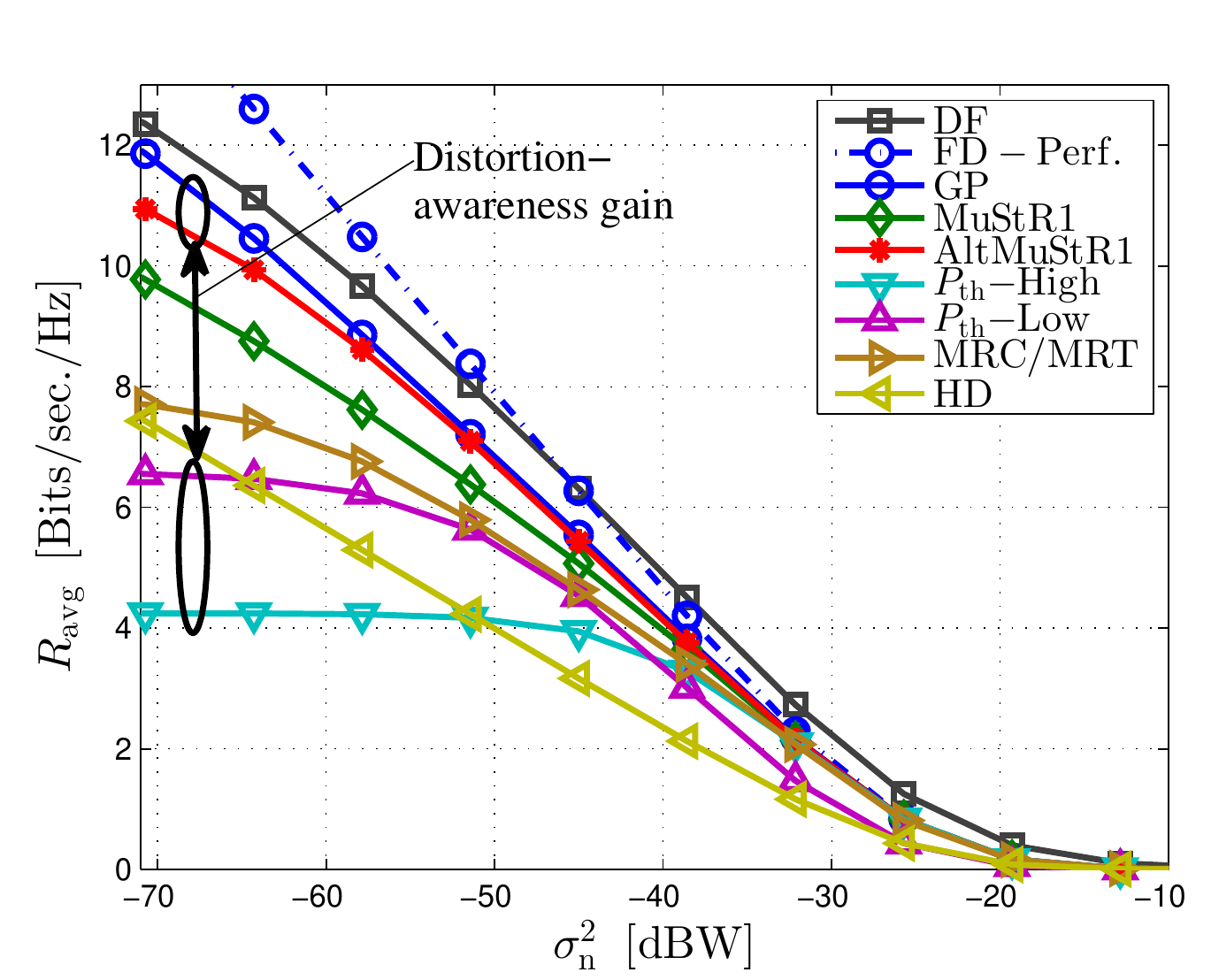}} \label{fig_conv2}
\hspace{-2.1cm}\subfigure[$R_{\text{avg}}$~vs.~Max.~Tx power]{\includegraphics[width = 0.666\columnwidth]{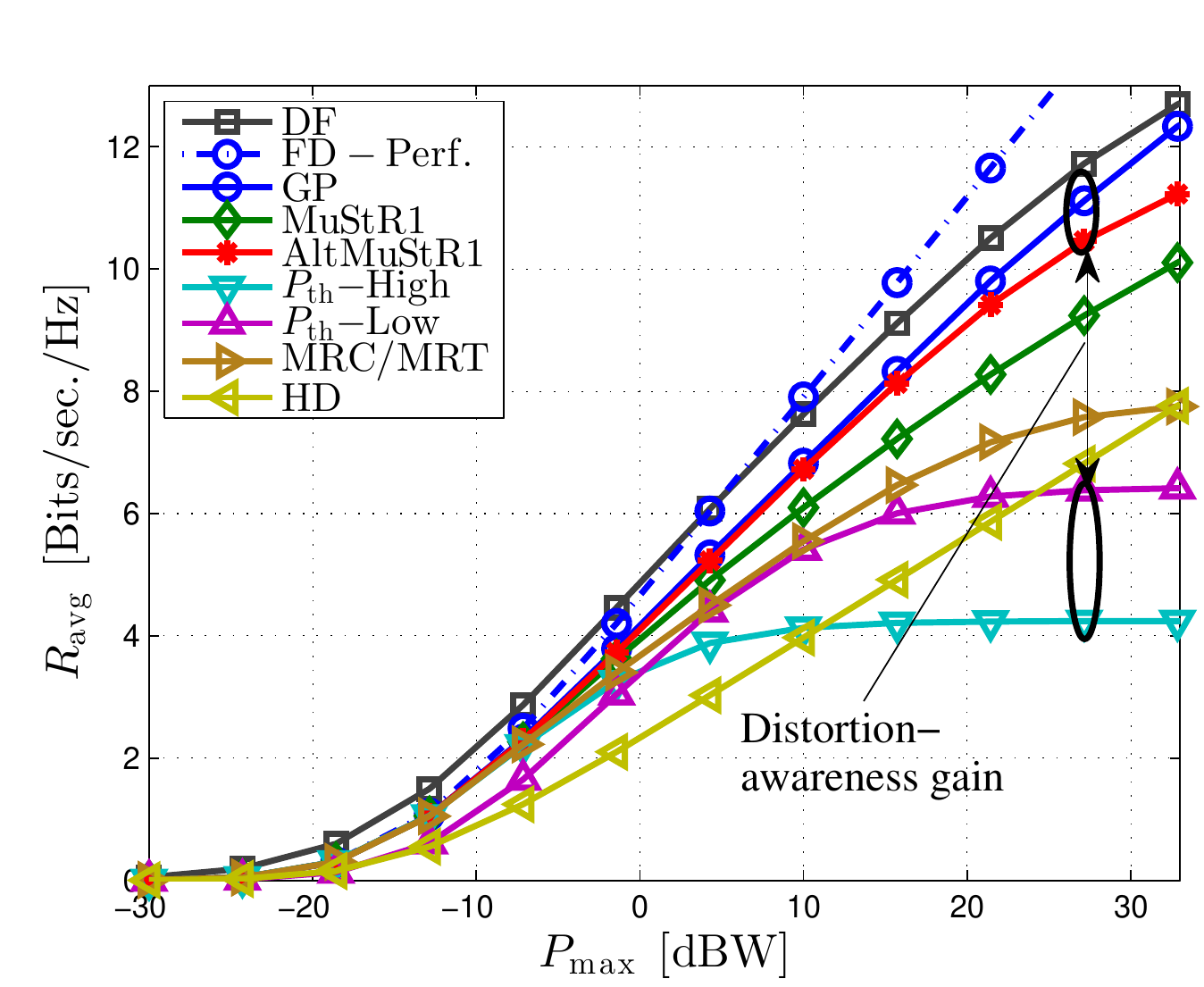}} \label{fig_conv3}
\subfigure[$R_{\text{avg}}$~vs.~$M$]{\includegraphics[angle = 0, width = 0.666\columnwidth]{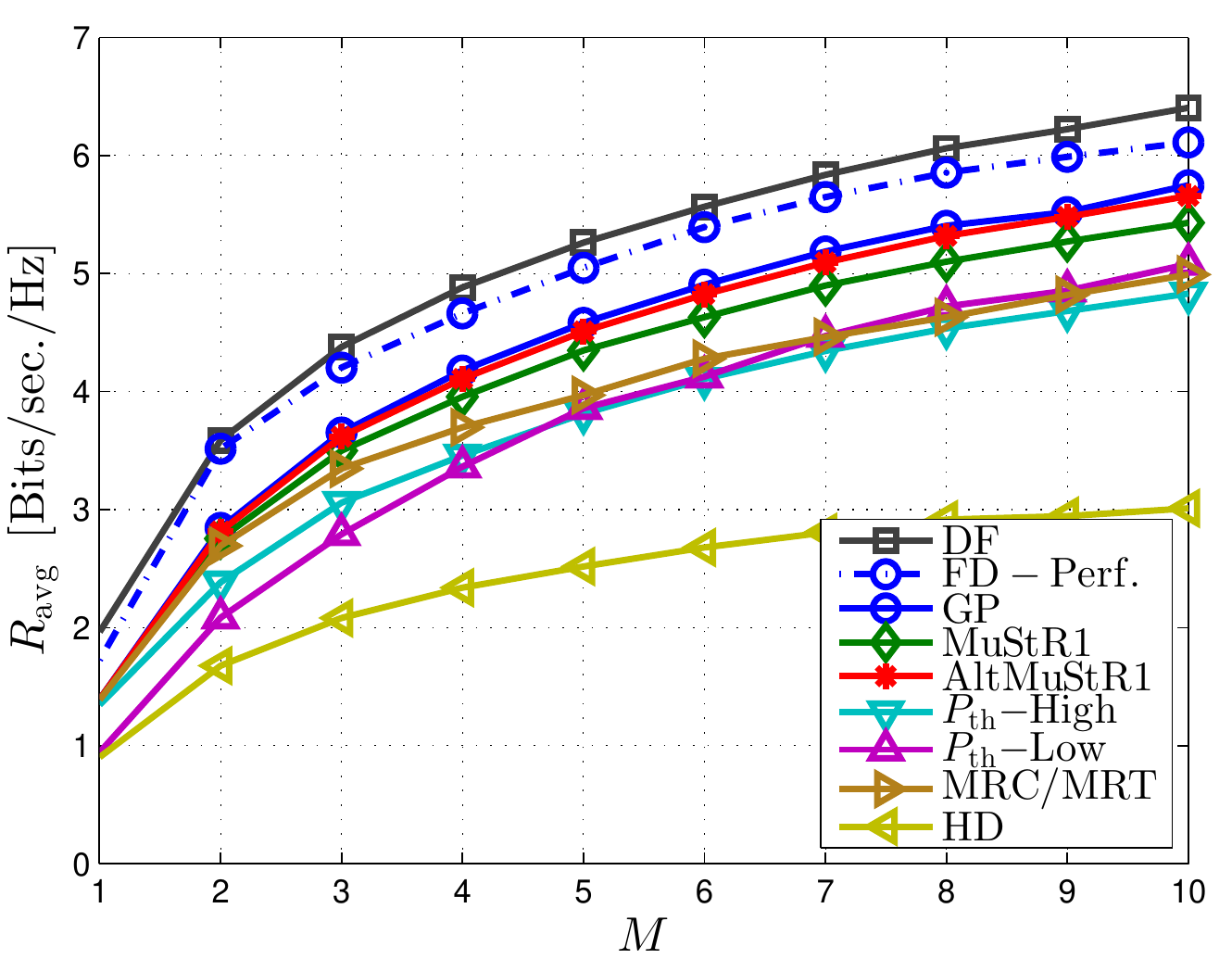}} \label{fig_conv1}
\subfigure[$R_{\text{avg}}$~vs.~Relay~position]{\includegraphics[width = 0.666\columnwidth]{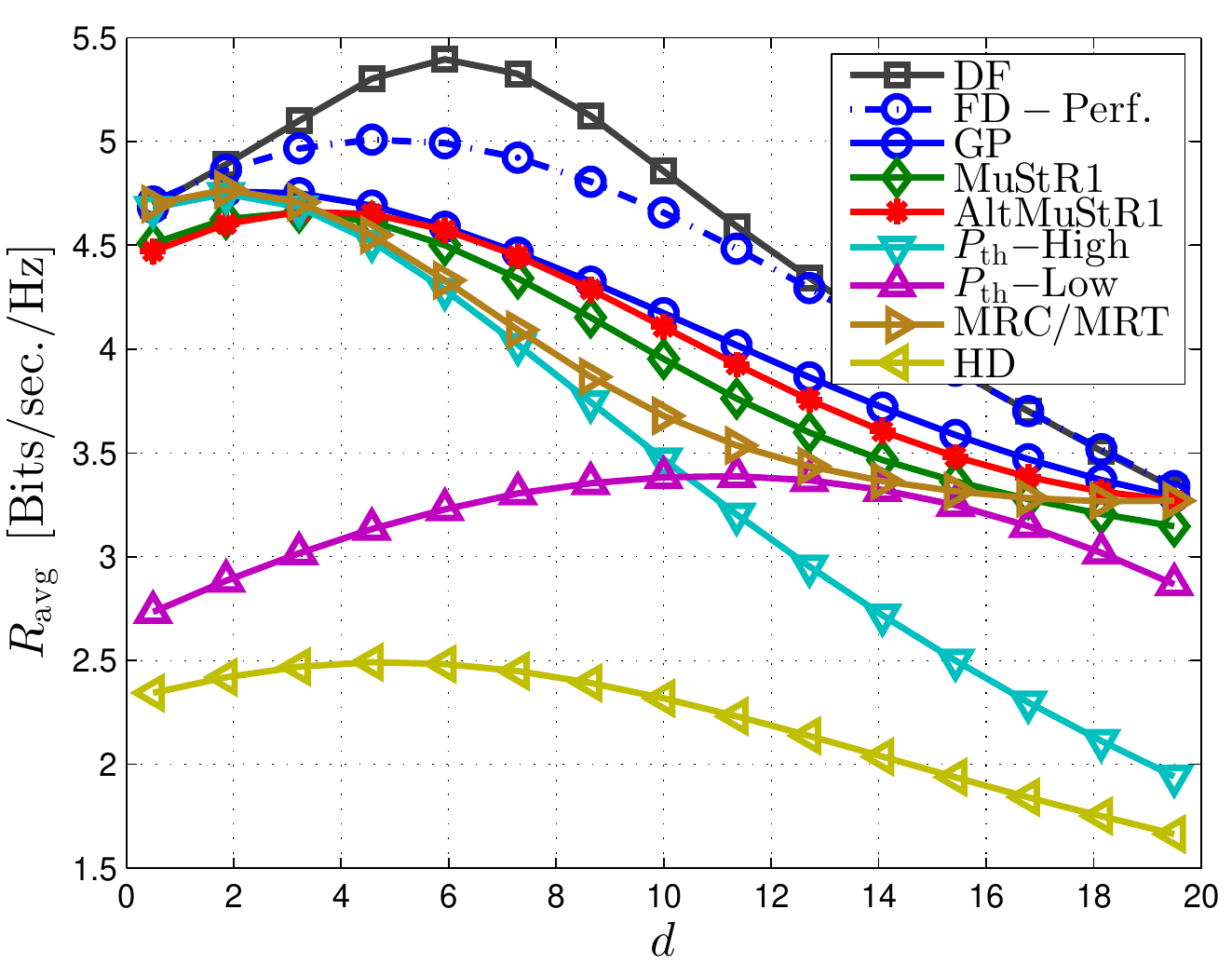}} \label{fig_conv2}
\hspace{0.5cm}\subfigure[$R_{\text{avg}}$~vs.~$\rho_{\text{rr}}$]{\includegraphics[width = 0.666\columnwidth]{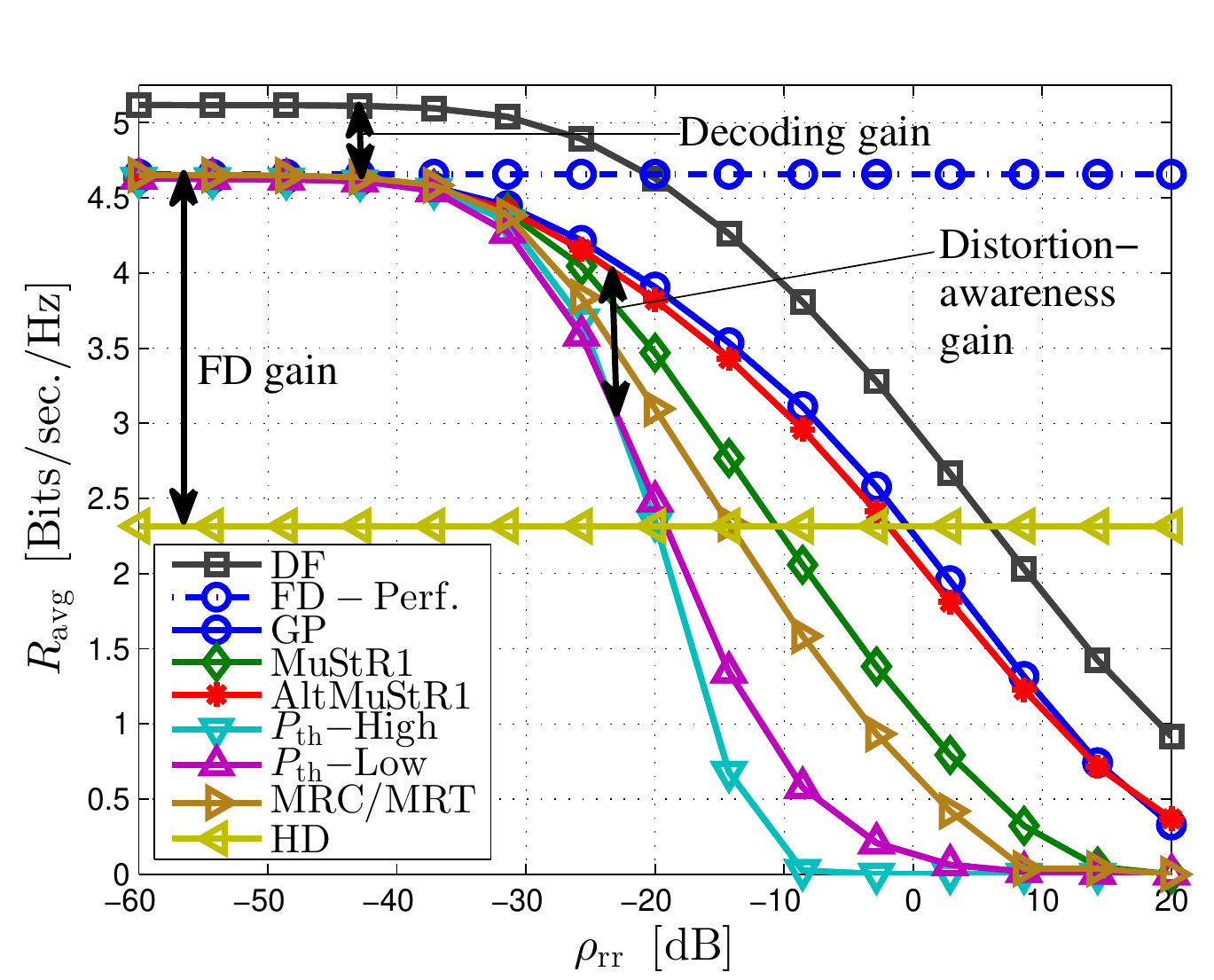}} \label{fig_conv3}
\caption{The comparison of average system performance $R_{\text{avg}}$ under different parameter ranges. }
\label{fig_visu}
\vspace{-7pt}
\end{figure*}

%
%\begin{figure}[!t] 

%%%%%%%%%%%%%%%%%%%%%%%%%%%%%%%%%%%%%%%%%%%%%%%%%%%%%%

\subsection{Performance comparison} \label{sim_performanceComparison}
In this part we evaluate the performance of the proposed designs, in comparison with the available designs in the literature. 
%In particular, we evaluate the performance of the obtained performance benchmark for FD AF relaying (GP), as well as the proposed methods in Section~\ref{section:channel_norm_1}, i.e., MuStR1 and AltMuStR1. Moreover, the performance of the equivalent HD relay system is also reported for comparison. 

\subsubsection{Available design approaches} We divide the relevant available literature on the FD AF relaying design into three main approaches. Firstly, as considered in \cite{KKMHPL:12, 4557197}, the SIC is purely relegated to the relay receiver end, via a combined time domain analog/digital cancellation techniques. The aforementioned approach imposes no design constraint on the self-interference power, i.e., $P_{\text{intf}} \leq \infty$, where $P_{\text{intf}}$ represents the self-interference power prior to analog/digital cancellation\footnote{This approach is equivalent to ignoring the impact of SIC in the beamforming design, as it has been usual in the earlier literature.}. Secondly, the SIC is purely done via transmit beamforming at the null space of the relay receive antennas, e.g., \cite{SKZYS:14, CP:12, ChunPark:12, SSWS:14, URW:15, 7558213}, hence imposing a zero interference power constraint for transmit beamforming design, i.e., $P_{\text{intf}} \leq 0$. Finally, as a generalization of the aforementioned extreme approaches, a combined transmit beamforming and analog/digital cancellation at the receiver is considered in \cite{Taghizadeh2016, KKC:14}. In the aforementioned case it is assumed that the received self-interference power should not exceed a certain threshold ($P_{\text{th}}$), i.e., $P_{\text{intf}} \leq P_{\text{th}}$. In all of the aforementioned cases, due to the perfect hardware assumptions, and upon imposition of the required self-interference power constraint, the SIC is assumed to be perfect. In our simulations, we evaluate the generalized approach in \cite{KKC:14,Taghizadeh2016} by once assuming a high self-interference power threshold, i.e., $P_{\text{th}}=P_{\text{r,max}}$, denoted as '$P_{\text{th}}$-High', and once assuming a low self-interference power threshold, i.e., $P_{\text{th}}=0.01 \times P_{\text{r,max}}$, denoted as '$P_{\text{th}}$-Low'\footnote{\revOmid{Note that the application of $P_{\text{th}}= 0 $, and $P_{\text{th}} = \infty$ is not feasible in our scenario.} This is since $P_{\text{th}}= 0$ strictly requires that $M_{\text{t}} > M_{\text{r}}$, and the $P_{\text{th}} = \infty$ often results in a non-stable relay function due to the impact of the distortion. Nevertheless, the chosen scenarios '$P_{\text{th}}$-High' and '$P_{\text{th}}$-Low' closely capture the nature of the aforementioned designs.}. \revOmid{Moreover, the proposed approach in \cite{7932472} is evaluated as a sub-optimal solution, where a power adjustment method is done at the relay, assuming a maximum ratio combining/transmission (MRC/MRT)\footnote{MRT corresponds to the utilization of the dominant eigenvector of $\ma{H}_{\text{rd}}$, when $M_{\text{d}} > 1$.}. The performance of an FD-AF relay with perfect hardware, i.e., $\kappa = 0$, is also illustrated as 'FD-Perf.'.} \par

\subsubsection{Decoding gain}
Other than the defined approaches for FD AF relaying, it is interesting to evaluate the impact of decoding in the studied system. This is since in a DF relay, the discussed distortion loop is significantly alleviated as the decoding process eliminates the inter-dependency of the received residual interference intensity to the relay transmit power. See Section~VI for the detailed implementation and performance evaluation of the DF process.

\subsubsection{\revOmid{Numerical results}}

In Figs.~\ref{fig_visu} (a)-(f) the average communication rate is evaluated under various system parameters.\par
In Fig.~\ref{fig_visu}~(a) the impact of the transceiver inaccuracy is depicted. It is observed that as $\kappa=\beta$ increase, the communication performance decreases for all methods. In this respect, the HD setup remains more robust against the hardware distortions, and outperforms the FD setup for large values of $\kappa=\beta$. This is since the strong self-interference channel, as the main cause of distortion, is not present for a HD setup. Relative to the benchmark performance for the FD-AF relaying (GP), a significant decoding gain is observed for the big values of $\kappa$, where the system performance is dominated by the impact of distortion loop, see Subsection~\ref{discussion:distortionLoop}. Moreover, it is observed that the proposed AltMuStR1 method performs close to the GP method for different values of $\kappa$. The performance of '$P_{\text{th}}$-High' reaches close to optimality for a small $\kappa$, where the '$P_{\text{th}}$-Low' reaches a relatively better performance as $\kappa$ increases. Nevertheless, both of the aforementioned methods degrade rapidly for higher values of $\kappa$. This is expected, as the impacts of hardware inaccuracies are not taken into account in the aforementioned approaches.   \par

In Fig.~\ref{fig_visu}~(b) and (c), the opposite impact of the thermal noise variance, $\sigma_{\text{n}}^2=\sigma_{\text{nr}}^2=\sigma_{\text{nd}}^2$, and the maximum transmit power, $P_{\text{max}} = P_{\text{s,max}} = P_{\text{r,max}}$, is observed on the average system performance. This is expected, as an increase (decrease) in $P_{\text{max}}$ ($\sigma_{\text{n}}^2$) increases the signal-to-noise ratio, while keeping the signal-to-distortion ratio intact. Furthermore, it is observed that in a low noise (high power) region the performance of the methods with perfect hardware assumptions saturate. This is since the role of hardware distortions become dominant for a high power or a low noise system.       \par

In Fig.~\ref{fig_visu}~(d) the resulting system performance is depicted with respect to the number of antennas. It is observed that the performance of all methods increase as the number of antennas increase. Moreover, the performance of the proposed (Alt)MuStR1 methods remain close to the benchmark GP performance. This is promising, considering the increasing computational complexity of the GP method as the number of antenna increases. \par

In Fig.~\ref{fig_visu}~(e) the impact of the relay position is observed. In this regard, it is assumed that the source is located with the distance $d_{\text{sr}}$ from the relay where the relay is located with the distance $d_{\text{rd}} = 20-d_{\text{sr}}$ from the source. The path loss values for each link is then obtained as $\rho_X = \frac{0.1}{d_X^2}$, $X\in \{\text{sr}, \text{rd} \}$. As expected, the decoding gain decreases when the relay is positioned close to the source or destination. Interestingly, the performance of the (Alt)MuStR1 methods are slightly dominated by '$P_{\text{th}}$-High', when relay is positioned very close to the source. The reason is that in such a situation the bottleneck shifts to the relay-destination path as the source-relay channel is very strong and is not degraded by the impact of distortion from the self-interference path. Nevertheless, the distortion awareness in (Alt)MuStR1 destructively limits the performance of the relay-destination path in order to avoid distortion on the relay receiver. It is worth mentioning that this mismatch does not appear for the GP method, since the final SDNR is considered as the optimization objective which remains relevant for any relay position.  \par

In Fig.~\ref{fig_visu}~(f) the impact of the self-interference channel intensity is depicted. It is observed that the performance of the FD relay operation, for all design methods, degrades as the $\rho_{\text{rr}}$ increase while the performance of the HD method is not changed. As expected, the performance of the methods with perfect hardware assumptions degrades faster compared to the proposed methods. Moreover, the performance of the proposed AltMuStR1 method remains close to that of GP, for different values of the self-interference channel intensity. \revOmid{It is observed that the MRC/MRT method suffers from a rapid degradation, when $\kappa$ or $\rho_{\text{rr}}$ increases, also see Fig.~\ref{fig_visu}~(a). This is expected, since the transmit/receive filters are designed with no consideration of the impact of distortion, e.g., the instantaneous CSI regarding the self-interference channel is not effectively used to control the impact of distortion.}

\revOmid{In Fig.~\ref{fig_kappabetatradeoff} the impact of the accuracy of transmit and receiver chains are studied, where $\kappa \text{[dB]} + \beta \text{[dB]}= \mathcal{A}_{\text{sum}}$, i.e., the sum-accuracy (in dB scale) is fixed. For instance, for an FD transceiver with massive antenna arrays where the utilization of analog cancelers is not feasible, and also the quantization bits are considered as costly resources, the value of $A_{\text{sum}}$ is related to the total number of quantization bits. The similar evaluation regarding the number of transmit/receive antennas is performed in Fig.~\ref{fig_MrMttradeoff}, where $M_{\text{t}} + M_{\text{r}} =  \mathcal{M}_{\text{sum}}$. It is observed that different available resources, i.e., $\mathcal{A}_{\text{sum}}$, $\mathcal{M}_{\text{sum}}$, result in different optimal allocations. However, as a general insight, it is observed that the performance is degraded when resources are concentrated only on transmit or receive side. Please note that similar approach can be used for evaluating different cost models for accuracy and antenna elements, relating different system setups, or SIC specifications\footnote{\revOmid{For instance, if the implemented SIC uses analog cancelers as proposed by \cite{Rice_1}, the total number of chains can be counted as $M_{\text{t}} + 2M_{\text{r}}$, or as $M_{\text{t}}M_{\text{r}} + M_{\text{t}} + M_{\text{r}}$, with the implementation in [6], c.f. [4]. In case of the antenna canceler proposed by [7], the total number of antennas are counted as $M_{\text{r}} + 2M_{\text{t}}$. If a setup with different number of transmit/receive chains is used, then the cost model for chain accuracy can be also modified as $\kappa M_{\text{t}} + \beta M_{\text{r}}  = A_{\text{sum}}$. Depending on the chosen cost model, different trade-off curves can be obtained.}}. 

In Table~\ref{tab_comparison} a compact comparison is made between the studied relaying strategies in terms of the algorithm computational complexity, processing complexity, as well as the resulting performance under different impairment conditions ($\kappa=\beta$). The percentage values indicate the performance improvement in the scale of the optimal FD-AF performance, compared to three common relaying schemes. This includes the HD-AF relaying, the FD-DF relaying, as well as the FD-AF relaying with simplified modeling, i.e., the best choice among '$P_{\text{th}}$-High' and '$P_{\text{th}}$-Low' schemes. As expected, the gain of a design with consideration of the distortion is invaluable for a large $\kappa$, and can be achieved with a reasonable computational complexity, via the utilization of the AltMuStR1 algorithm.}

\revOmid{
\begin{table*}[!t]  
	\renewcommand{\arraystretch}{0.8}
  \caption{Complexity-performance tradeoff for different relaying strategies.} \vspace{-5pt} \label{tab_comparison}
  \centering
  \response{\begin{tabular}[t]{||c|c|c|c|c|c|c||}
	\hline
   Alg. & Complexity (alg.) & Complexity (proc.) & \scriptsize{$\kappa=-60$ dB } & \scriptsize{$\kappa=-30$ dB }   & \scriptsize{$\kappa=-10$dB}  & Compared to \\
   \hline
	 GP & $ \mathcal{O} \left(  16/3 (\gamma_1 + \gamma_2) M^6 \right)  $& $2 M^2 - M$ & \begin{tabular}{@{}c@{}}-10\% \\ +49\% \\ 0.1 \%   \end{tabular} &  \begin{tabular}{@{}c@{}} -19\% \\ +38\% \\ +31 \% \end{tabular} & \begin{tabular}{@{}c@{}} -48\% \\ -15\% \\ +99 \%\end{tabular}  & \begin{tabular}{@{}c@{}} FD-DF\\ HD-AF \\ FD-Simple \end{tabular}\\
	 \hline 
	 AltMuStR1 & $\mathcal{O} \left((\gamma_3  35/3) M^3  \right)$& $3M$ & \begin{tabular}{@{}c@{}} -10\% \\ +49\% \\ 0.1 \% \end{tabular} & \begin{tabular}{@{}c@{}} -21\% \\ +36\% \\ +29 \% \end{tabular} & \begin{tabular}{@{}c@{}}-79\% \\ -38\% \\ +76 \% \end{tabular}  & \begin{tabular}{@{}c@{}} FD-DF \\ HD-AF \\  FD-Simple \end{tabular}\\
	\hline
  \end{tabular} }
\end{table*} 

\begin{figure}[!t] 
\begin{center}
        \includegraphics[angle=0,width=0.7\columnwidth]{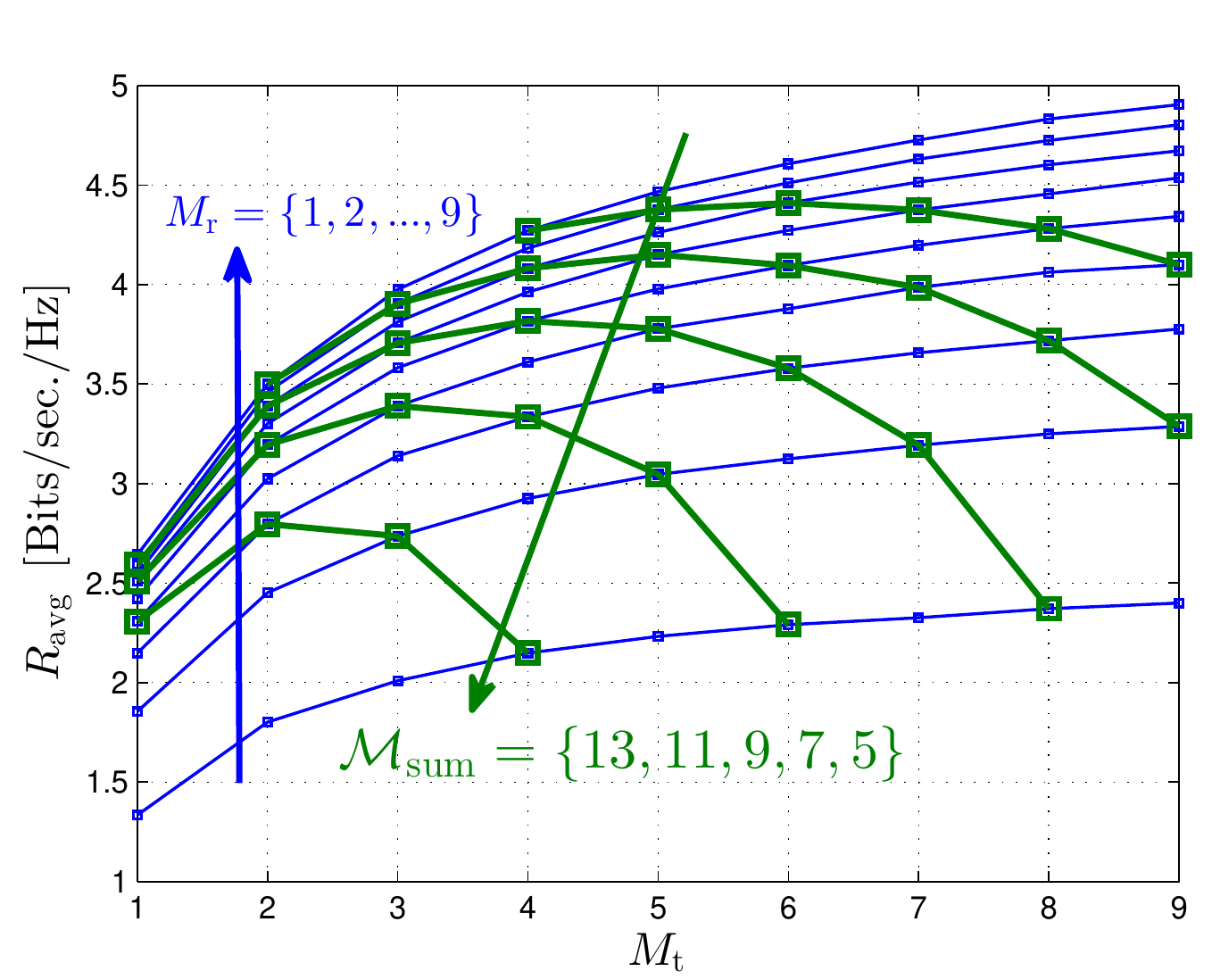}
				        %\fbox{model_rect6.pdf}
    \caption{\revOmid{Impact of the number of antennas, $M_{\text{t}} + M_{\text{r}} =  \mathcal{M}_{\text{sum}}$. $M_{\text{d}}=1$. } } \label{fig_MrMttradeoff}
    \end{center} \vspace{-4mm} 
\end{figure} 

\begin{figure}[!t] 
\begin{center}
        \hspace{5mm}  \includegraphics[angle=0,width=0.76\columnwidth]{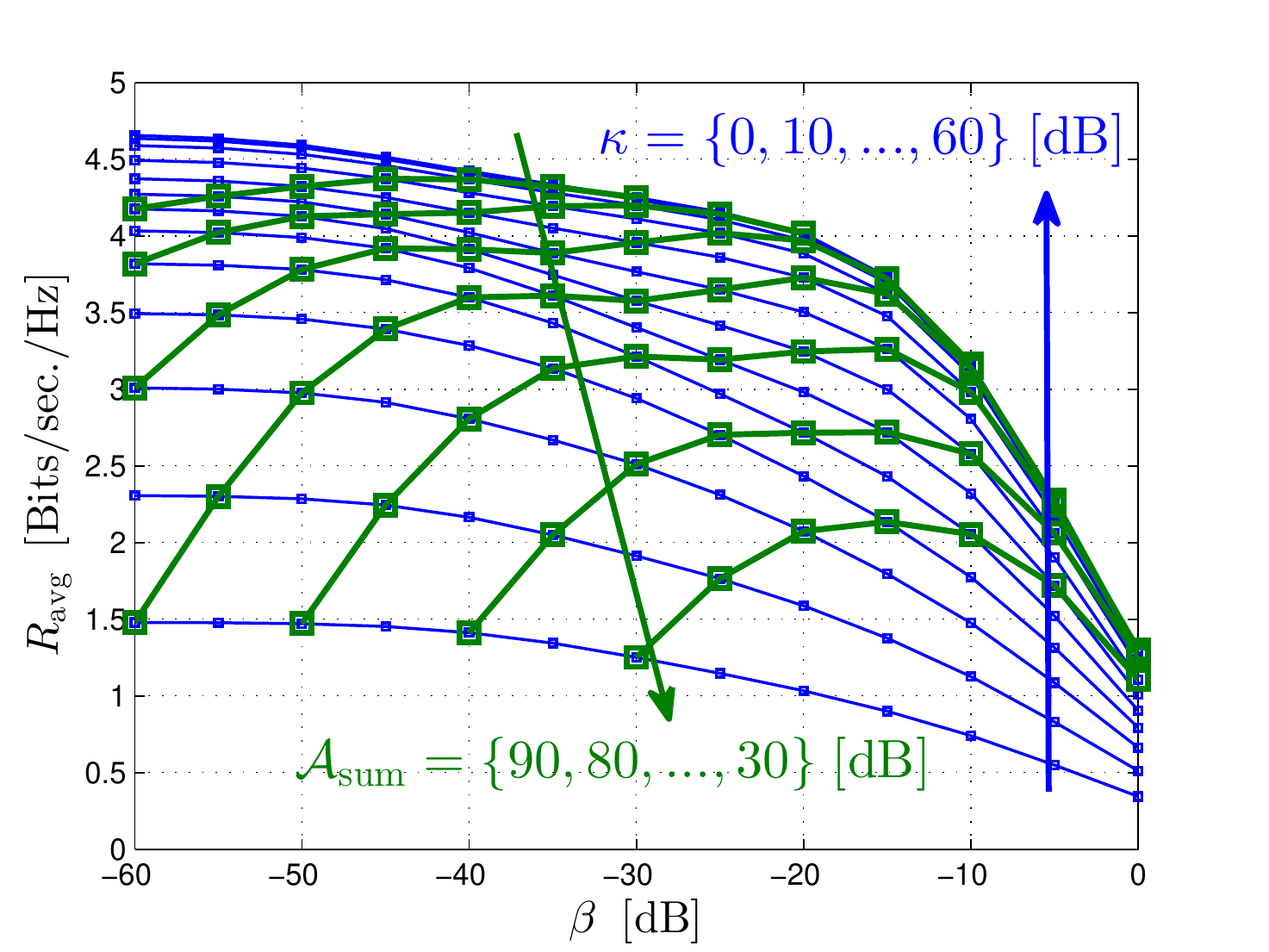}
				        %\fbox{model_rect6.pdf}
    \caption{\revOmid{Impact of the transmit/receive chain accuracy, $\kappa  + \beta = \mathcal{A}_{\text{sum}}$. } } \label{fig_kappabetatradeoff}
    \end{center} \vspace{-4mm} 
\end{figure}

}
 
\section{Conclusion}
The impact of hardware inaccuracies is of particular importance for an FD transceiver, due to the high strength of the self-interference channel. In particular, for an FD-AF relaying system, such impact is significant due to the inter-dependency of the relay transmit power, as well as the residual self-interference, which results in a distortion loop effect. In this work, we have analytically observed the aforementioned effect, and proposed optimization strategies to alleviate the resulting degradation. It is observed that the proposed GP algorithm can be considered as the performance benchmark, though, imposing a high computational complexity. On the other hand, the proposed (Alt)MuStR1 methods provide a significant reduction in the complexity, at the expense of a slightly lower performance. In particular, the comparison to the available schemes in the literature reveals that for a system with a small thermal noise variance, or a high power or transceiver inaccuracy, the application of a distortion-aware design is essential. \revOmid{Moreover, it is observed that an FD-DF relay is more robust against the increase of hardware distortions, compared to an FD-AF relay. This is expected, since the observed distortion loop for FD AF relays does not exist for an FD-DF relay, due to decoding.} 
%\input{main_conclusion}
%\section{Appendix}
%\input{main_Appendix}
\appendices
\section{Optimal $P_{\text{s}}$ as a function of $\ma{W}$} \label{appendix:PsLemmaProof}

Let $\ma{W}$ and $\ma{z}$ be the fixed (given) relay amplification and receive filter, respectively. From (\ref{eq_loop_A})-(\ref{eq_loop_P_error}), the SDNR at the destination can be written as a function of $P_{\text{s}}$
\begin{align} \label{app_ps_concavefunction}
\text{SDNR}(P_{\text{s}}) = \frac{\alpha_1 P_{\text{s}}}{\alpha_2 P_{\text{s}} + \alpha_3}, 
\end{align}    
where $\alpha_1,\alpha_2,\alpha_3 \in \real^+$, such that
\begin{align} \label{appendix_Ps_alphas_def}
\alpha_1 &:= \ma{z}^{{H}} \ma{H}_{{\text{rd}}} \ma{W} \ma{h}_{{\text {sr}}} \ma{h}_{{\text{sr}}}^{{H}} \ma{W}^{{H}} \ma{ H}_{{\text{rd}}}^{{H}} \ma{z},  \nonumber \\
\alpha_2 &:=  \ma{q}\left( \ma{W}, \ma{z} \right) \text{vec}\left(\ma{h}_{\text{sr}} \ma{h}_{\text{sr}}^H \right) +  \ma{z}^{{H}} \ma{h}_{\text{sd}} \ma{h}_{{\text{sd}}}^{H}  \ma{z} -\alpha_1 ,\nonumber \\
\alpha_3 &:=  \ma{z}^{{H}} \ma{z} \sigma_{\text{nd}}^2  + \ma{q}\left( \ma{W}, \ma{z} \right)\text{vec}\left(\sigma_{\text{nr}}^2 \ma{I}_{M_{\text{r}}} \right)  , \nonumber \\
\ma{q}\left( \ma{W}, \ma{z} \right) :&=  \left( \ma{z}^T \otimes \ma{z}^H \right) \left(\ma{H}_{\text{rd}}^{*} \otimes \ma{H}_{\text{rd}}  \right) {\ma{\Theta} \big(  \ma{W}, \ma{H}_{\text{rr}} , \kappa, \beta \big).}
 %\left( \ma{I}_{M_{\text{t}}^2} + \kappa\ma{S}_{\text{D}}^{M_{\text{t}}} \right) \nonumber \\ & \hspace{-12mm} \times  \Big( \ma{I}_{M_{\text{t}}^2} - \left( \ma{W}^{*} \otimes \ma{W} \right) \ma{C} \Big)^{-1}  \left( \ma{W}^{*} \otimes \ma{W} \right) \left(\ma{I}_{M_{\text{r}}^2} + \beta \ma{S}_{\text{D}}^{M_{\text{r}}}   \right).  
\end{align}    
It is observed, by taking the first and second order derivatives of the obtained function in (\ref{app_ps_concavefunction}), that SDNR is an increasing and concave function over $P_{\text{s}}$, which concludes the proof.

%\section{Minimum Euclidean distance projection} \label{appendix_EucklideanDistance}
%\input{main_appendic_min_Eucklidean_Dist}
\section{Equivalent Transmit Distortion Channel Expression} \label{appendix:DistortionChannels}
Via the application of $\ma{w}_{\text{tx}}$, the collective received distortion power due to the relay transmission, here denoted as $\theta_1$, is written as 
\begin{align} \label{eq_Tx_distortion_channel_calculation} 
{\theta_1} & = 
  P_{\text{r,max}} \kappa \text{tr}\bigg( \hspace{-0.5mm} \ma{H}_{\text{rr}} \text{diag} \left( \ma{w}_{\text{tx}}\ma{w}_{\text{tx}}^H\right) \ma{H}_{\text{rr}}^H \hspace{-0.5mm}  + \hspace{-0.5mm} \ma{H}_{\text{rd}} \text{diag} \left( \ma{w}_{\text{tx}}\ma{w}_{\text{tx}}^H \right) \ma{H}_{\text{rd}}^H  \hspace{-0.5mm} \bigg) \nonumber \\  
	&+  P_{\text{r,max}} \beta \text{tr}\bigg(   \text{diag} \left(\ma{H}_{\text{rr}} \ma{w}_{\text{tx}}\ma{w}_{\text{tx}}^H   \ma{H}_{\text{rr}}^H \right)   \bigg) \nonumber \\ 
	& = \sum_{ i \in \mathbb{F}_{ M_{\text{t} }} }  \sum_{ X \in \{\text{rr,~rd}\} }  P_{\text{r,max}} \kappa \text{tr}\big(   \ma{H}_X \ma{\Gamma}_i^{M_{\text{t}}} \ma{w}_{\text{tx}}\ma{w}_{\text{tx}}^H  {\ma{\Gamma}_i^{M_{\text{t}}}}^H   \ma{H}_X^H   \big) \nonumber \\  
	&+  \sum_{ i \in \mathbb{F}_{ M_{\text{r} }} }  P_{\text{r,max}} \beta \text{tr}\big( \ma{\Gamma}_i^{M_{\text{r}}}  \ma{H}_{\text{rr}} \ma{w}_{\text{tx}}\ma{w}_{\text{tx}}^H \ma{H}_{\text{rr}}^H  {\ma{\Gamma}_i^{M_{\text{t}}}}^H    \big) \nonumber \\ 
	& = \kappa P_{\text{r,max}} \sum_{i \in \mathbb{F}_{ M_{\text{t} }}}  \sum_{ X \in \{\text{rr,~rd}\} } \left\|  \ma{H}_X \ma{\Gamma}_i^{M_{\text{t}}} \ma{w}_{\text{tx}} \right\|_{2}^2 \nonumber \\ & + \beta P_{\text{r,max}} \sum_{i \in \mathbb{F}_{ M_{\text{t} }}} \left\|  \ma{\Gamma}_i^{M_{\text{r}}} \ma{H}_{\text{rr}}  \ma{w}_{\text{tx}} \right\|_{2}^2 \nonumber \\ 
%& = \left\| \left\lceil \sqrt{\gamma} \ma{W}_1 \ma{H}_{\text{rr}} \ma{\Gamma}_i \ma{Q}^{1/2} \; \sqrt{\beta} \ma{W}_1 \ma{\Gamma}_i \ma{H}_{\text{rr}}  \ma{Q}^{1/2}  \right\rceil_{i \in \{1 \cdots M_{\text{r}}\} } \right \|_{\text{fro}}^2
& = P_{\text{r,max}} \Bigg\| \left[   \begin{array}{c} 
\lfloor  \sqrt{\kappa} \ma{H}_X \ma{\Gamma}_i^{M_{\text{t}}} \ma{w}_{\text{tx}} \rfloor_{i \in \mathbb{F}_{ M_{\text{t} }},\; X \in \{\text{rr,~rd}\}}\\
\lfloor \sqrt{\beta } \ma{\Gamma}_i^{M_{\text{r}}} \ma{H}_{\text{rr}}  \ma{w}_{\text{tx}} \rfloor_{i \in \mathbb{F}_{ M_{\text{r} }} } \end{array} \right] \Bigg \|_{2}^2  \nonumber \\ 
& = P_{\text{r,max}} \Bigg\| \underbrace{ \left[   \begin{array}{c} 
\lfloor  \sqrt{\kappa } \ma{H}_X \ma{\Gamma}_i^{M_{\text{t}}} \rfloor_{i \in \mathbb{F}_{ M_{\text{t} }} ,\; X \in \{\text{rr,~rd}\}}\\
\lfloor \sqrt{\beta } \ma{\Gamma}_i^{M_{\text{r}}} \ma{H}_{\text{rr}}   \rfloor_{i \in \mathbb{F}_{ M_{\text{r} }} } \end{array} \right] }_{=: \ma{H}_{\text{D,tx}}} \ma{w}_{\text{tx}} \Bigg \|_{2}^2,  \nonumber \\ 
\end{align}
where $\ma{\Gamma}_i^M$ is an $M \times M$ all-zero matrix, except for the $i$-th diagonal element equal to one, and $\ma{H}_{\text{D,tx}}$ is viewed as the equivalent distortion channel. 
 %distortion channel and $\tilde{\theta}$ represents the distortion power included in the signal ${\tilde{\ma{r}}}_{\text{supp}}$. Note that ${\ma{H}_{\text{eq},2}}$ acts as the aforementioned equivalent distortion channel at the relay transmitter, and justifies the used expressions in (\ref{eq_ch_norm_W2_OP}) and (\ref{eq_ch_norm_H2}).
%\input{main_appendix_distortion_channel_Tx}
%\section{Re-structuring (\ref{eq_ch_norm_W1_OP}) and (\ref{eq_ch_norm_W2_OP}) as generalized Rayleigh Quotient problems} \label{appendix:GenRaylQutient}
%\input{main_appendix_generalized_Rayl_Qotient}

\section{Derivation of (\ref{eq_ch_norm_omega_equation_sinr})-(\ref{eq_ch_norm_omega_equation_power}) and the coefficients (\ref{eq_ch_norm_coeffs_1})-(\ref{eq_ch_norm_coeffs_4})} \label{appendix:coefficients}
The desired signal power at the destination prior to the application of $\ma{z}$, here denoted as $\theta_2$, can be calculated applying the known matrix equalities \cite[Eq. (486), (487), (496)]{matrixcookbook} as
\begin{align} \label{eq:proof_appendix_p_des}
%& \text{ derired received power at destination} \nonumber \\
\theta_2 &= P_{\text{s}} \text{tr} \left(\ma{H}_{\text{rd}} {\ma{W}} \ma{h}_{\text{sr}}  \ma{h}_{\text{sr}}^H {\ma{W}}^H \ma{H}_{\text{rd}}^H \right) \nonumber \\
& = {\omega}  P_{\text{s}} \text{tr} \left(\ma{H}_{\text{rd}} \ma{w}_{\text{tx}}\ma{w}_{\text{rx}}^H \ma{h}_{\text{sr}}  \ma{h}_{\text{sr}}^H \left(\ma{w}_{\text{tx}}\ma{w}_{\text{rx}}^H\right)^H \ma{H}_{\text{rd}}^H \right) \nonumber \\
& ={\omega}  P_{\text{s}} \ma{d}_{M_{\text{t}}}^T \Big(\left( \ma{H}_{\text{rd}}^* (\ma{w}_{\text{tx}}\ma{w}_{\text{rx}}^H)^* \right) \otimes \left( \ma{H}_{\text{rd}} \ma{w}_{\text{tx}}\ma{w}_{\text{rx}}^H \right) \Big) \text{vec}\left( \ma{h}_{\text{sr}}  \ma{h}_{\text{sr}}^H\right) \nonumber \\
& = {\omega}  P_{\text{s}} \ma{d}_{M_{\text{t}}}^T \left( \ma{H}_{\text{rd}}^* \otimes \ma{H}_{\text{rd}}  \right) \tilde{\ma{W}} \text{vec}\left( \ma{h}_{\text{sr}}  \ma{h}_{\text{sr}}^H\right) \nonumber \\
& = {\omega} a_{\text{d}},
\end{align} 
where $a_{\text{d}}$ and $\ma{d}_{M_{\text{t}}}$ are respectively defined in (\ref{eq_ch_norm_coeffs_1}) and immediately after (\ref{eq_ch_norm_coeffs_4}), and $\tilde{\ma{W}} := (\ma{w}_{\text{tx}}\ma{w}_{\text{rx}}^H)^{*} \otimes (\ma{w}_{\text{tx}}\ma{w}_{\text{rx}}^H)$. Similarly, following (\ref{eq_loop_vec_Q})-(\ref{eq_loop_a}) and the matrix identity \cite[Eq. (186)]{matrixcookbook} the noise+interference power at destination, here denoted as $\theta_3$, is calculated as
\begin{align}
\theta_3 
& \;\; = N - a_{\text{d}}{\omega}  + \text{tr} \left(\ma{H}_{\text{rd}} \mathbb{E} \{ \ma{r}_{{\rm out}} \ma{r}_{{\rm out}}^{{H}} \} \ma{H}_{\text{rd}}^H \right) \label{eq:appendix_coeff_noise_interference_1} \\
& \;\; = N  - a_{\text{d}}{\omega}  + \ma{d}_{M_{\text{d}}}^T \left(\ma{H}_{\text{rd}}^{*} \otimes \ma{H}_{\text{rd}}  \right) \text{vec} \left(  \mathbb{E} \{ \ma{r}_{{\rm out}} \ma{r}_{{\rm out}}^{{H}} \} \right) \nonumber  \\
& \;\; = N - a_{\text{d}}{\omega}  + \ma{d}_{M_{\text{d}}}^T \left(\ma{H}_{\text{rd}}^{*} \otimes \ma{H}_{\text{rd}}  \right) \left( \ma{I}_{M_{\text{t}}^2} + \kappa\ma{S}_{\text{D}}^{M_{\text{t}}} \right)  \nonumber \\ 
& \quad\quad\quad\quad\quad \quad\quad\quad\quad\quad\quad  \times \Big( \ma{I}_{M_{\text{t}}^2} - {\omega}\tilde{\ma{W}}  \ma{C} \Big)^{-1}  {\omega}\tilde{\ma{W}} \ma{c} \nonumber \\
& \;\;=  N - a_{\text{d}}{\omega}  + \ma{d}_{M_{\text{d}}}^T \left(\ma{H}_{\text{rd}}^{*} \otimes \ma{H}_{\text{rd}}  \right) \left( \ma{I}_{M_{\text{t}}^2} + \kappa\ma{S}_{\text{D}}^{M_{\text{t}}} \right) \nonumber \\
& \quad\quad\quad\quad\quad\quad\quad\quad\quad \quad\quad \times \sum_{k\in\{0\cdots\infty\}} \left({\omega} \tilde{\ma{W}}  \ma{C} \right)^k {\omega} \tilde{\ma{W}} \ma{c} \label{eq:appendix_coefficients_series_interference_excact} \\
& \;\;\approx N - a_{\text{d}}{\omega}  + \sum_{ k \in  \mathbb{F}_K }   \ma{d}_{M_{\text{d}}}^T \left(\ma{H}_{\text{rd}}^{*} \otimes \ma{H}_{\text{rd}}  \right) \left( \ma{I}_{M_{\text{t}}^2} + \kappa\ma{S}_{\text{D}}^{M_{\text{t}}} \right) \nonumber \\
& \quad\quad\quad\quad\quad \quad\quad\quad\quad\quad\quad\times  \left(\tilde{\ma{W}}  \ma{C} \right)^{k-1}  \tilde{\ma{W}} \ma{c} {\omega}^k \label{eq:appendix_coefficients_sqries_interference_approximation} \\
& \;\;\approx a_0 + \sum_{ k \in \mathbb{F}_K } a_k {\omega}^k,
\end{align} 
where $K$ represents the approximation order, $N:=\sigma_{\text{nd}}^2 M_{\text{d}} + P_{\text{s}} \|\ma{h}_{\text{sd}}\|_2^2$, and $a_k$ is defined in (\ref{eq_ch_norm_coeffs_2}) and (\ref{eq_ch_norm_coeffs_3}). Note that the identity in (\ref{eq:appendix_coefficients_series_interference_excact}) holds for any feasible relay transmit strategy, see (\ref{eq_loop_optimization_problem_b}). This stems from the fact that the effect of the distortion components are attenuated after passing through the loop process, i.e., $\omega \tilde{\ma{W}} \ma{C}$, in each consecutive symbol duration\footnote{Otherwise, the impact of distortion is accumulated in time, leading to an infinite distortion power and instability.}. Following the same arguments as in (\ref{eq:appendix_coeff_noise_interference_1})-(\ref{eq:appendix_coefficients_sqries_interference_approximation}) we calculate the relay transmit power as 
\begin{align}
& \text{tr} \left(\mathbb{E} \{ \ma{r}_{{\text{out}}} \ma{r}_{{\rm out}}^{{H}} \}  \right)  \\
& \;\; =  \ma{d}_{M_{\text{t}}}^T \left( \ma{I}_{M_{\text{t}}^2} + \kappa\ma{S}_{\text{D}}^{M_{\text{t}}} \right)  \Big( \ma{I}_{M_{\text{t}}^2} - {\omega}\tilde{\ma{W}}  \ma{C} \Big)^{-1}  {\omega}\tilde{\ma{W}} \ma{c} \nonumber \\
& \;\;= \ma{d}_{M_{\text{t}}}^T  \left(\ma{I}_{M_{\text{t}}^2} + \kappa\ma{S}_{\text{D}}^{M_{\text{t}}}\right)\sum_{k\in\{0\cdots\infty\}} \left({\omega} \tilde{\ma{W}}  \ma{C} \right)^k {\omega} \tilde{\ma{W}} \ma{c} \\
& \;\;\approx   \sum_{k \in \mathbb{F}_K} \ma{d}_{M_{\text{t}}}^T \left(\ma{I}_{M_{\text{t}}^2} + \kappa\ma{S}_{\text{D}}^{M_{\text{t}}} \right) \left(\tilde{\ma{W}}  \ma{C} \right)^{k-1}  \tilde{\ma{W}} \ma{c} {\omega}^k \\
& \;\;\approx b_0 +  \sum_{k\in \mathbb{F}_K} b_k {\omega}^k,
\end{align} 
where $b_k$ is defined in (\ref{eq_ch_norm_coeffs_4}). 
%\section{Derivation of (\ref{eq_ch_norm_omega_equation_sinr_alternating}) and the coefficients (\ref{eq_ch_norm_coeffs_5})-(\ref{eq_ch_norm_coeffs_7})} \label{appendix:coefficients_alternating}
%\input{main_appendix_coefficients_calculate_alternating}
\section*{Acknowledgment}
The authors would like to thank B.Sc.~Tianyu~Yang, for his assistance in numerical implementations, Subsection~\ref{sim_performanceComparison}. We would like to thank him for his efforts and commitment.

{

}

% that's all folks

\end{document}